\pdfoutput=1
\documentclass[11pt,a4paper,twoside,openright]{report}
\usepackage{amsmath,amssymb,color,graphicx}
\usepackage[figuresright]{rotating}
\usepackage{tikz}

\newcommand{\be}{\begin{equation}}
\newcommand{\ee}{\end{equation}}
\newcommand{\bea}{\begin{eqnarray}}
\newcommand{\eea}{\end{eqnarray}}
\newcommand{\ba}[1]{\begin{array}{#1}}
\newcommand{\ea}{\end{array}}
\newcommand{\im}{\text{Im}}
\newcommand{\re}{\text{Re}}

\makeatletter
\renewcommand{\@makecaption}[2]{
   \vskip\abovecaptionskip
   \sbox\@tempboxa{#1. #2}%
   \ifdim \wd\@tempboxa >\hsize
      \hspace*{1cm}\begin{minipage}{14cm}{\bf #1.} #2\end{minipage}
   \else
     \global \@minipagefalse
     \hb@xt@\hsize{\hfil\box\@tempboxa\hfil}%
   \fi
   \vskip\belowcaptionskip}
\makeatother

\begin{document}


\noindent\includegraphics[scale=0.5]{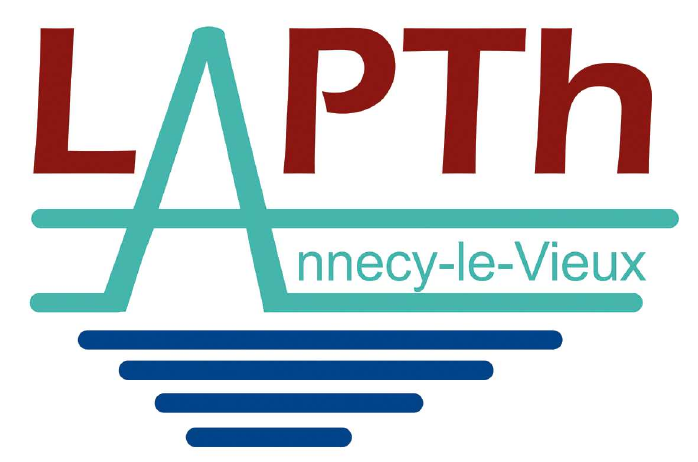}\hfill
\raisebox{2mm}{\includegraphics[scale=2.]{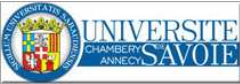}}

\fontfamily{phv}\selectfont
\begin{center}

\vspace*{2cm}
{\fontfamily{phv}\fontseries{b}\fontshape{n}\fontsize{15 pt}{5pt}\selectfont
Quantum integrable systems. Quantitative methods in biology\\[1cm]
Syst\`emes int\'egrables quantiques. M\'ethodes quantitatives en biologie
}

\vspace{2cm}
M\'emoire d'habilitation \`a diriger les recherches
pr\'esent\'e par

\vspace{2cm}
{\LARGE Giovanni Feverati}

\vspace{2cm}
Universit\'e de Savoie

\vspace{2cm}
13 D\'ecembre 2010\\
14H00
\end{center}

\vspace{2cm}

\noindent
Membres du jury:\\[2mm]

\begin{tabular}{lll}
Jean Avan & LPTM, CNRS/Universit\'e de Cergy-Pontoise & rapporteur\\
Michele Caselle & Dip. fisica teorica, Universit\`a di Torino & rapporteur\\
Luc Frappat & LAPTH, Universit\'e de Savoie &pr\'esident\\
Francesco Ravanini &Dip. fisica, Universit\`a di Bologna& rapporteur\\
Jean-Marc Victor &LPTMC, CNRS/Universit\'e de Paris VI \\
Laurent Vuillon &LAMA, Universit\'e de Savoie
\end{tabular}

\thispagestyle{empty}
\fontfamily{ptm}\selectfont

\newpage\mbox{}\thispagestyle{empty}

\newpage
\vspace*{1cm}
\paragraph{Abstract.}
Quantum integrable systems have very strong mathematical properties that allow an exact description of 
their energetic spectrum. From the Bethe equations, I formulate the Baxter ``T-Q'' relation, that is the
starting point of two complementary approaches based on nonlinear integral equations. The first one is 
known as thermodynamic Bethe ansatz, the second one as  Kl\"umper-Batchelor-Pearce-Destri-de Vega. 
I show the steps 
toward the derivation of the equations for some of the models concerned. I study the infrared and 
ultraviolet limits and discuss the numerical approach. Higher rank integrals of motion can be obtained, 
so gaining some control on the eigenvectors. After, I discuss the Hubbard model in relation to the 
$N=4$ supersymmetric gauge theory. The Hubbard model describes hopping electrons on a lattice. 

\mbox{~~}

In the second part, I present an evolutionary model based on Turing machines. The goal is to describe 
aspects of the real biological evolution, or Darwinism, by letting evolve populations of algorithms. 
Particularly, with this model one can study the mutual transformation of coding/non coding parts in 
a genome or the presence of an error threshold.

\mbox{~~}

The assembly of oligomeric proteins is an important phenomenon which interests the majority 
of proteins in a cell. I participated to the creation of the project ``Gemini'' which has for 
purpose the 
investigation of the structural data of the interfaces of such proteins. The objective is to 
differentiate the role of amino acids and determine the presence of patterns characterizing 
certain geometries.

\vspace{2cm}
\paragraph{R\'esum\'e.}
Les syst\`emes int\'egrables quantiques ont des propri\'et\'es math\'ematiques qui permettent la 
d\'etermination exacte de leur spectre \'energ\'etique. A partir des \'equations de Bethe, je pr\'esente 
la relation de Baxter \guillemotleft T-Q\guillemotright. Celle-ci est \`a l'origine des deux 
approches que j'ai 
prioritairement employ\'e dans mes recherches, les deux bas\'es sur des \'equations int\'egrales non 
lin\'eaires, celui de l'ansatz de Bethe thermodynamique et celui des \'equations de 
Kl\"umper-Batchelor-Pearce-Destri-de Vega. Je montre le chemin qui permet de d\'eriver les \'equations \`a partir 
de certain mod\`eles sur r\'eseau. J'\'evalue les limites infrarouge et ultraviolet et je discute 
l'approche num\'erique. D'autres constantes de mouvement peuvent \^etre \'etablies, ce qui permet 
un certain contr\^ole sur les vecteurs propres. Enfin, le mod\`ele d'Hubbard, qui d\'ecrit des 
\'electrons
interagissants sur un r\'eseau, est pr\'esent\'e en relation \`a la th\'eorie de jauge 
supersym\'etrique $N=4$.

Dans la deuxi\`eme partie, je pr\'esente un mod\`ele d'\'evolution darwinienne bas\'e sur les machines 
de Turing. En faisant \'evoluer une population d'algorithmes, je peut d\'ecrire certains aspects de 
l'\'evolution biologique, notamment la transformation entre parties codantes et non-codantes dans un 
g\'enome ou la pr\'esence d'un seuil d'erreur. 

L'assemblage des prot\'eines oligom\'eriques est un aspect important qui int\'eresse la majorit\'e 
des prot\'eines dans une cellule. Le projet \guillemotleft Gemini\guillemotright\ que j'ai 
contribu\'e \`a cr\'eer a pour finalit\'e d'explorer les donn\'es structuraux des interfaces des 
dites prot\'eines pour diff\'erentier le r\^ole des acides amin\'es et d\'eterminer la pr\'esence 
de patterns typiques de certaines g\'eom\'etries.
\newpage\mbox{}\thispagestyle{empty}

\tableofcontents
\addcontentsline{toc}{part}{Preface}

\chapter*{Preface}

The text that I present in the next pages aims at giving some flavour of the researches I have 
carried on after my degree in physics, obtained in 1995. I tried to give to these notes the style 
of a comprehensible presentation of the ideas that have animated my researches, with emphasis  
on the unity of the development. The single steps are here presented in a correlated view. 
The calculation details are usually available on my original papers, therefore they have been 
omitted here considering space and time constraints too.

For many years, I have been interested in quantum integrable systems. They are physical models
with very special properties that allow to evaluate observable quantities with exact calculations.
Indeed, exact calculations are seldom possible in theoretical physics. For this reason,
it is instructive to be able to perform exact evaluations in some specific model.
B.~Sutherland entitles ``Beautiful models'' his book \cite{suth} to express the elegant physical
and mathematical properties of integrable models.
Thus, in the Introduction to Part I, I define
and present, in a few examples, a number of basic properties of quantum integrable systems. 
These examples will be used in the following three chapters. I will describe the work 
I did on Destri-de Vega equations, in the first chapter, on the thermodynamic Bethe ansatz in 
the second one, on the Hubbard model in the third one. 
In each chapter I also give one or few proposals for the future, to show that the respective 
field is an active domain of research.  
Of course, I had to make a choice of the subjects I presented and, forcedly, others were excluded 
to keep the text into a readable size. 
Particularly, I regret I could mention very little of the physical combinatorics of TBA
quasi-particles, work that I have carried on with P. Pearce and that would have required several 
additional pages.

Around 2006, I started to follow lectures and seminars delivered by people that, coming 
from a theoretical 
physicist background, were starting to work on genome, proteins and cells. 
Two colleagues of my laboratory, L. Frappat and P. Sorba, were working on a quantum group 
model for the genetic code. I was curious: how can someone even think to apply quantum groups or
let say integrable systems, to the genetic code? Now I know that beyond the application of 
the apparatus of theoretical physics to biology, it is important to find the new
ideas, the new equations, the new models that are needed to better capture the properties of biological 
systems. In the Part II of this text I will clarify this attitude, especially with the
motivations at page~\pageref{introbio}.
After a series of lectures by M. Caselle, I started to experiment with an evolutionary model
based on Turing machines. The model, created by a colleague of mine, F. Musso, and myself,
will be presented in Chapter~\ref{c:turing}.
Near the end of 2007, Paul Sorba was contacted by a biologist interested in finding 
theoretical physicists for collaboration. This was an unusual request so Paul organized a meeting
with C. Lesieur, to listen to her researches and projects. I immediately accepted to 
participate and the team ``Gemini'' was created. A few months later a regular collaboration
was on,  
especially after my primitive but successful attempts to use the art of computer programming 
to search for the protein interfaces. 
The two projects on biophysics are now my main research activities, and the time I dedicate to
integrable systems has been considerably reduced. 

I think the changes I made in my activities reflect more than a personal event and highlight
the new horizons theoretical physics is called to explore.


\newcommand{\D}{\mbox{\boldmath $D$}}
\newcommand{\xilatt}{\xi_{\text{latt}}}
\newcommand{\DD}{\D}
\newcommand{\td}{\tilde{d}}
\newcommand{\tD}{\tilde{D}}
\newcommand{\dd}{\mbox{\boldmath $d$}}
\newcommand{\Q}{\mbox{\boldmath $Q$}}
\newcommand{\T}{\mbox{\boldmath $T$}}
\newcommand{\I}{\mbox{\boldmath $I$}}
\newcommand{\hd}{\hat{d}}
\newcommand{\hy}{\hat{y}}
\newcommand{\pv}{\diagup\hspace{-4mm}\int_{-\infty}^{\infty}}
\newcommand{\m}{\boldsymbol{m}}
\newcommand{\n}{\boldsymbol{n}}

\part{Integrable models}

\chapter{Introduction}
The study of integrable models is the study of physical systems that are too elegant to be true but too 
physical to be useless. 

Take water in a shallow canal ($\phi$ is the wave amplitude) and you will find the known example
of the Korteweg-de Vries equation (KdV, formalized in 1895 but the first observation of solitary waves in a canal dates to 1844 by J. Scott-Russel)
\be\label{KdV}
\partial_t\phi+\partial^3_x\phi+6\phi\partial_x\phi=0.
\ee
This equation is nonlinear thus different waves are expected to interact each other.
Its speciality is that it admits ``solitonic'' solutions, namely wave packets in which each component 
emerges undistorted after a scattering event. This rare property is similar to free waves motion, in which
different wave components move independently, but is dramatically 
broken when interactions are switched on, unless there are some special constraints that 
forbid the distortion. For the sake of precision, notice that the KdV equation has also ``normal''
dispersive waves. 
The wave propagation conserves an infinite number of integrals of motion. This makes more 
clear the presence of the constraints that force the unusual solitonic behaviour.
It is a general theorem of Hamiltonian mechanics that if a (classical) system of coordinates 
$q_i,p_i\,,\quad i=1,\ldots,N$ and Hamiltonian $H(q,p)$ possesses $N$ independent functions 
$I_i(q,p)$ such that
\be\label{involuz}
\{H,I_i\}=0=\{I_i,I_j\}
\ee
then there exist $N$ angle-action variables $\phi_i,I_i$. 
The Hamiltonian is a function of the $I_i$ only $H(I_i)$ and the equations of motion can be 
explicitly solved by just one integration.
This is the origin of the name of integrable models.

This theorem is lost when $N \rightarrow \infty$ or for a quantum system, but somehow its ``spirit''
remains: the presence of several integrals of motion, as is (\ref{involuz}), over-constraints the
scattering parameters of waves or particles and special behaviours appear.

\begin{figure}
\hspace*{50mm}\begin{tikzpicture}
\draw (0.2,4.3) -- (5,1.8);\draw(0,4.3) node{$i_1$};\draw(5.2,1.8) node{$j_1$};
\draw (1.,4.7) -- (2.8,0);\draw(0.9,4.9) node{$i_2$};\draw(2.8,-0.3) node{$j_2$};
\draw (3,3.9) -- (0.5,0.1);\draw(3.1,4.1) node{$i_3$};\draw(0.4,-0.1) node{$j_3$};
\draw (5,2.5) -- (-0.1,0.5);\draw(5.3,2.5) node{$i_4$};\draw(-0.3,0.3) node{$j_4$};
\draw(0.5,3.8) node{$\theta_1$}; \draw(1.,4.2) node{$\theta_2$};
\draw(3,3.3) node{$\theta_3$}; \draw(4.5,2.7) node{$\theta_4$};
\draw(2,3.7) node{$k_1$}; \draw(1.3,2.9) node{$k_2$};
\draw(2.42,2.55) node{$k_3$}; \draw(3.8,1.6) node{$k_4$};
\draw(3.7,2.7) node{$l_1$}; \draw(2.3,1.8) node{$l_2$};
\draw(1.3,1.9) node{$l_3$}; \draw(1.8,0.9) node{$l_4$};
\end{tikzpicture}
\caption{\label{f:scatt}A four particles scattering: $i_1+i_2+i_3+i_4\rightarrow j_1+j_2+j_3+j_4$.
Incoming and outgoing momenta do coincide. Time flows downward.}
\end{figure}
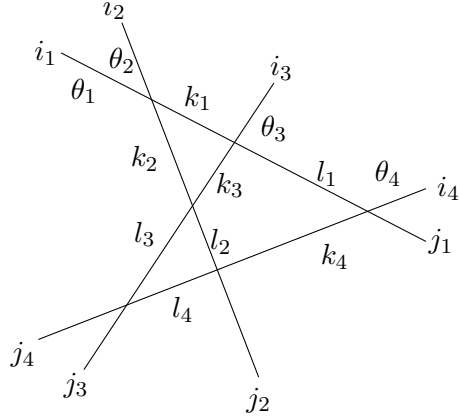

In $1+1$ quantum field theory, this has been made precise by showing \cite{zam1979, parke} the absence 
of particle production and the factorization of the scattering matrix when there are at least two 
local conserved charges that are integrals of Lorentz tensors of rank two or higher.
This theorem has very strong consequences. It implies for example that the scattering is elastic, namely the set of incoming 
momenta coincides with the set of outgoing momenta. As an example,
the factorization is written here for a four particles scattering
\be \label{fattor}
S_{i_{1}...i_{4}}^{j_{1}...j_{4}}= 
\sum_{k_{1}k_{2}k_{3}k_{4}l_{1}l_{2}l_{3}l_{4}}
S_{i_{1}i_{2}}^{k_{1}k_{2}} S_{k_{1}i_{3}}^{l_{1}k_{3}} 
S_{k_{2}k_{3}}^{l_{2}l_{3}} S_{l_{1}i_{4}}^{j_{1}k_{4}}
S_{l_{2}k_{4}}^{j_{2}l_{4}} S_{l_{3}l_{4}}^{j_{3}j_{4}}
\ee
but the generalization is simple \cite{zam1979}. The sum goes over internal indices
as in Figure~\ref{f:scatt}.

The message is clear: an $N$ particles scattering 
factorizes in $N(N-1)/2$ two-particle interactions. This means that a scattering event always 
decomposes into independent two-particle events, without multi-particle effects. 
Internal indices can only appear when there are particles within the same mass multiplet, 
otherwise the conservation of momenta forces the conservation of the type of particle 
$i_1=k_1=l_1=j_1$ and so on. This means that particle annihilation or creation are forbidden
outside a mass multiplet.

The factorization comes from the presence of higher rank integrals of motion and from the peculiar 
property of a two-dimensional plane that non-parallel lines always meet \cite{parke}.
In a Minkowski space, integrals of motion that are integrals of Lorenz tensors act by 
parallel shifting trajectories.  For example, they parallel shift lines in Figure~\ref{f:scatt}.
In $1+1$ dimensions, two non parallel straight trajectories will always have a cross point but in higher 
dimensions a parallel movement can suppress the cross point. 
This geometrical fact indicates that the constraints imposed in higher dimensions are stronger that 
in $1+1$ and the theory will be a free one, as shown by Coleman and Mandula \cite{colemanmandula}.
Therefore, factorization is a very strong property that has no equal in a general field theory.
The example of the sine-Gordon model \cite{zam1979} will be presented later, in which the full S-matrix is known.

In order to construct a scattering formalism, we need to use asymptotic states namely we need 
the so called IN states ($t \rightarrow -\infty$) and OUT states ($t \rightarrow \infty$).

A two-particles scattering then takes the form 
\be  \label{commut}
|A_{i}(\theta_{1}) A_{j}(\theta_{2})\rangle_{\text{in}} =\sum_{k,l} S_{ij}^{kl}(\theta_1-\theta_2) 
|A_{k}(\theta_{1}) A_{l}(\theta_{2})\rangle_{\text{out}}
\ee 
where it has been taken into account that Lorentz boosts shift 
rapidities\footnote{$E=m\cosh \theta\,,\quad p=m\sinh \theta\,,\quad \sinh \theta=\frac{v}{\sqrt{1-v^2}}$} 
of a constant amount so the amplitude depends on the difference only. 

By parallel shifting lines in Figure~\ref{f:scatt}, it is possible to appreciate that there are two
possible factorizations for a $3\rightarrow 3$ particles scattering. Their consistency
implies the following equation known as Yang-Baxter equation or factorization equation
\be \label{YB}
\sum_{k_1,k_2,k_3}S_{i_{1}i_{2}}^{k_{1}k_{2}}(\theta_{1}-\theta_2) 
S_{k_{1}i_{3}}^{j_{1}k_{3}}(\theta_{1}-\theta_3) 
S_{k_{2}k_{3}}^{j_{2}j_{3}}(\theta_{2}-\theta_3)=\sum_{k_1,k_2,k_3}
S_{i_{2}i_{3}}^{k_{2}k_{3}}(\theta_{2}-\theta_3)
S_{i_{1}k_{3}}^{k_{1}j_{3}}(\theta_{1}-\theta_3) S_{k_{1}k_{2}}^{j_{1}j_{2}}(\theta_{1}-\theta_2)
\ee  
This equation characterizes quantum integrability. It first appeared 
in the lattice case as the star-triangle relation obtained in the context of the Ising and 
six-vertex models (see for example \cite{baxter}).
In the lattice context, scattering amplitudes are replaced by Boltzmann weights.

A fully general definition of integrable theories is difficult because integrable models are found
in a variety of cases and contexts from lattice models to continuum theories, from classical to
quantum dynamics.
Therefore, rather than trying to give a general definition, I prefere to indicate the most relevant 
features. Indeed, 
the three key ingredients of an integrable theory, both apparent in the 
KdV and in the sine-Gordon case (see after), are:
\begin{enumerate}
\item[P1] incoming parameters of waves or particles are left unchanged by the scattering event, apart for time shifts,
\item[P2] there are infinite integrals of motion in involution,
\item[P3] a Yang-Baxter equation holds.
\end{enumerate}
The first one expresses the conservation of the incoming momenta. The second characterization generalizes the
original notion of integrability for classical Hamiltonian systems with finite degrees of freedom.
The third property expresses the mathematics of integrability.

\section{The sine-Gordon model}
The sine-Gordon model will be used here as a complete example of several ``integrable'' ideas. 
Later it will be used to introduce the nonlinear integral equation of type 
Kl\"umper-Pearce-Destri-de Vega.

The Lagrangian density is
\be\label{senogordone}
\mathcal{L}[\phi]=
\frac12\ \partial_{\mu}\phi\ \partial^{\mu}\phi+\frac{\mu^2}{\beta^2} (\cos\beta\phi-1)
\ee
and will be considered in 1+1 dimensions (signature of the metric $(1,-1)$).
The corresponding equation of motion is
\be\label{senogordone2}
\frac{\partial^2\phi}{\partial t^2} -\frac{\partial^2\phi}{\partial x^2} 
=-\frac{\mu^2}{\beta} \sin\beta \phi
\ee
At small $\beta$ this model appears as a deformation of the Klein-Gordon equation in which $\mu$ plays 
the role of a mass. Expanding the cosinus function in the Lagrangian (or the sinus in the equations of motion)
the coupling $\beta$ first appears with the fourth order term $-\beta \phi^4$, while $\beta=0$ is precisely the 
Klein-Gordon equation.
The sine-Gordon equation (\ref{senogordone2}) admits solitonic solutions, satisfying property P1, 
that are distinct in three types\footnote{It is usual to interpret as equivalent those fields that
differ by multiples of $2\pi/\beta$.}
\begin{enumerate}
\item the solitons, characterized by $\phi(+\infty,t)-\phi(-\infty,t)=\frac{2\pi}{\beta} m$, $m>0$, 
integer; 
\item the antisolitons, $\phi(+\infty,t)-\phi(-\infty,t)=\frac{2\pi}{\beta} m$, $m<0$, integer;
\item the breathers, with $\phi(+\infty,t)=\phi(-\infty,t)=0$.
\end{enumerate}
Solutions that combine an arbitrary number of these three elementary types do exist and they are 
all known \cite{faddeevtak}. They all behave as indicated in property P1. Precisely for this reason, 
one can think the soliton as an entity ``in its own'': it is recognizable and well identified 
even if it participates to a multicomponent wave.

The name, soliton or antisoliton, suggests that these two waves are distinct because they have 
opposite sign of the ``topological charge'': $\phi(+\infty,t)-\phi(-\infty,t)$. Having the breather 
zero topological charge, it can be interpreted as a bound state of soliton and antisoliton. 

The single soliton state at rest is 
\be
\phi_s(x)=\frac{4}{\beta}\ \text{atan} \exp (\mu x )
\ee
and the single antisoliton is simply given by $\phi_a=-\phi_s$. By Lorentz boost, the single soliton 
at speed $u$ is $\phi_s((x-ut)/\sqrt{1-u^2})$

An example of soliton-antisoliton state is given by 
\be\label{santis}
\phi_{sa}(x,t)=\frac{4}{\beta}\ \text{atan} \frac{\sinh \frac{\mu ut}{\sqrt{1-u^2}}}{{u\cosh\frac{\mu x}{\sqrt{1-u^2}}}}
\ee
This state is not the breather (see later). Indeed, at large $|t|$ this state decomposes into a soliton and 
an antisoliton solution travelling in opposite directions (and non bounded)
\begin{eqnarray}\label{decomp}
\phi_{sa}(x,t)&\xrightarrow[t\rightarrow -\infty]{}& \phi_s\Big(\frac{x+ut}{\sqrt{1-u^2}}+\log u\Big)+
\phi_a\Big(\frac{x-ut}{\sqrt{1-u^2}}-\log u\Big)\\
\phi_{sa}(x,t)&\xrightarrow[t\rightarrow \infty]{}& \phi_s\Big(\frac{x+ut}{\sqrt{1-u^2}}-\log u\Big)+
\phi_a\Big(\frac{x-ut}{\sqrt{1-u^2}}+\log u\Big)\nonumber
\end{eqnarray}
Notice that each wave maintains its initial speed, as indicated by property P1, just experiencing a 
phase shift of $-2\sqrt{1-u^2}\,(\log u)/u>0$, being $0\leq u\leq 1$. The phase shift is positive namely the 
two interacting waves accelerate with respect to their asymptotic motion.
This acceleration indicates an attraction, consistently with the idea that solitons and antisolitons
have opposite charge.

The simplest breather-like solution 
\be
\phi_{b}(x,t)=\frac{4}{\beta}\ \text{atan} \frac{\sin \frac{\mu ut}{\sqrt{1+u^2}}}{{u\cosh\frac{\mu x}{\sqrt{1+u^2}}}}
\ee
is a time periodic solution that takes its name from the fact that it resembles a mouth that opens and closes. 
Curiously, it can be formally obtained from the soliton-antisoliton state (\ref{santis}) by rotating to an 
imaginary speed $u\rightarrow i\,u$.
This type of solution can be interpreted as a bound state of a soliton and an antisoliton because it has zero 
topological charge and because the soliton and antisoliton can attract each other.
It is significantly different from the soliton-antisoliton state because, asymptotically, it does not  
decompose into two infinitely separated wases as it does the soliton-antisoliton state (\ref{decomp}).

Finally, the sine-Gordon model admits an infinite number of conserved integrals of motion in involution 
(property P2).

The sine-Gordon model discussed so far is strictly classical namely the field $\phi$ is a real function
of space and time. Nevertheless, the model can be quantized with fields becoming operators on an Hilbert 
space leading to a scattering theory of quantum particles. 
Notice that the various solutions given so 
far do not survive the limit $\beta\rightarrow 0$ namely they aren't 
perturbative solutions of the Klein-Gordon equation (see after \ref{senogordone2}). 
The coupling $\beta$ is not very important in the classical theory and could be removed 
by redefinition of the field and the space-time coordinates. On the contrary, in the quantum theory, 
it will play a true physical role.

In \cite{coleman1975},\cite{dashen1975},\cite{korepin1975} some interesting steps of 
the quantization procedure are performed. 
In particular, the need to remove ultraviolet divergences and the existence of a lower
bound for the ground state energy lead to observe that outside the range
\be\label{range}
0\leq \beta^2 \leq 8\pi
\ee
the theory seems not well defined, missing a lower bond for the Hamiltonian. 
Within the range, the theory describes two particles, 
charge conjugated, that carry the same name of the classical counterparts, soliton and antisoliton, 
and other particles, corresponding to the breathers. 

Within this interval, the theory shows up to coincide with the massive Thirring model,
in the sector of even number of solitons plus antisolitons (``even sector''):
\be\label{thirring}
\mathcal{L}[\psi]=
\bar\psi i\gamma_{\mu}\partial^{\mu}\psi -m_F\bar{\psi}\psi -\frac12\ g\ j_{\mu} j^{\mu}\,, 
\quad \mbox{with} \quad j_{\mu}=\bar{\psi} \gamma_{\mu}\psi
\ee
The equivalence of the two models is better stated by saying that they have the same correlation 
functions in the even sector, provided the respective coupling constants are identified by
\be\label{coupling}
\frac{\beta^2}{4\pi}=\frac{1}{1+g/\pi}
\ee
Another useful coupling will be
\be\label{gamma}
\gamma=\frac{\beta^2}{1-\frac{\beta^2}{8\pi}}
\ee
Notice that there is no equivalence outside the even sector because the soliton does not correspond 
to the fermion \cite{mand1975}: the transformation between the two is highly nonlocal.
In other words, there are states of sine-Gordon that do not exist in Thirring and vice versa.

The relation between the Thirring and sine-Gordon couplings reveals that the special point 
$g=0$ or $\beta^2=4\pi$ describes a free massive Dirac theory. This free point 
separates two distinct regimes
\be\begin{array}{lc@{\hspace{7mm}}c@{\hspace{7mm}}c}
\mbox{repulsive}  & -\frac{\pi}2<g<0 & 8\pi>\beta^2 >4\pi & \infty>\gamma>8\pi\\
\mbox{attractive} & 0<g<\infty & 4\pi>\beta^2 >0 & 8\pi>\gamma>0\\
\end{array}\label{repattr}
\ee
The repulsive regime is so called because no bound state of the Thirring fermions or sine-Gordon bosons
is observed. Vice versa, in the attractive regime the quantum fields corresponding to the classical
breathers describe bound states between solitons and antisolitons.
The attractive regime admits small values of beta, where the theory is close to a $\phi^4$ theory
(\ref{senogordone}) but with an unusual attractive sign
\be\label{senogordoneappr}
\mathcal{L}[\phi]=
\frac12\ \partial_{\mu}\phi\ \partial^{\mu}\phi-\frac{\mu^2}{2} \phi^2+\frac{\mu^2}{4!} \beta^2\phi^4\ldots
\ee
The mass of the breathers is given by the exact expression 
\be\label{breathermass}
M_n=2M\sin \frac{n \gamma}{16}\,,\qquad n=1,2,\ldots,<\frac{8\pi}{\gamma}\,,
\ee
where $M$ is the mass of the soliton. In the repulsive regime, no integer is in the range, indicating 
that breathers do not exist; this mass formula makes sense in the attractive regime only.
The interpretation of breathers as bound states comes also from the fact that the breather masses 
are below the threshold $M_n<2 M$. 
These breathers originate in the quantization of the classical breather solutions and, from 
(\ref{senogordoneappr}), 
they correspond to the perturbation of the Klein-Gordon particles. 
Indeed, given that the soliton mass at leading order is
\be\label{solitonmass}
M=\frac{8\mu}{\gamma}
\ee
the smallest breather mass $M_1$ for in the weak coupling $\beta^2\rightarrow 0$ is
\be
M_1= 2\frac{8\mu}{\gamma} \frac{\gamma}{16}=\mu
\ee
So, the lowest breather originates in the perturbation of the Klein-Gordon boson. Notice that the
breather is a bound state while the Klein-Gordon model has no bound states at all. This is true 
even for the $n$th breather
\be\label{breathern}
M_n=n\, \mu
\ee
so the Klein-Gordon free multiparticle states become bound states in sine-Gordon.
Differently from the breather, the soliton doesn't emerge from the Klein-Gordon theory:
its mass diverges in this limit (\ref{solitonmass}) so this particle is considered decoupled from
the theory.

The relations (\ref{coupling}, \ref{repattr}) indicate a strong/weak duality between 
sine-Gordon and Thirring: strong interactions in one model correspond to weak interactions in the 
other. Can we see a physical track of this? Yes, for example in the weak sine-Gordon regime 
$\beta^2\rightarrow 0$. Indeed, the $n$th breather appears to be a 
bound state of $n$ Klein-Gordon particles (\ref{breathern}). It is a stable state that takes
its stability from the strong fermionic coupling $g \gg 0$ of Thirring. Moving to higher 
values of $\beta^2$, the fermionic coupling decreases therefore we expect to have less and less
stable breathers, consistently with the mass expression (\ref{breathermass}). 
In the same weak regime $\beta^2\rightarrow 0$, the soliton is decoupled from the
theory as it has an almost infinite mass (\ref{solitonmass}). It is strongly coupled in the
repulsive regime where its mass is small. 

The most important feature is that the quantum sine-Gordon model is still integrable.
This was first seen by showing that conservation laws do survive perturbative quantization.
Now this result is known beyond perturbation theory \cite{sasaki1987} and grants the already 
discussed necessary conditions for the factorization of scattering (\ref{fattor}). Thus, all
the properties P1, P2, P3 hold. 

Particle annihilation and creation are forbidden. Consequently, all the bound states in (\ref{breathermass})
are stable particles even when there are breather states above the creation threshold 
$M_n>2\, M_1$. This happens for some $n$ and for sufficiently small $\gamma$. Notice that in the attractive 
regime the lowest breather is always the lightest particle. 
Moving toward the repulsive regime, one observes that the $n$th breather disappears into a 
soliton-antisoliton state when $8\pi/\gamma$ is a positive integer
\be
\lim_{\gamma\rightarrow (\frac{8\pi}{n})^{-}} M_n =2\, M
\ee
The lowest breather disappears at the free fermion point $\gamma=8\pi$.

If one can show the existence of conserved charges as required in the factorization theorem, 
the two particles scattering amplitudes can be evaluated on the basis of their symmetries. 
In other words, the Yang-Baxter equation (\ref{YB}) supplemented 
with usual analytic properties (poles from mass spectrum), unitarity and crossing symmetry, 
is (often) enough to find the scattering amplitudes. This avoids a much more
lengthy calculation based on the evaluation of Feynman diagrams to all orders.
For the sine-Gordon model, this has been done in \cite{zam1979}.

The notation in (\ref{commut}) is now used to write down the amplitudes. For the solitonic part only, 
there are three 
$2\rightarrow2$ particle processes 
\begin{eqnarray}
A_s+A_s&\rightarrow& A_s+A_s  \nonumber \\
A_s+A_{\bar{s}}&\rightarrow& A_s+A_{\bar{s}}  \label{scattssbar}\\
A_{\bar{s}}+A_{\bar{s}}&\rightarrow& A_{\bar{s}}+A_{\bar{s}} \nonumber
\end{eqnarray}
where $A_s$ ($A_{\bar{s}}$) indicates a soliton (antisoliton) momentum state.
Charge conjugation symmetry makes the first and the last processes to
have identical amplitude. Using $\theta=\theta_1-\theta_2$, we have 
\newcommand{\ket}[1]{|{#1}\rangle} 
\begin{eqnarray}\nonumber
\ket{A_s(\theta_1)A_s(\theta_2)}_{\text{in}} &=& S(\theta)\, \ket{A_s(\theta_1)A_s(\theta_2)}_{\text{out}}\\
\ket{A_s(\theta_1)A_{\bar s}(\theta_2)}_{\text{in}} &=& 
S_T(\theta)\, \ket{A_s(\theta_1)A_{\bar s}(\theta_2)}_{\text{out}}+
S_R(\theta)\, \ket{A_{\bar s}(\theta_1)A_s(\theta_2)}_{\text{out}} \\ \nonumber
\ket{A_{\bar s}(\theta_1)A_{\bar s}(\theta_2)}_{\text{in}} &=&  S(\theta)\,
\ket{A_{\bar s}(\theta_1)A_{\bar s}(\theta_2)}_{\text{out}}
\end{eqnarray}
There are just three independent amplitudes to be determined, that we organize in a $4\times 4$
matrix to be used in the Yang-Baxter equation (\ref{YB}) 
\be
S=\begin{pmatrix} S(\theta)\\ &S_T(\theta)&S_R(\theta)\\
&S_R(\theta)&S_T(\theta)\\&&& S(\theta)\end{pmatrix}
\ee 
Notice that, in the Yang-Baxter equation, the conservation of the set of momenta forbids amplitudes describing 
particles with different mass to mix each other. In sine-Gordon, there are just two particles with identical mass, 
the soliton and the antisoliton, with interactions listed in (\ref{scattssbar}).
This means that the scattering processes involving a breather do no mix with those in
(\ref{scattssbar}). As the breathers have different mass, the following processes 
are of pure transmission, reflection being forbidden
\begin{eqnarray}\nonumber
A_s+B_n &\rightarrow& A_s+B_n\\
A_{\bar s}+B_{n} &\rightarrow&  A_{\bar s}+B_{n}\\
B_{m}+B_{n} &\rightarrow&  B_{m}+B_{n}\nonumber
\end{eqnarray}
The whole knowledge of the scattering amplitudes is not needed. The soliton part is given by
\be\label{matriceS}
S=\begin{pmatrix} -i\sinh\Big(\frac{8\pi}{\gamma}(i\pi-\theta)\Big)\\
& -i\sinh\Big(\frac{8\pi}{\gamma}\theta\Big) & \sin \frac{8\pi^2}{\gamma}\\
& \sin \frac{8\pi^2}{\gamma} & -i\sinh\Big(\frac{8\pi}{\gamma}\theta\Big) \\
&&&-i\sinh\Big(\frac{8\pi}{\gamma}(i\pi-\theta)\Big)
\end{pmatrix} U(\theta)
\ee
where $U(\theta)$ is a known factor. 
The expression of the scattering matrix will be useful soon, in relation to the six-vertex model.

\section{The six-vertex model}

It is a two dimensional classical statistical mechanics model in which interactions are associated  
with a vertex: the four bonds surrounding a vertex fix the Boltzmann weight associated with it.
In the present model the possible vertices are those shown here: 
\begin{center}\includegraphics[scale=0.5]{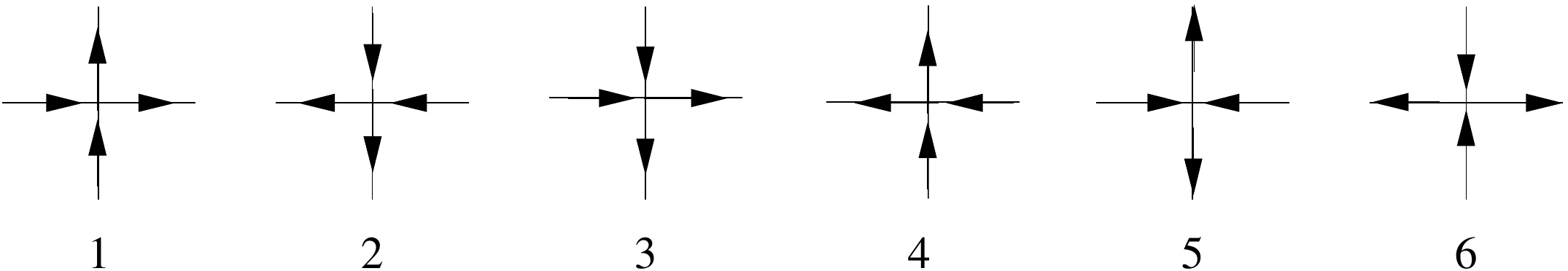}\end{center}
A bond can therefore be in either one of two states, that will be indicated by ``0'' or ``1'' 
(0 associated to up and right, 1 associated to down and left).
Initially, this model was introduced as a two dimensional idealization of an ice crystal and called ice-type model.
Indeed, the vertex represents the oxygen atom and the four bonds connected to it represent two covalent bonds and 
two hydrogen bonds. The arrows indicate to which oxygen the hydrogen atom is closer, thus differentiating the 
covalent bonds from the hydrogen bonds.

The Boltzmann weights for a vertex are nonnegative values indicated by 
$w_{1},w_{2},w_{3},w_{4},w_{5},w_{6}$. Hereafter I will put 
$w_{1}=w_{2}=a$, $w_{3}=w_{4}=b$ and $w_{5}=w_{6}=c$, as in \cite{baxter}.
Their product on the whole lattice vertices is summed 
on all the configurations to build up the partition function
\begin{equation}
\mathcal{Z} 
= \sum_{{\rm conf}}\, w_{1}^{N_{1}}\, w_{2}^{N_{2}}\,w_{3}^{N_{3}}\, w_{4}^{N_{4}}\,w_{5}^{N_{5}}\, w_{6}^{N_{6}}
\label{partizione}
\end{equation}
where $N_i$ is the number of occurrences of the type $i$ vertex in the lattice.
Periodic boundary conditions will now be uses in the vertical and horizontal directions.
The expression for the partition function takes a useful form if one introduces the transfer matrix
and the so called R matrix.
The transfer matrix {\bf T} is a $2^{N}\times 2^{N}$ matrix that describes how the system 
``evolves'' from a row to the next one of the lattice. The R matrix somehow summarizes the possible 
behaviours on a single vertex or lattice site. On a given row, the vertical bond at site $i$ is associated 
with a local vector space $V_i=\mathbb{C}^2$. Also, $A$ represents an auxiliary space $A=\mathbb{C}^2$.
The R matrix is a $4\times 4$ matrix acting on $A\otimes V_i$ (or else on $A\otimes A$)
\bea\label{rmatrix}
R &=&\left(\begin{array}{cccc} a&&&\\&b&c\\ &c&b\\&&&a\end{array}\right)  \\
&& \hspace{5mm} {\scriptstyle 00\hspace{2.2mm}  01\hspace{2.2mm}  10\hspace{2.2mm}  11} \nonumber
\eea
where the lower line indicates how the entries are interpreted with respect to the two possible
bond configurations. 
The transfer matrix acts on the physical vector space $\mathcal{V}$ and is a product of R 
matrices
\be\label{trasf}\ba{l} 
\mathcal{V}=\displaystyle \mathop{V}_{\scriptscriptstyle 1}\otimes\mathop{V}_{\scriptscriptstyle 2}
\otimes\ldots\otimes \mathop{V}_{\scriptscriptstyle N}\\[4mm]
{\bf T}:\mathcal{V} \rightarrow \mathcal{V}\\[4mm]
{\bf T}=\displaystyle\mathop{\text{Tr}}_A\  R_{A\,1}R_{A\,2}\ldots R_{A\,N}
\ea\ee
where the trace is taken on the auxiliary space $A$\footnote{Here, the 
standard notation of of lattice integrable systems is used such that the lower
indices of the matrix do not indicate its entries but the spaces on which the matrix acts
(namely the auxiliary space and one of the horizontal lattice sites, enumerated fro 1 to $N$).}.
The partition function can be written as the trace of the product of $M$ transfer matrices 
(if $M$ is the number of rows of the lattice)
\begin{equation}
\mathcal{Z} = {\rm Tr}\,{\rm {\bf T}}^{M}
\label{transfer}
\end{equation}
having now taken the trace on $\mathcal{V}$, namely on all the horizontal sites. It turns out that 
if the Boltzmann weights have an appropriate form, the transfer matrix generates an integrable system. 
The following parametrization makes the game
\newcommand{\gammasv}{\gamma_{\text{6v}}}
\begin{equation}
\label{R_matrix_entries}
a=a(\theta)=\sinh \frac{\gammasv}{\pi}(\theta+i\pi),\quad b=b(\theta)=
\sinh\frac{\gammasv}{\pi} \theta ,\quad 
c=c(\theta)=i\sin \gammasv 
\end{equation}
and gives an R matrix function of the spectral parameter\footnote{The 
spectral parameter is a complex number that is used to describe a sort of off-shell physics; usually it is fixed
to a specific value or interval to construct a physical model.} 
$\theta$, $R(\theta)$, and also of the coupling $\gammasv$. The R~matrix satisfies a Yang-Baxter equation 
(see later).

It turns out that this parametrization is very much the same as in (\ref{matriceS}), except for 
the identification of the couplings that requires some care and will be done later. 
This means that the integrable sine-Gordon model and the six-vertex model have something in common. 
Anticipating a later discussion, one can use the six-vertex model for a lattice regularization of the sine-Gordon.
In other words, sine-Gordon appears as a certain continuum limit of the six-vertex model, provided
a mass scale is introduced.

The disadvantage of the parametrization (\ref{R_matrix_entries}) is that is introduces complex Boltzmann 
weights and this looks
odd in statistical mechanics. However, this is not a serious problem, first because the 
statements that concern integrability do hold for arbitrary complex parameters, second because it is easy
to get a real transfer matrix, simply by using an imaginary value for $\theta$.

The Yang-Baxter equation satisfied by the R matrix (\ref{R_matrix_entries}) is
\be\label{YBlattice}
R_{12}(\lambda-\mu)R_{13}(\lambda)R_{23}(\mu)=R_{23}(\mu)R_{13}(\lambda)R_{12}(\lambda-\mu)
\ee
where $\lambda,\mu$ are arbitrary complex spectral parameters. Thus, property P3 above also holds for 
the lattice case. From this equation, a 
very general construction shows that the transfer matrix forms a commuting family
\be
[{\bf T}(\theta),{\bf T}(\theta')]=0
\ee
for arbitrary values of the spectral parameters. Now, any expansion of the transfer matrix 
produces commuting objects. In particular, 
it is possible to evaluate the logarithmic derivative of the transfer matrix
\be\label{hamilt}
H=\mathcal{A}\left. \frac{d\log{\bf T}(\theta)}{d\theta}\right|_{\frac{i\pi}{2}}
\ee
and all the higher derivatives. The operators obtained with this procedure are local 
and commutative therefore we conclude that property P2 is satisfied for the lattice model.
The logarithmic derivative is manageable, at least for the six-vertex model, and leads to a 
very interesting expression (the overall factor $\mathcal{A}$ is easy to evaluate but not very important)
\bea\nonumber
H&=&-\sum_{i=1}^{N-1}\Big[
\sigma_i^1\sigma_{i+1}^1+\sigma_i^2\sigma_{i+1}^2+\Delta \big(1+\sigma_i^3\sigma_{i+1}^3)\Big]\\
\Delta&=&\cos\gammasv \label{xxz}
\eea

The $\sigma_i^{j}\,,\ j=1,2,3$ are Pauli matrices acting on the site $i$, where, by definition, 
different site matrices always commute.
This one-dimensional lattice quantum Hamiltonian is known as XXZ model. The more general version 
with three different coefficients and three spatial dimensions was introduced by 
W. Heisenberg (1928) as a natural physical description of magnetism in solid state physics.
Indeed, the Heisenberg idea was to consider, on each lattice site, a quantum magnetic
needle of spin $\frac{1}{2}$ fully free to rotate.
The magnetic needle is assumed sensitive to the nearest neighbor needles
with the simplest possible coupling of magnetic dipoles. 
At $\gammasv=0$ it is fully isotropic with rotational $su(2)$ symmetry.
As soon as $\gammasv\neq 0$ is introduced, the model acquires an anisotropy. 

The XXX Hamiltonian is free of couplings apart from the overall sign.
Given the present sign choice, it is apparent that adjacent parallel spins
lower the energy. This explains the name ``ferromagnetic'' attributed to
the Hamiltonian in (\ref{xxz}), if $\Delta=1$. The Hamiltonian with opposite sign is known as ``antiferromagnetic''.

The presence of $\Delta\neq 1$ in the XXZ model spoils this distinction
because the ferromagnetic or antiferromagnetic behavior depends by the
coupling and the name cannot be attached to the Hamiltonian itself but to the phases it describes. 
The phases of the two models are indicated in table~\ref{xxzcoupl}.
\begin{table}\begin{center}
\begin{tabular}{c|l@{\hspace*{12mm}}l}
 &XXZ& six-vertex  \\ \hline
$\Delta >1$ &  ferromagnetic  & ferroelectric, vertex: one of 1,2,3,4\\
$-1<\Delta<1$ & \multicolumn{2}{l}{critical case, multi-degenerate ground state}\\
$\Delta <-1$ & antiferromagnetic & antiferroelectric, vertices 5 and 6 alternate
\end{tabular}\end{center}
\caption{\label{xxzcoupl}The thermodynamic phases of the XXZ and six-vertex model are indicated. 
The XXZ model being defined by the Hamiltonian, the phases are understood at zero temperature, while
the six-vertex phases are read from the transfer matrix and contain a temperature, hidden in the parameters $a,b,c$.
Moreover, one-dimensional models do not break symmetries: phase transitions can only occur at zero temperature.}
\end{table}

Finally, as the XXZ model is embedded in XYZ, the six-vertex is embedded in the more general eight-vertex model,
that is still integrable. 

The XXZ model is presented in the review \cite{defk} and also in the book \cite{suth}.

\subsection{The Baxter T-Q relation\label{s:baxter}}
\noindent The one-dimensional XXZ model has the merit of having inaugurated the studies of quantum 
integrable systems and of the methods known as Bethe ansatz, in the celebrated Bethe paper \cite{bethe}.

Indeed, his idea was to try to guess, or ansatz, the appropriate eigenfunctions for the Hamiltonian (\ref{xxz}),
from a trial form, then show that the guess is correct.
This approach is called coordinate Bethe ansatz and produces a set of constraints 
on the parameters of the wave function known as Bethe equations. 
Baxter \cite{baxter}, from the Bethe equations, was able to show the existence of a T-Q relation 
(\ref{TQbordo}) for the transfer matrix. 
After, he could reverse the approach and, with a more direct construction, he derived the T-Q relation 
from the Yang-Baxter equation. Therefore, he obtained the Bethe equations from the T-Q relation. The presentation 
will follow the second approach.

Following Baxter, the transfer matrix satisfies a functional equation~\cite{baxter} 
for periodic boundary conditions on a row of $N$ sites. This means that there exist a matrix $\Q(u)$
such that
\be\label{TQperiod}
\T(u)\Q(u)=f\left(u+\frac{\lambda}{2}\right)
\Q(u-\lambda)+f\left(u-\frac{\lambda}{2}\right)\Q(u+\lambda)
\ee
where we have used
\be
f(u)=\left(\frac{\sin u}{\sin \lambda}\right)^{N}
\ee
The coupling and spectral parameters are related to the previous ones by
\be
\lambda=\gammasv\,,\qquad u=-i\frac{\gammasv}{\pi}\theta
\ee
The new operator $\Q(u)$ forms a family of matrices that commute each other and with the 
transfer matrix $[\Q(u),\Q(v)]=[\Q(u),\T(v)]=0$. This implies that the same functional equation (\ref{TQperiod})
holds true also for the eigenvalues $T(u)$ and $Q(u)$. Moreover, all these operators have the 
same eigenvectors independent of $u$. The eigenvalues of $Q$ are given by 
\be
Q(u)=\prod_{j=1}^M \sin(u-u_j) \label{BaxQ}
\ee
where $u_j$ are the Bethe roots and appear now as zeros of the eigenvalues of $\Q(u)$. 
The Bethe ansatz equations result by imposing that the transfer matrix eigenvalues on the left are entire 
functions. Indeed, when $Q(u)=0$, being $T(u)$ entire, the right hand side must vanish. This forces the 
constraints (Bethe equations)
\be\label{betheTQ}
\left(\frac{\sin\left(u+\frac{\lambda}{2}\right)}{\sin\left(u-\frac{\lambda}{2}\right)}\right)^{N}=-
\frac{Q(u+\lambda)}{Q(u-\lambda)}
\ee
The T-Q relation (\ref{TQperiod}) shows that the columns of $\Q(u)$ are eigenvectors of $\T$ so this
equation actually provides both information on eigenvalues and eigenvectors. 

The Bethe equations have a finite number of solutions
in the periodicity strip
\be
\re (u)\in [0,\pi]
\ee
This is easily seen because they can be transformed into algebraic equations in the new variables  
$$z_j=\exp(i\ u_j)$$
Notice that the lattice model also has a finite number of states: indeed, 
$$
\dim \mathcal{V}=2^N
$$
This is the size of the transfer matrix and is also the number of expected 
solutions of the Bethe equations. Indeed, it has been possible to show  
that the Bethe equations have the correct number of solutions and that the corresponding Bethe eigenvectors form a base for $\mathcal{V}$, 
see \cite{yy1966_2}, \cite{kirillov1987} and references there. This is referred to as the 
``completeness'' of the Bethe ansatz.
One feature observed is that Bethe roots satisfy a Pauli-like principle, 
in the sense that they are all distinct: there is no need to consider 
solutions with $u_j=u_k$ for different $j\neq k$.

\section{Conformal field theories\label{sect:CFT}}
Conformal field theories (CFT) are scale invariant quantum field theories. 
They were introduced for two main reasons, one being the study of continuum phase transitions 
and the other being the interest of describing quantum strings on their world sheet, 
a two-dimensional surface in a ten-dimensional space. This second point strongly motivated 
the treatment of the two dimensional case, initiated in the fundamental work 
\cite{bpz}. Curiously, the subsequent development of the two-dimensional case, instead, 
was much more 
statistical mechanics oriented. Two-dimensional conformal field theories are very close to 
the subject of 
quantum integrability because they also are integrable theories and, often, they appear 
in certain limits of lattice or continuum integrable theories.
These topics and some connections between conformal field theory and integrability 
will be discussed later, in relation to several of the investigations that I have 
carried on: nonlinear Destri-de Vega equations, thermodynamic Bethe ansatz equations and so on.

Four-dimensional conformal field theories are studied in the Maldacena gauge/string duality 
framework. In particular, the superconformal gauge theory $\mathcal{N}=4$
appears in relation to some integrable theories, after the work \cite{MZ}. In particular,
in that paper the XXX model appeared. As the paper \cite{bpz} created the bridge
between integrable models and two-dimensional field theories, \cite{MZ} inaugurated
the interchange between (some) aspects of integrable models and four-dimensional
superconformal field theories.

In the two-dimensional case, the generators of the conformal symmetry are the modes
of the Virasoro algebra  (\( {\cal V} \))
\be
\label{virasoro}
[L_{m},L_{n}]=(m-n)L_{m+n}+\frac{c}{12}(m^{3}-m)\delta _{m+n,0}
\ee
 where the constant \( c \) that appears in the central extension term is called 
\emph{conformal anomaly} or often \emph{central charge}.
For a physical theory on the Minkowski plane or on a cylinder geometry in which the space is
periodic and the time flows in the infinite direction, the algebra of the full conformal group 
is the tensor product 
of two copies of (\ref{virasoro}) $ {\mathcal{V}}\otimes {\bar{\mathcal{V}}}$.
For other geometries, it can be different. For example, on a strip (finite space, 
infinite time) there is a single copy. 
According to this, the Hamiltonian is \( L_{0}+\bar{L}_{0} \) for the plane
or cylinder and is $L_0$ in a strip.
We need this distinction because, later on, we will use both types of space-time. 
Somehow, the presence of two copies in the plane and in the cylinder with periodic space is justified 
because there are two types of movers: left and right movers, namely massless particles 
moving at the speed of light toward left or right\footnote{Conformal invariance implies that
particles are massless and move at the speed of light.}. 
In a strip, corresponding to a finite space with spatial borders, movement is not allowed, 
thus just one copy remains.

All the states of a CFT must lie in some irreducible
representation of the algebra (\ref{virasoro}). 
Physical representations must have the Hamiltonian spectrum
bounded from below, i.e. they must contain a so called \emph{highest weight 
state}
(HWS) \( |\Delta \rangle  \) for which \begin{equation}
\label{hws}
L_{0}|\Delta \rangle =\Delta |\Delta \rangle \qquad ,\qquad L_{n}|\Delta 
\rangle =0\qquad ,\qquad n>0
\end{equation}
 These representations are known as \emph{highest weight representations} (HWR).
The irreducible representations of \( {\cal V} \) are labelled by two numbers, 
namely the central charge \( c \) and the conformal dimension \( \Delta  \). 
We shall denote the HWRs of
\( {\mathcal{V}} \) by \( {\mathcal{V}}_{c}(\Delta ) \). For a given theory,
the Hilbert space \( {\mathcal{H}} \) of the theory is built up of all possible representations \( 
{\mathcal{V}}_{c}(\Delta ) \) with the same \( c \), each one with a certain multiplicity: 
\begin{equation}
\label{hilbert}
{\mathcal{H}}=\bigoplus _{\Delta ,\bar{\Delta }}{\mathcal{N}}_{\Delta 
,\bar{\Delta }}{\mathcal{V}}_{c}(\Delta )\otimes 
{\bar{\mathcal{V}}}_{c}(\bar{\Delta })
\end{equation}
 If a certain \( {\mathcal{V}}_{c}(\Delta )\otimes 
{\bar{\mathcal{V}}}_{c}(\bar{\Delta }) \)
does not appear, then simply \( {\mathcal{N}}_{\Delta ,\bar{\Delta }}=0 \).
The numbers \( {\mathcal{N}}_{\Delta ,\bar{\Delta }} \) count the multiplicity
of each representation in \( {\mathcal{H}} \), therefore they must always
be non negative integers. They are not fixed by conformal invariance alone
as they depend on the geometry and on possible boundary conditions. 

Every HWS (\ref{hws}) in the theory
can be put in one-to-one correspondence with a field through the formula \( 
|\Delta,\bar{\Delta}\rangle =A_{\Delta,\bar{\Delta}}(0,0)|0,0\rangle  \),
where the vacuum \( |0,0\rangle  \) is projective (i.e. \( 
L_{0},\bar{L}_{0},L_{\pm 1},\bar{L}_{\pm 1} \))
invariant. In particular the HWS (\ref{hws}) correspond to some fields \( \phi 
_{\Delta ,\bar{\Delta }}(z,\bar{z}) \)
that transform under the conformal group as \begin{equation}
\label{primary}
\phi _{\Delta ,\bar{\Delta }}(z,\bar{z})=\left( \frac{\partial z'}{\partial 
z}\right) ^{\Delta }\left( \frac{\partial \bar{z}'}{\partial \bar{z}}\right) 
^{\bar{\Delta }}\phi _{\Delta,\bar{\Delta } }(z',\bar{z}')
\end{equation}
They are called \emph{primary fields}. Non primary fields (secondaries) do have
much more involved transformations. A basis for the states  
can be obtained by applying strings of \( L_{n},n<0 \) to \( |\Delta \rangle  
\).
The commutation relations imply \begin{equation}
\label{secondaries}
L_{0}L_{n}^{k}|\Delta \rangle =(\Delta +nk)L_{n}^{k}|\Delta \rangle 
\end{equation}
 Therefore \( L_{0} \) eigenvalues organize the space \( 
{\mathcal{V}}_{c}(\Delta ) \)
(called a \emph{module}) so that the states lie on a {}``stair{}'' whose
\( N \)-th step (called the \( N \)-th \emph{level}) has \( L_{0}=\Delta +N 
\)\begin{equation}
\label{levels}
\begin{array}{ccc}
\mbox {states} & \mbox {level} & L_{0}\\
...... & ... & ...\\
L_{-3}|\Delta \rangle \, ,\, L_{-2}L_{-1}|\Delta \rangle \, ,\, 
L_{-1}^{3}|\Delta \rangle  & 3 & \Delta +3\\
L_{-2}|\Delta \rangle \, ,\, L_{-1}^{2}|\Delta \rangle  & 2 & \Delta +2\\
L_{-1}|\Delta \rangle  & 1 & \Delta +1\\
|\Delta \rangle  & 0 & \Delta 
\end{array}
\end{equation}
 All the fields corresponding to the HWR \( {\mathcal{V}}_{c}(\Delta ) \) are
said to be in the \emph{conformal family} \( [\phi _{\Delta }] \) generated
by the primary field \( \phi _{\Delta } \).

For the following, the most important conformal models will be those knows 
as minimal models, characterized by the central charge
\be\label{minimal}
c=1-\frac{6}{p(p+1)}
\ee
with $p=3,4,\ldots$
These models are all unitary and all have a finite number of primary fields.  
The first one, $c=\frac12$, is the universality class of the Ising model. The next one 
$c=\frac7{10}$, is the tricritical Ising model, namely an Ising model with vacancies.
After, we find the universality class of the three-states Potts model and so.
The limit $p\rightarrow \infty$ is also a CFT; it is one point of the class of the free massless boson
with $c=1$.

Indeed, $c=1$ is a wide class of unitary conformal field theories, all derived from a free massless
boson compactified in the following way
\be
\phi \equiv \phi +2\pi m R\,,\quad m\in\mathbb{Z}
\ee
and radially quantized. A full description of this theory would be very long. A sketch is presented
in \cite{tesidott}. The theory turns out to be characterized by 
certain vertex operators
\be
\label{vertex_operators}
V_{(n,\, m)}(z,\overline{z})=:\exp i (p_{+} \phi (z)+p_{-}\bar{\phi }(\overline{z})):  ,
\qquad p_{\pm}=\frac{n}{R}\pm \frac12 m R.
\ee
with conformal weight $\Delta=p_{+}^2/2,\,\bar{\Delta}=p_{-}^2/2$.

Each pair $(n,m)$ describes a different sector of the theory; its states are obtained by the 
action  
of the modes of the fields, $\partial _z \phi$ and $\bar{\partial_z}\bar{\phi}$, in a standard 
Fock space construction.

It is important to stress that a particular \( c=1 \) CFT is specified by giving
the spectrum of the quantum numbers \( (n,m) \) (and the compactification radius
\( R \)) such that the corresponding set of vertex operators (and their descendants)
forms a \emph{closed and local} operator algebra. The locality requirement is
equivalent to the fact that the operator product expansions of any two such
local operators is single-valued in the complex plane of \( z \). 

By this requirement of locality, it was proved in \cite{km93} that there are
only two maximal local subalgebras of vertex operators: \( {\cal A}_{b} \), purely bosonic,
generated by the vertex operators 
\be
V_{(n,\, m)}:\, n,\, m\in \mathbb {Z}
\ee
and \( {\cal A}_{f} \), fermionic,  generated by
\be
V_{(n,\, m)} : \, n\in \mathbb{Z},\, m\in 2\mathbb{Z} \quad \mbox{ or }\quad n\in \mathbb{Z}+\frac{1}{2},\, 
m\in 2\mathbb{Z}+1 .
\ee
Other sets of vertex operators can be built, but the product of two of them gives a nonlocal expression
(namely the operator product expansion is multi-valued).
\begin{figure}[h]
\hspace{50mm}\includegraphics[scale=0.5]{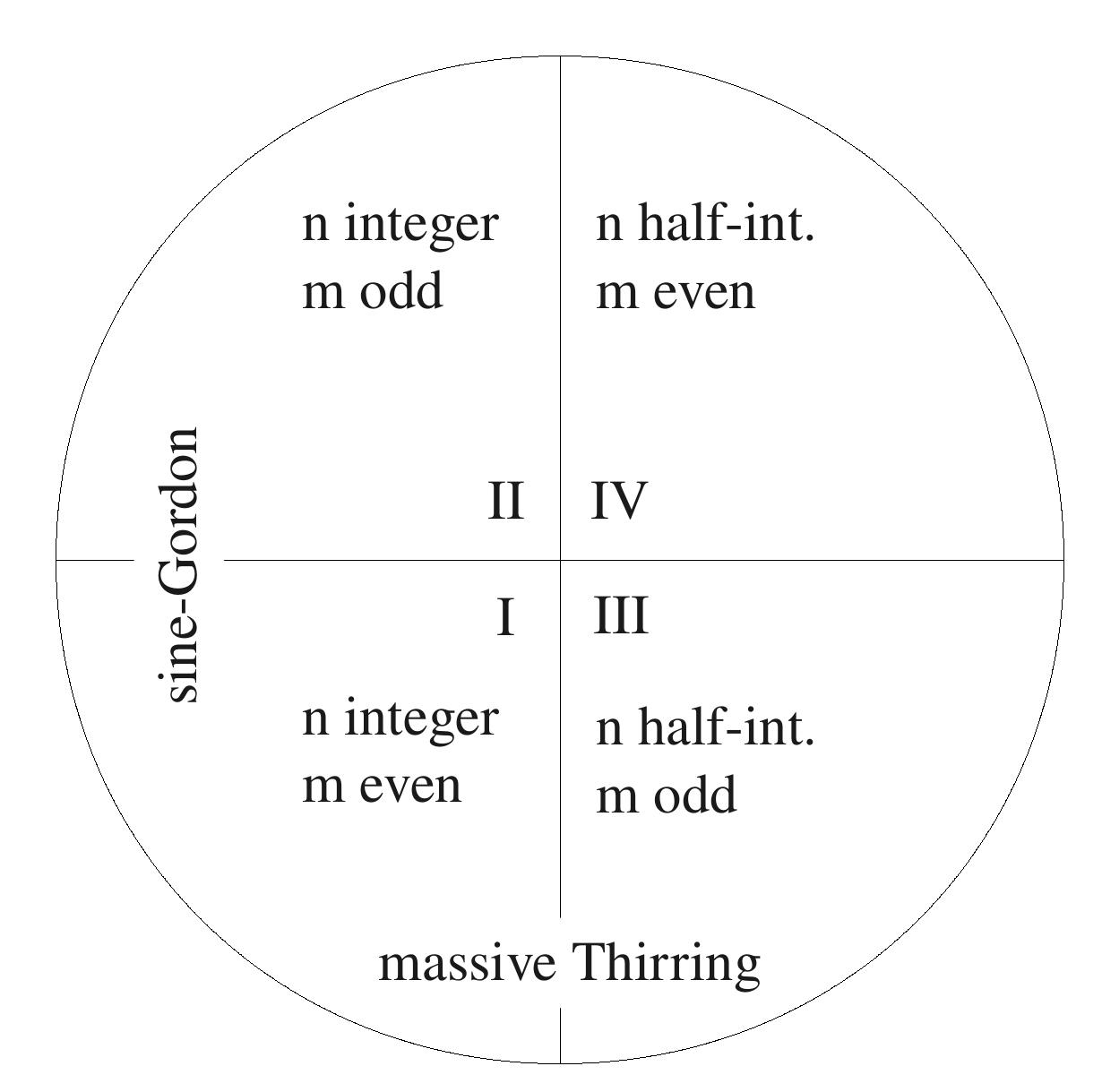}
\caption{The family of vertex operators $ V_{(n,\, m)}$ with $ n\in \mathbb Z/2$ 
and $ m\in \mathbb Z$. Sectors \textbf{I} and \textbf{II} are ${\cal A}_{b}$, namely the 
ultraviolet limit of the sine-Gordon model. Sectors \textbf{I} and \textbf{III} are ${\cal A}_{f}$, 
that defines the UV behaviour of massive Thirring. Sector \textbf{I} is the one where sine-Gordon 
and massive Thirring are equivalent.
\textbf{IV} is a sector of non mutually local vertex operators.\label{4sectors.eps}}
\end{figure}

The sine-Gordon model appears as a integrable perturbation of the $c=1$ free boson by an operator with 
scaling 
dimensions $(\Delta,\bar{\Delta})=(\beta^2/8\pi,\beta^2/8\pi)$. The corresponding unperturbed 
algebra is ${\cal A}_{b}$ while the algebra ${\cal A}_{f}$ can be perturbed to give rise to the 
massive Thirring model. The compactification radius amd the sine-Gordon coupling are related by
$$R^2=\frac{4\pi}{\beta^2}$$.

\section{Perturbed conformal field theory}
We may think to define a quantum field theory as a deformation of a conformal field theory by some 
operators \cite{Zam-adv}, i.e. to perturb the action of a CFT as in the following expression
\begin{equation}
\label{pcft-action}
S[\Phi_i]=S_{CFT}+\sum _{i=1}^{n}\lambda _{i}\int d^{2}x\ \Phi _{i}(x)
\end{equation}
Of course, the class of two-dimensional field theories is larger than the one described by
this action. Nevertheless, this class of \emph{perturbed conformal field theories} has a 
special role because it describes the vicinity of critical points in the theory of critical 
phenomena.
The main goal is to be able to compute off-critical correlation functions by
\be
\langle X\rangle =\int {\cal D}\varphi \, Xe^{-S[\Phi ]}=\int {\cal 
D}\varphi \, X \exp \big[-S_{CFT}-\lambda \int d^{2}x\ \Phi (x)\big]
\ee
Indeed, expanding in powers of \( \lambda  \) one can express \( \langle X\rangle  \)
as a series of conformal correlators (in principle computable by conformal field theory
techniques).
The perturbed theory is especially important if it maintains the integrability of
the conformal point. If so, the perturbed theory has a factorized scattering.

The possible perturbing fields are classified with respect to their renormalization group action 
as
\begin{itemize}
\item \textbf{relevant} if \( \Delta <1 \). If such a field perturbs a conformal 
action,
it creates exactly the situation described above, i.e. the theory starts to
flow along a renormalization group trajectory going to some IR destiny.
\item \textbf{irrelevant} if \( \Delta >1 \). Such fields correspond to
perturbations which describe the neighborhood of non trivial IR fixed points. 
It is more appropriate to refer to them as \emph{attraction} fields because the perturbation
is not able to move off the critical point; it actually always returns to it.
We shall not deal with this case in the following, but the interested reader
may consult, for example, \cite{Fev-Quattr-Rav} to see some possible 
applications of this situation.
\item \textbf{marginal} if \( \Delta =1 \). Their classification requires investigation of
derivatives of the beta function.
\end{itemize}

\section{Conclusion}
A typical phenomenon of integrable systems emerges, namely the fact that 
different models can transform the one into the other in some conditions and 
also shown to be equivalent.  
Firstly, the equivalence of the bosonic sine-Gordon model with the fermionic massive Thirring
has been presented. After, the correspondence of the six-vertex model and the XXZ model 
has been shown. Moreover, these lattice models share the same R or S matrix as the sine-Gordon 
model. 
At this point, it is natural to expect that a proper continuum limit on the 
six-vertex model could produce sine-Gordon; indeed, this is the case and will be discussed later
in the context of the nonlinear integral equation.
This ``game'' of models that are related one to the other can be pushed forward. For example,
if $\Delta=0$, the XXZ model reduces to the XX model, that can also be written as a lattice
free fermion (one fermionic species). 
If $\Delta\rightarrow \pm \infty$, one obtains the one-dimensional Ising model. 
More important for what follows, the XXX model emerges in the high coupling limit of the 
Hubbard model, that is a lattice quantum model of two fermionic species (namely, spin up, 
spin down).

The deep reason of these strong connections between different models is that, for a given  
size of the R matrix, there are very few solutions of the Yang-Baxter equation. In other words,
there are very few classes of integrability, classified by the solution of the Yang-Baxter equation.

\chapter{A nonlinear equation for the Bethe ansatz\label{c:nlie}}

\section{Light-cone lattice\label{s:lightcone}}
In this section I present a lattice regularization of the Sine-Gordon model
which is particularly suitable to study finite size effects. It is well known to 
lattice theorists that the same continuum theory can often be obtained as  
limit of many different lattice theories. 
This means that there are many possible regularizations of the same theory 
and it is customary to choose the lattice action possessing 
the properties that best fit calculational needs. In the present context
the main goal is to have a lattice discretization of the Sine-Gordon model that preserves
the property of integrability. The following light-cone lattice construction
is a way (not the unique!) to achieve this goal.

\begin{figure}[h]
\hspace*{50mm}\includegraphics[scale=0.7]{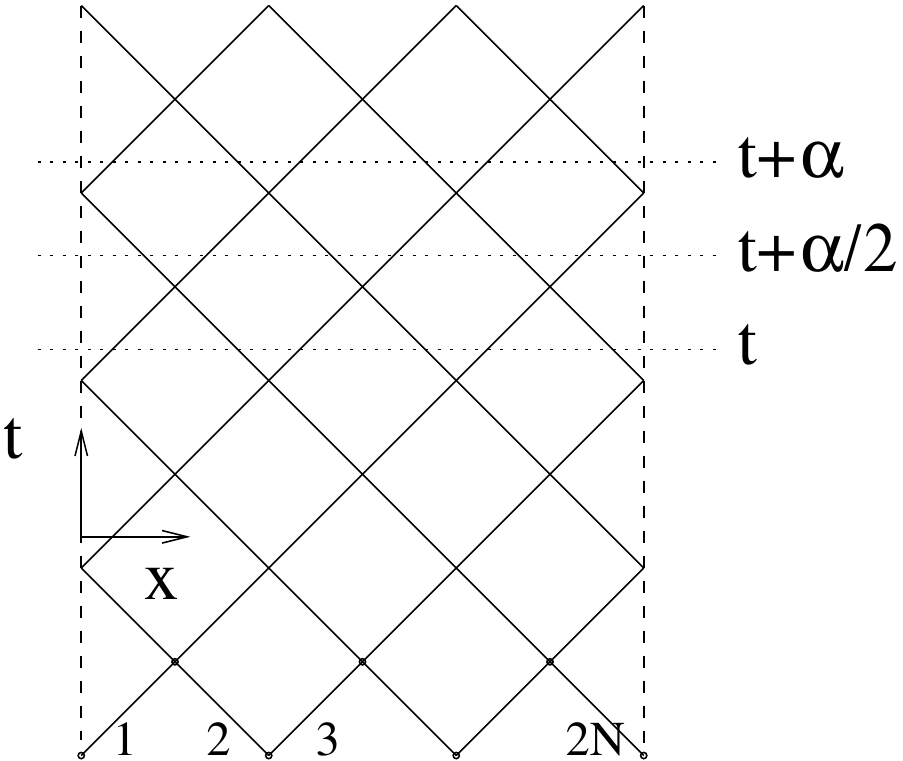}
\caption{Light-cone lattice with periodic boundary conditions in space direction; 
states are associated to edges, enumerated from 1 to $2N$.
\label{lcl1}}
\end{figure}

\begin{figure}[h]
\hspace*{18mm}\includegraphics[scale=0.7]{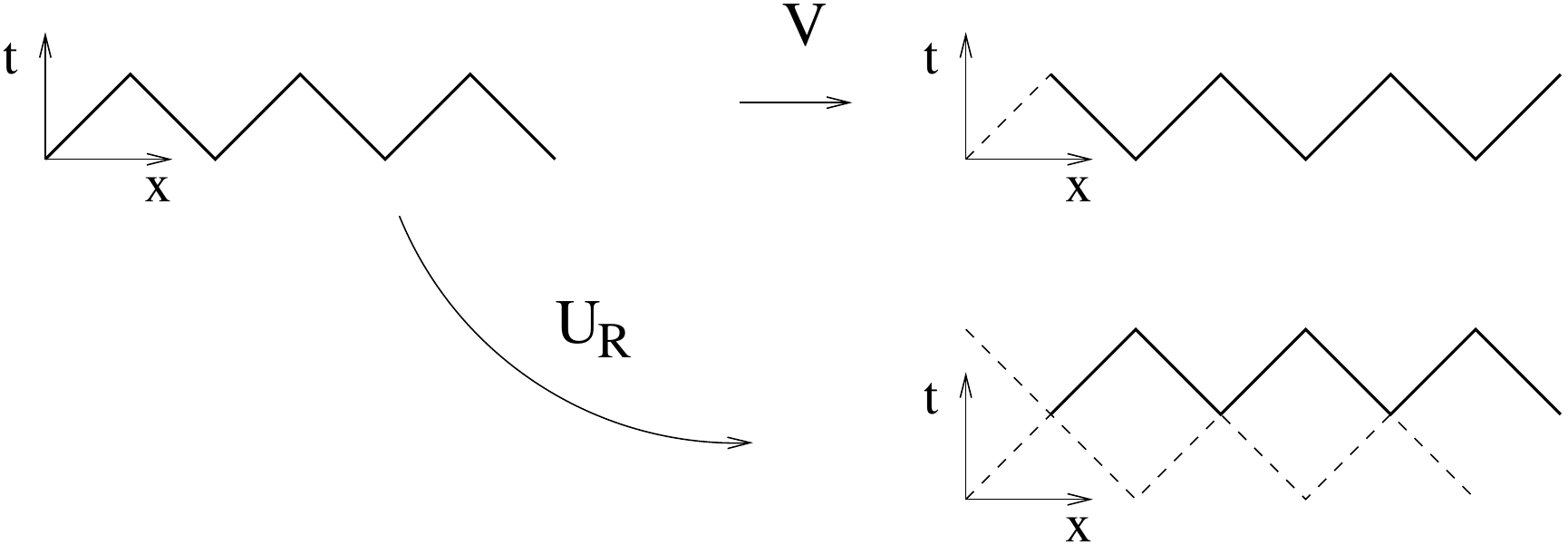}
\caption{The action of the operators $V$ and $U_R$ is illustrated. The operator $U_L$ 
acts as $U_R$ but with leftward movement.\label{lcl2}}
\end{figure}

In two dimensions, the most obvious approach would be to use a rectangular lattice with
axes corresponding to space and time directions. Here, a different approach \cite{ddv87} 
is adopted where space-time is discretized along light-cone directions.
Light-cone coordinates in Euclidean or Minkowski space-time are
\be
x_{+}=x+t\,,\quad x_{-}=x-t
\ee
When discretized, they define a light-cone lattice of ``events'' as in figure \ref{lcl1}.
Then, any rational and not greater than \( 1 \) value is permitted as particle speed,
in an infinite lattice. The shortest displacement of the particle (one lattice
spacing) is realized at light speed \( \pm 1 \) and corresponds, from the statistical
point of view, to nearest neighbors interactions. Particles are therefore massless
and can be right-movers (R) or left-movers (L) only.
Smaller speeds can be obtained with displacements longer than the fundamental cell and 
correspond to higher neighbors interactions. They will not be used here.
With only nearest neighbor interactions, the evolution from one row to the next one, 
as in figure \ref{lcl1}, is governed by a transfer matrix.
Here there are four of them. Two act on the light-cone, $U_{R}$ and $U_{L}$, 
the first one shifting the state of the system one step forward-right, the other one 
step forward-left. The remaining $U$ and $V^2$ act in time and space directions respectively
\be
U=e^{-i\,\alpha\,H}=U_RU_L\,,\qquad V^2=e^{-i\,\alpha\,P}=U_RU_L^{\dagger}
\ee
so they actually correspond to the Hamiltonian (forward shift) and the total momentum (right shift). 
Their action is pictorially suggested in figure~\ref{lcl2}. Also, the relations hold
\be\label{EP}
U_R=e^{-i\,\frac{\alpha}{2}(H+P)}\,,\qquad U_L=e^{-i\,\frac{\alpha}{2}(H-P)}
\ee
Much more details are given in \cite{tesidott} and on the original papers cited there.
I now construct the alternating transfer matrix 
\be\label{atm}
\mathbf{T}(\theta,\Theta,\gammasv)=\mathop{\mbox{Tr}}_A
L_{1}(\theta-\Theta)L_{2}(\theta+\Theta)\ldots L_{2N-1}(\theta-\Theta)L_{2N}(\theta+\Theta)
\ee
where the operators $L_i$ are associated to a vertex. To remove ambiguity, one has to associate 
odd numbers to the lower vertex, even numbers to the upper vertex, compare with Figure~\ref{lcl1}.
$\Theta$ for the moment is a free parameter. Its presence corresponds to make the lattice 
model inhomogeneous such that the interaction on each site is tuned by the presence of the 
inhomogeneity. This does not change integrability. 
The standard construction of (\ref{trasf}) suggest to take  
\be
L_i(\theta)=R_{Ai}(\theta)
\ee
using the R matrix of the six-vertex model (\ref{R_matrix_entries}).
The forward-right and the forward-left operators are obtained by
\be
U_R=\mathbf{T}(\Theta,\Theta,\gammasv)\,,\qquad 
U_L=\mathbf{T}(-\Theta,\Theta,\gammasv)
\ee
One can interpret these expressions by noticing that to switch from a right mover to a left mover 
one has to change the sign of rapidity. 
The transfer matrices depend on $\Theta$ and on the coupling
$\gammasv$, whose values are not yet specified.
The methods of Bethe ansatz can be used to diagonalize these operators as in 
the subsection~\ref{s:baxter}. 
The specific case is treated in \cite{devega89} and gives the following results.
The eigenvalues of the transfer matrix are given by
\be\label{eigenT}
\begin{array}{c}
\displaystyle T (\theta ,\Theta )=\left[ a(\theta -\Theta )\, a(\theta +\Theta )\right] ^{N}\prod ^{M}_{j=1}\frac{\sinh \displaystyle\frac{\gammasv }{\pi }\left[ i\frac{\pi }{2}+\vartheta _{j}+\theta \right] }{\sinh \displaystyle\frac{\gammasv }{\pi }\left[ i\frac{\pi }{2}-\vartheta _{j}-\theta \right] }+\\[7mm]
+\displaystyle \left[ b(\theta -\Theta )\, b(\theta +\Theta )\right] ^{N}\prod ^{M}_{j=1}\frac{\sinh \displaystyle\frac{\gammasv }{\pi }\left[ i\frac{3\pi }{2}-\vartheta _{j}-\theta \right] }{\sinh \displaystyle\frac{\gammasv }{\pi }\left[ -i\frac{\pi }{2}+\vartheta _{j}+\theta \right] }
\end{array}
\ee
provided the values of $ \vartheta _{j} $ satisfy the set of coupled nonlinear
equations called Bethe equations
\begin{equation}
\label{bethe}
\left( \displaystyle\frac{\sinh \displaystyle\frac{\gammasv }{\pi }\left[ \vartheta _{j}+\Theta +\displaystyle\frac{i\pi }{2}\right] \sinh \displaystyle\frac{\gammasv }{\pi }\left[ \vartheta _{j}-\Theta +\displaystyle\frac{i\pi }{2}\right] }{\sinh \displaystyle\frac{\gammasv }{\pi }\left[ \vartheta _{j}+\Theta -\displaystyle\frac{i\pi }{2}\right] \sinh \displaystyle\frac{\gammasv }{\pi }\left[ \vartheta _{j}-\Theta -\displaystyle\frac{i\pi }{2}\right] }\right) ^{N}=-\prod _{k=1}^{M}\displaystyle\frac{\sinh \displaystyle\frac{\gammasv }{\pi }\left[ \vartheta _{j}-\vartheta _{k}+i\pi \right] }{\sinh \displaystyle\frac{\gammasv }{\pi }\left[ \vartheta _{j}-\vartheta _{k}-i\pi \right] }
\end{equation}
where $M$ can by any integer from $0$ to $N$ included. The complex numbers $\{ \vartheta _{j} \}$ are called \emph{Bethe roots}. These equations are a modification of (\ref{betheTQ}).
Because of the periodicity 
\begin{equation}
\label{periodicita'}
\vartheta _{j}\, \rightarrow \, \vartheta _{j}+\displaystyle\frac{\pi ^{2}}{\gammasv }i
\end{equation}
further analyses can be restricted to a strip around the real axis 
\begin{equation}
\label{strisciafisica}
\vartheta _{j}\in \mathbb {R}\times i\, \left] -\displaystyle\frac{\pi ^{2}}{2\gammasv },\displaystyle\frac{\pi ^{2}}{2\gammasv }\right] .
\end{equation}
Details on Bethe equations and Bethe roots were given in subsection~\ref{s:baxter}.

Another form of the Bethe equations can be obtained by taking the logarithm of the previous one.
It is important to fix and consistently use a logarithm determination: here the fundamental one
will be used. The equations become
\bea
2\pi I_j&=&
N\log \displaystyle\frac{\sinh \displaystyle\frac{\gammasv }{\pi }\left[ \vartheta _{j}+\Theta +\frac{i\pi }{2}\right]}{\sinh \frac{\gammasv }{\pi }\left[ \vartheta _{j}+\Theta -\displaystyle\frac{i\pi }{2}\right]}+N\log\frac{\sinh \displaystyle\frac{\gammasv }{\pi }\left[ \vartheta _{j}-\Theta +\frac{i\pi }{2}\right] }{ \sinh \displaystyle\frac{\gammasv }{\pi }\left[ \vartheta _{j}-\Theta -\frac{i\pi }{2}\right] }
\nonumber \\
&-&\sum _{k=1}^{M}\log\displaystyle\frac{\sinh \displaystyle\frac{\gammasv }{\pi }\left[ \vartheta _{j}-\vartheta _{k}+i\pi \right] }{\sinh \displaystyle\frac{\gammasv }{\pi }\left[ \vartheta _{j}-\vartheta _{k}-i\pi \right] }\label{bethelog}
\eea
where their nature of quantization conditions is now explicit: the $I_j$ are quantum numbers,
taken half-integers or integers according to the parity of the number of Bethe roots
\be
I_j\in \mathbb{Z}+\frac{1+M}{2}
\ee
The energy \( E \) and momentum \( P \) of a state can be read out from (\ref{EP}) and (\ref{eigenT})
\begin{equation}
\label{autovalori}
\displaystyle e^{i\displaystyle\frac{\alpha}{2}(E\pm P)}=\prod ^{M}_{j=1}\frac{\sinh \displaystyle\frac{\gammasv }{\pi }\left[ i\displaystyle\frac{\pi }{2}-\Theta \pm \vartheta _{j}\right] }{\sinh \displaystyle\frac{\gammasv }{\pi }\left[ i\displaystyle\frac{\pi }{2}+\Theta \mp \vartheta _{j}\right] }
\end{equation}
or by the same equation in logarithmic form 
\be\label{autovalorilog}
E\pm P=-i\frac{2}{\alpha}\sum ^{M}_{j=1}\log \frac{\sinh \displaystyle\frac{\gammasv }{\pi }\left[ i\displaystyle\frac{\pi }{2}-\Theta \pm \vartheta _{j}\right] }{\sinh \displaystyle\frac{\gammasv }{\pi }\left[ i\displaystyle\frac{\pi }{2}+\Theta \mp \vartheta _{j}\right] }
\ee
The logarithmic forms reveal an interesting aspect of the Bethe ansatz namely that energy, 
momentum, spin (see later) and all the higher integrals of motion have 
an additive structure in which Bethe roots resemble rapidities of independent particles 
\be\label{quasipart}
I_l=\sum^{M}_{j=1} f_l(\vartheta_j)
\ee
called ``quasiparticles''. Quasiparticles are usually distinct from physical particles. They are degrees of 
freedom that do not appear in the Hamiltonian (\ref{hamilt}) but are ``created'' by the Bethe ansatz
and incorporate the effects of the interactions. Indeed, their dispersion relation is not Galilean nor 
relativistic\footnote{Galilean: $E=\frac{p^2}{2m}$, relativistic $E=\sqrt{p^2 c^2+m^2 c^4}$.}.
Quasiparticles can have complex ``rapidities''.

With this quasiparticle nature of Bethe roots in mind, the left hand side of the Bethe equations
(\ref{bethe}) is precisely the $j$th momentum term that one can extract from (\ref{autovalorilog}).
The right hand side represents the interaction of pairs of quasi-particles. 

In this Bethe ansatz description, the third component of the spin is given by
\be\label{spinz}
S_z=N-M
\ee
where the reference state for the algebraic Bethe ansatz 
is taken with all spins up or all spins down and it is described by $M=0$ in (\ref{bethe}): 
it is the ferromagnetic state.
Then, every Bethe root corresponds to overturning a spin: it is a ``magnon'', because it 
carries a unit of ``magnetization''. It is also called spin wave. It is the smallest 
``excitation'' of the ferromagnetic state.
When $M=N$, all roots are real and one has the antiferromagnetic state, that is an ordered state
with zero total spin but with a nontrivial local spin organization. Here, with the present sign 
conventions, it is also the ground state.
Its actual expression is complicated. Just to give an idea of what it can look like, 
in the simplest case of an homogeneous ($\Theta=0$) XXX ($\gammasv=0$) model for a two-site chain 
it is
\be
|\uparrow\downarrow\rangle-|\downarrow\uparrow\rangle
\ee
while for four sites it is
\be
|\uparrow\uparrow\downarrow\downarrow\rangle+|\downarrow\uparrow\uparrow\downarrow\rangle+
|\downarrow\downarrow\uparrow\uparrow\rangle+|\downarrow\uparrow\uparrow\downarrow\rangle-2
(|\uparrow\downarrow\uparrow\downarrow\rangle+|\downarrow\uparrow\downarrow\uparrow\rangle)
\ee
Excitations of the antiferromagnetic state are: (1) ``real Bethe holes'', namely real positions 
corresponding to real roots in the antiferromagnetic state but excluded in the excitation,
(2) complex roots. 

In what follows, only the antiferromagnetic ground state and its excitations will be considered,
because it has one important property: in the thermodynamic limit 
\( N\, \rightarrow \, \infty  \) it can be interpreted as a Dirac sea and the its excitations, 
holes and complex roots, behave as particles. 

For later convenience, the coupling constant \( \gammasv  \) is expressed 
in terms of a different variable \( p \):
\be\label{pp}
p=\frac{\pi }{\gammasv }-1,\qquad 0<p<\infty 
\ee
and I will work in the range of $0<\gammasv<\pi $. Notice that, in (\ref{xxz}) and in 
table~\ref{xxzcoupl}, this is the choice that corresponds to the critical regime.
In this new parameter, the strip becomes
\be
\label{striscia}
\vartheta _{j}\in \mathbb {R}\times i\, \left] -\displaystyle\frac{\pi (1+p)}{2},
\displaystyle\frac{\pi (1+p)}{2}\right] .
\ee
This new parameter is related to the sine-Gordon ones by 
\be\ba{c}
p=\frac{\beta^2}{8\pi-\beta^2}\\
0<p<1  \quad \mbox{attractive regime};\qquad 1<p <\infty\quad  \mbox{repulsive regime}
\ea\ee
see also (\ref{repattr}). With this parameter, the relation between the six-vertex and sine-Gordon coupling is
\be
\gamma=8\pi p=8\pi\left(\frac{\pi}{\gammasv}-1\right)
\ee
Thus, the XXX chain is characterized by $\gammasv=0$ that means $\beta^2=8\pi$. This is the strongest point in the 
repulsive regime of sine-Gordon. Afther that point, the quantum sine-Gordon model seems to loose meaning.

\section{A nonlinear integral equation from the Bethe ansatz}

In this section the fundamental nonlinear integral equation driving sine-Gordon scaling
functions will be presented. In the literature it is known with several names, following the different
formulations that have been done: Kl\"umper-Batchelor-Pearce equation or Destri-de Vega equation. It
has also been indicated with the nonspecific tetragram NLIE (nonlinear integral equation). 
It was first obtained in \cite{klumper} for the vacuum (antiferromagnetic ground state) 
scaling functions of XXZ then, with different methods, in \cite{ddv95} 
for XXZ and sine-Gordon. I will follow the Destri-de Vega approach applied to the sine-Gordon model.

The treatment of excited states was pioneered in \cite{fioravanti} and refined
in \cite{ddv97,noi PL1}, to arrive to the final form in \cite{noiNP,tesidott}.

It is important to stress that this formalism is equivalent to the Bethe equations but
is especially indicated to the antiferromagnetic regime. In general, it adapts to regimes
where the number of Bethe roots is of the order of the size of the system.
Indeed, the key idea is to sum up a macroscopically large number of Bethe roots 
for the ground state or for the reference state 
and replace them by a small number of holes to describe deviations near
the reference state, as holes in a Dirac sea.

\subsection{Counting function\label{section:count-funct}}

First, it is possible to write the Bethe equations (\ref{bethelog}) in terms of a 
\emph{counting function} \( Z_{N}(\vartheta ). \) 
I introduce the function 
\[
\phi _{\nu }(\vartheta )=i\log \frac{\sinh \frac{1}{p+1}(i\pi \nu +\vartheta 
)}{\sinh \frac{1}{p+1}(i\pi \nu -\vartheta )},\qquad \phi _{\nu }(-\vartheta 
)=-\phi _{\nu }(\vartheta )\]
The ``oddity'' on the analyticity strip around the real axis defines a precise
logarithmic determination. The counting function is defined by 
\begin{equation}
\label{def.Zn}
\displaystyle Z_{N}(\vartheta )=N[\phi _{1/2}(\vartheta +\Theta )+\phi 
_{1/2}(\vartheta -\Theta )]-\sum _{k=1}^{M}\phi _{1}(\vartheta -\vartheta 
_{k})
\end{equation}
The logarithmic form of the Bethe equations (\ref{bethelog}) takes now 
a simple form in terms of the counting function
\begin{equation}
\label{quantum}
\displaystyle Z_{N}(\vartheta _{j})=2\pi I_{j}\, ,\quad I_{j}\in \mathbb 
{Z}+\frac{1+\delta }{2},\quad \delta =(M) \text{mod}\, 2=(N-S_z) \text{mod}\, 2\in 
\left\{ 0,1\right\}
\end{equation}
Notice that the counting function is not independent of the Bethe roots. 
Said differently, one cannot separate the construction of $Z_N$ and the 
solution of (\ref{quantum}). 
Now it is possible to give a formal definition of ``holes'':
they are solutions of (\ref{quantum}) that do not appear in (\ref{def.Zn}).
I will not make use of nonreal holes.
Bethe roots and holes are zeros of the equation
\be
\label{zero-condition}
1+(-1)^{\delta }e^{iZ_{N}(\vartheta _{j})}=0
\ee
once the counting function is known. More, they are simple zeros because Bethe roots/holes 
exclude each other.

\subsection{Classification of Bethe roots and counting 
equation\label{classificazione}}

From Bethe Ansatz it is known that a solution of (\ref{bethelog}), namely a Bethe state,
is uniquely characterized
by the set of quantum numbers \( \left\{ I_{j}\right\} _{j=1,...,M}\, ,\quad 0\leq 
M\leq N \)
that appear in (\ref{quantum}). Notice that \( M\leq N \) means \( S\geq 0. \)

Bethe roots can either be real or appear in complex conjugate pairs. Complex conjugate pairs 
grant the reality of the energy, momentum and  transfer matrix.
In the specific case (\ref{bethe}), there is another possibility, due to the periodicity
(\ref{periodicita'}): if a complex solution has imaginary part \( \text{Im}\, 
\vartheta =\frac{\pi }{2}(p+1) \), it can appear alone (its complex conjugate is not required).
A root with this value of the imaginary part is called \emph{self-conjugate root}. 

From the point of view of the counting function, a more precise classification of roots is required:

\begin{itemize}
\item \emph{real roots}; they are real solutions of (\ref{quantum}); their number is \( M_{R} \); 
\item \emph{holes}; real solutions of (\ref{quantum}) that do appear in the 
ground state but not in the excitation under examination; in practice, they are real 
of (\ref{quantum}) that do not enter in the counting function (\ref{def.Zn}); 
their number is \( N_{H} \); 
\item \emph{special roots or holes} (special objects); they are real roots or 
holes whose counting function derivative \( Z_{N}'(\vartheta _{j}) \) is negative,
contrasted with normal roots or holes, whose derivative is positive; 
their number is \( N_{S} \); they must be counted both as {}``normal{}''
and as {}``special{}'' objects; 
\item \emph{close pairs}; complex conjugate solutions with imaginary part in 
the range \( 0<|\im\, \vartheta |<\pi \min (1,p) \); this range is dictated by the first 
singularity (essential singularity) of the function $\phi_1(\theta)$, when moving of the real axis;
their number is \( M_{C} \); 
\item \emph{wide roots in pairs}: complex conjugate solutions with imaginary 
part belonging to the range \( \pi \min (1,p)<|\text{Im}\, \vartheta |<\pi \frac{p+1}{2} \)
namely after the first singularity of $\phi_1(\theta)$;
\item \emph{self-conjugate roots}: complex roots with imaginary part \( \text{Im}\, 
\vartheta =\pi \frac{p+1}{2} \); they are wide roots but miss the complex conjugate
so they are single; their number is \( M_{SC} \).
\end{itemize}
The total number of wide roots appearing in pairs or as single is \( M_{W} \).
The following notation will be used, sometimes, for later convenience, to 
indicate
the position of the solutions: \( h_{j} \) for holes, \( y_{j} \) for special
objects, \( c_{j} \) for close roots, \( w_{j} \) for wide roots.

\begin{figure}[h]
{\par\centering 
\resizebox*{0.9\textwidth}{0.37\textheight}{\includegraphics{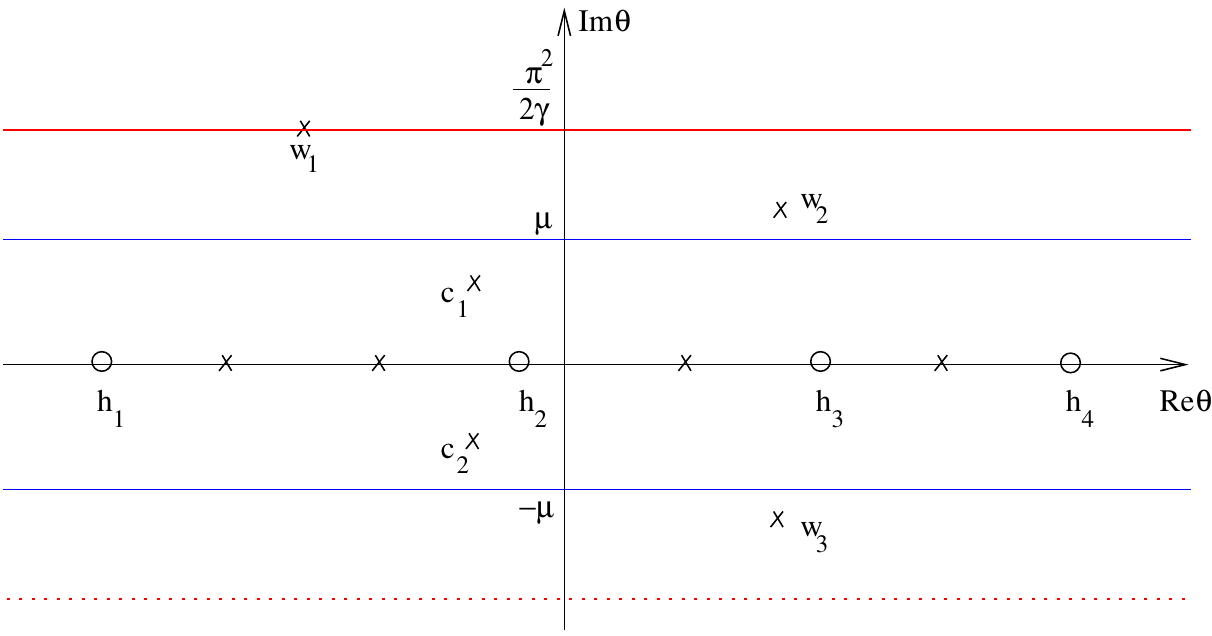}} 
\par}
\caption{The different types of roots and holes and their position in the complex 
plane. We denote $\mu=\pi \min (1,p) $.
The red line at \protect\( \protect\frac{\pi ^{2}}{2\gammasv }=
\protect\frac{\pi}{2}(p+1)\protect\)
is the self-conjugate one. Close roots are located within the blue lines, wide roots lie outside.
The holes are indicated by the circles on the real axis.\label{radici.eps}}
\end{figure}

Complex roots with imaginary part larger than the self-conjugates are not 
required
because of the periodicity of Bethe equations (\ref{strisciafisica}). 
A graphical representation of
the various types of solutions is given in figure \ref{radici.eps}. 

The function \( Z_{N} \) (\ref{def.Zn}) has a number of branch point singularities produced by the 
presence of the logarithms. 
The largest horizontal strip containing the real axis and free of singularities is
bounded by the singularities of the various terms \( \phi _{\nu }(\vartheta ) \).
The strip is at the largest size when no complex roots are introduced, otherwise 
it is narrower because the imaginary part of the complex roots in 
\(\phi _{\nu }(\vartheta-\vartheta_k ) \) displaces the position of the singularities.

An important property follows from this classification: the \( Z_{N} \) 
function is \emph{real analytic} in a strip that contains the real axis
\begin{equation}
\label{analiticita_{r}eale}
Z_{N}\left( \vartheta ^{*}\right) =\left( Z_{N}(\vartheta )\right) ^{*}
\end{equation}

By considering asymptotic values of \( \phi _{\nu }(\vartheta ) \) and \( 
Z_{N}(\vartheta ) \)
for \( \vartheta \rightarrow \pm \infty  \), it is possible to obtain an 
equation
relating the numbers of all the various types of roots. I refer the reader
interested in the details of the derivation to \cite{ddv97,tesidott}.
Here I only mention the final result, in the form where the continuum limit
\( N\rightarrow \infty  \), \( a\rightarrow 0 \) and \( L=Na \) finite, has 
already been taken\begin{equation}
\label{counting-eq}
N_{H}-2N_{S}=2S_z+M_{C}+2\, \theta (p-1)\, M_{W}
\end{equation}
 where \( \theta (x) \) is the step function: \( \theta (x)=0 \) for \( x<0 \)
and \( \theta (x)=1 \) for \( x>0 \). 
Recall that \( S_z \) is a nonnegative integer and the right hand side only contains nonnegative 
values. From this, it turns out that \( N_{H} \) is even (\( M_{C} \) is the number of 
close roots, and is even).
This \emph{counting equation} expresses the fact that
the Bethe equations have a finite number of solutions only. There are also other constraints, 
once $N$ is fixed:
\be
S_z\leq N\,,\qquad M_C+M_W\leq N\,,\qquad N_H\leq N.
\ee
Moreover, the various types of roots/holes do respect the mutual exclusion principle. 
This means that, in order to accommodate complex roots, one has to ``create'' space by inserting holes,
or vice versa. 

Notice also that in the attractive regime the wide roots do not participate to
the counting and that at the free fermion point $p=1$, or $\beta^2=4\pi$, 
they do exist as self-conjugate only, namely there are no wide roots in pairs. This suggests that 
the role of wide roots is different in the two regimes.

The most important fact is that the number of real roots does not appear in
this equation: they have been replaced by the number of holes. This, together to what will be 
explained in the next paragraph, allows to consider the real roots as a sea of particles, or Dirac sea, 
and all other types of solutions, holes and complex roots, as excitations above it.

\subsection{Nonlinear integral equation\label{section:NLIE_1}}

\begin{figure}[h]
\begin{center}\includegraphics[scale=0.7]{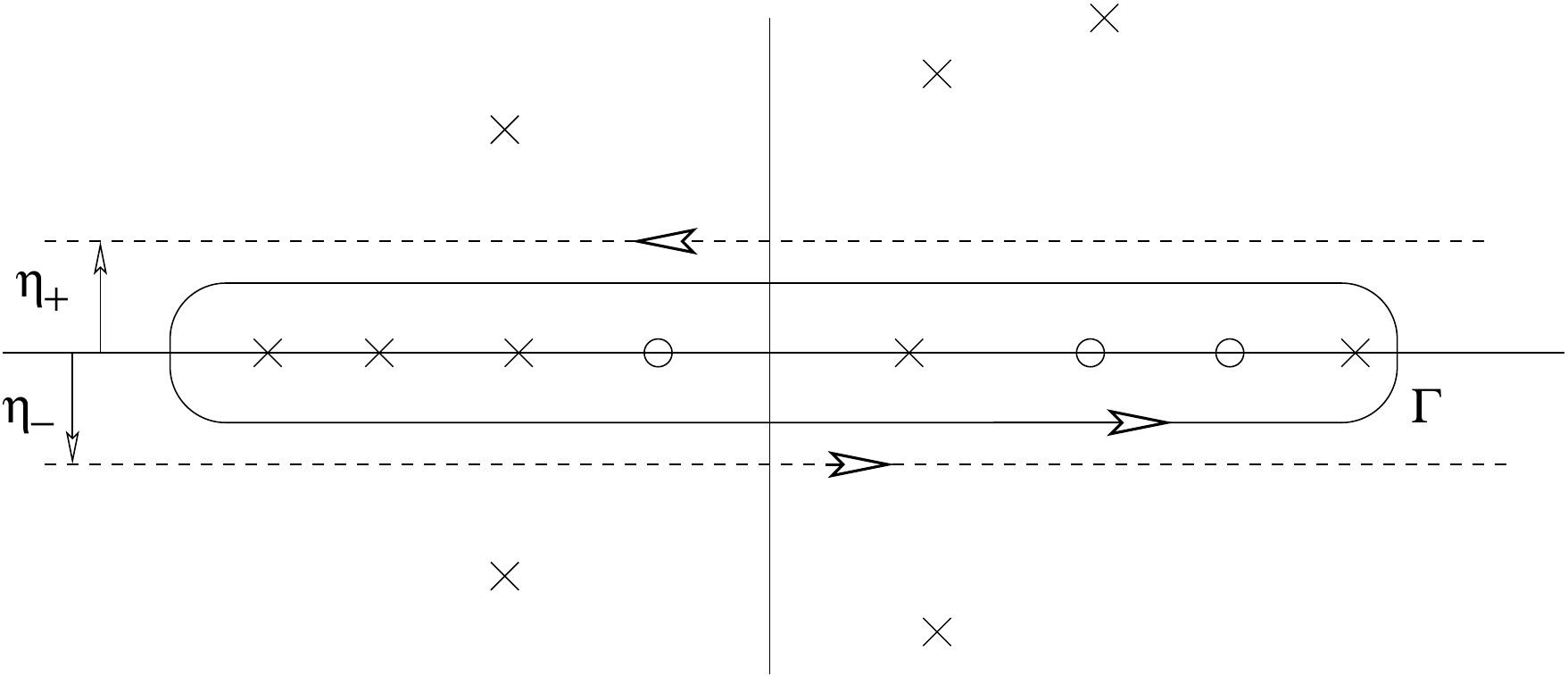}\end{center}
\caption{The integration curve encircles real solutions only. The crosses represent roots while 
the circles represent holes. \label{curvagamma.eps}}
\end{figure}

\noindent Let \( \hat{x} \) be a real solution of the Bethe equation. Thanks
to Cauchy's integral formula, an holomorphic function \( f(x) \) 
admits the following representation 
\begin{equation}
\label{cauchy}
\displaystyle f(\hat{x})=\oint _{\Gamma _{\hat{x}}}\frac{d\mu }{2\pi 
i}\frac{f(\mu )}{\mu -\hat{x}}=\oint _{\Gamma _{\hat{x}}}\frac{d\mu }{2\pi 
i}f(\mu )\frac{(-1)^{\delta }e^{iZ_{N}(\mu )}iZ_{N}'(\mu )}{1+(-1)^{\delta 
}e^{iZ_{N}(\mu )}}
\end{equation}
where \( \Gamma _{\hat{x}} \) is an anti-clockwise simple path encircling \( \hat{x} \), namely 
one of the real holes or complex roots, and avoiding all the others, see (\ref{zero-condition}).
In the region where \( \phi _{1}(\vartheta ) \) is holomorphic, (\ref{cauchy}) can be used to write
an expression for all the real roots and real holes
\begin{equation}
\label{integr_{g}amma}
\begin{array}{c}
\displaystyle \sum ^{M_{R}+N_{H}}_{k=1}\phi _{1}(\vartheta -x_{k})=\sum 
^{M_{R}+N_{H}}_{k=1}\oint _{\Gamma _{x_{k}}}\frac{d\mu }{2\pi i}\phi 
_{1}(\vartheta -\mu )\frac{(-1)^{\delta }e^{iZ_{N}(\mu )}iZ_{N}'(\mu 
)}{1+(-1)^{\delta }e^{iZ_{N}(\mu )}}=\\
\displaystyle =\oint _{\Gamma }\frac{d\mu }{2\pi i}\phi _{1}(\vartheta -\mu 
)\frac{(-1)^{\delta }e^{iZ_{N}(\mu )}iZ_{N}'(\mu )}{1+(-1)^{\delta 
}e^{iZ_{N}(\mu )}}
\end{array}
\end{equation}
 The sum on the contours has been modified to a unique curve \( \Gamma  \) 
encircling all the real solutions \( {x_{k}} \) (real root plus holes), and avoiding the complex 
Bethe solutions as in the Figure~\ref{curvagamma.eps}; this is possible because they are finite 
in number.

Clearly the \( \Gamma  \) curve must be contained in the strip \[
0<\eta _{+},\eta _{-}<\min \{\pi ,\pi p,|\text{Im}\, c_{k}|\, \forall \, k\}\]
 Without loss of generality, assume that \( \eta _{+}=\eta _{-}=\eta  \), and
deform \( \Gamma  \) to the contour of the strip characterized by \( \eta  \).
The regions at \( \pm \infty  \) do no contribute because, as the lattice size is finite, those
regions are free of root or holes. Moreover, in those regions, $ Z_N' $ vanishes
therefore the integral can be just evaluated on the lines \( \mu =x\pm i\eta  \),
where \( x \) is real. After algebraic manipulations involving integrations
by parts and convolutions (for details see \cite{tesidott}) one arrives
at a nonlinear integral equation for the counting function \( Z_{N}(\vartheta) \)
\be
\begin{array}{rl}
\displaystyle Z_{N}(\vartheta ) & 
\displaystyle =2\,N\arctan \frac{\sinh \vartheta }{\cosh 
\Theta }+\sum ^{N_{H}}_{k=1}\chi (\vartheta -h_{k})-2\sum ^{N_{S}}_{k=1}\chi 
(\vartheta -y_{k})-\\
 & \displaystyle -\sum ^{M_{C}}_{k=1}\chi (\vartheta -c_{k})
 -\sum ^{M_{W}}_{k=1}\chi (\vartheta -w_{k})_{II}+\\
 & \displaystyle +2\, \im \int ^{\infty 
}_{-\infty }d\rho\, G(\vartheta -\rho -i\eta )\log \left( 1+(-1)^{\delta 
}e^{iZ_{N}(\rho +i\eta )}\right) 
\end{array}\ee
The kernel 
\be
\label{funzioneG}
G(\vartheta )=\frac{1}{2\pi }\int ^{+\infty }_{-\infty }dk\, e^{ik\vartheta 
}\frac{\sinh \frac{\pi (p-1)k}{2}}{2\sinh \frac{\pi pk}{2}\, \cosh \frac{\pi 
k}{2}}
\ee
presents a singularity at the same place where \( \phi _{1}(\vartheta ) \)
does: $\vartheta=i\pi\min\{1,p\}$. 
An analytic continuation outside the fundamental strip \( 0<|\im\, \vartheta |<\pi \min \{1,p\} \)
(I determination region) must take this fact into account.
The source terms are given by \[
\chi (\vartheta )=2\pi \int _{0}^{\vartheta }dx\ G(x)\]
and 
\be
\chi (\vartheta )_{II}=\left\{ 
\begin{array}{ll}
\chi (\vartheta )+\chi \left( \vartheta -i\,\pi\ \mathrm{sign}\left( \im\,\vartheta \right) \right) \, ,& p>1\, ,\\
\chi (\vartheta )-\chi \left( \vartheta -i\,p\,\pi\ \mathrm{sign}\left( \im\,\vartheta \right) \right) \, , & p<1\, .
\end{array}
\right. 
\ee
is a modification of the source term due to the analytic continuation over the
strip \( 0<|\im\, \vartheta |<\pi \min\{1,p\} \), i.e. in the so called II
determination region.

Such equation, together with the quantization condition (\ref{quantum}), is equivalent to the 
original Bethe Ansatz (\ref{bethe}).
The NLIE for $Z_N$ is not independent of the Bethe roots: it and the quantization conditions must be 
solved at the same time. 
Once the Bethe roots are known, one can use them into eqs.(\ref{autovalori})
to compute the energy and momentum of a given state.

\subsection{Continuum limit\label{section:limite_cont.}}

Although such NLIE is already a precious tool for the lattice model itself,
its importance becomes essential when a continuum limit is done.

The continuum limit has the objective to transform a lattice system into a continuum model.
As already mentioned, one has to take the lattice size \( N\rightarrow \infty \) (that would be 
the normal thermodynamic limit of statistical mechanics) and the lattice edge \( \alpha\rightarrow 0 \)
simultaneously, in such a way that the product \( L=N\,\alpha \) stays finite. 
In this way one obtain a continuum theory with finite size space, namely a cylindrical geometry. 
However, the lattice spacing is not present in the Boltzmann weights and in the 
transfer matrix (\ref{atm}, \ref{R_matrix_entries}) so we have no way to use it. 
Moreover, one can convince himself, by explicit calculations, that if the
limit is taken by keeping the \( \Theta  \) parameter fixed, the lattice NLIE
blows up to infinity and looses meaning. This reflects the fact that the number
of roots increases as \( N \) in the thermodynamic limit. However, as shown in ref.\cite{ddv87}, 
if one assumes a dependence of \( \Theta  \) on \( N \) of the form
\begin{equation}
\label{teta_{n}}
\displaystyle \Theta \approx \log \frac{4N}{{\mathcal{M}}L}.
\end{equation}
it is possible to get a finite limit out of the lattice NLIE. This limit is exactly 
the one that was used in \cite{ddv87} to bring a lattice fermion field into the Thirring fermion field 
on the continuum. 
Notice that sending \( \Theta \rightarrow \infty  \)
in this way naturally introduces a renormalized physical mass \( {\mathcal{M}} \). This is the 
deep reason of the use of the light-cone lattice and of the inhomogeneity in 
(\ref{atm}). In other words, without the inhomogeneity, the continuum system would be a 
critical one, massless, with central charge $c=1$ (\cite{klumper}).

The \emph{continuum counting function} is defined by: \begin{equation}
\label{def_{Z}}
\displaystyle Z(\vartheta )=\lim _{N\, \rightarrow \, \infty }Z_{N}(\vartheta )
\end{equation}
and appears in a continuum NLIE 
\begin{equation}
\label{nlie-cont}
\displaystyle Z(\vartheta )=l\sinh \vartheta +g(\vartheta |\vartheta 
_{j})+2 \im \int ^{\infty }_{-\infty }dx\,G(\vartheta -x-i\eta )\log 
\left( 1+(-1)^{\delta }e^{iZ(x+i\eta )}\right) 
\end{equation}
where \( l={\cal M}L \). The first term on the right hand side is a momentum term.
The second one, \( g(\vartheta |\vartheta _{j}) \), is a source term, in the sense that is adapts 
to the different combinations of roots and holes
\be\label{source}
\displaystyle g(\vartheta |\vartheta _{j})=\sum ^{N_{H}}_{k=1}\chi (\vartheta 
-h_{k})-2\sum ^{N_{S}}_{k=1}\chi (\vartheta -y_{k})-\sum ^{M_{C}}_{k=1}\chi 
(\vartheta -c_{k})-\sum ^{M_{W}}_{k=1}\chi (\vartheta -w_{k})_{II}
\ee
The positions of the sources \( \{\vartheta _{j}\}\equiv \{h_{j}\, ,\, y_{j}\, ,\, c_{j}\, ,\, w_{j}\} \)
are fixed by the Bethe quantization conditions
\be\label{quantum2}
Z(\vartheta _{j})=2\pi I_{j}\quad ,\quad I_{j}=\mathbb Z+\frac{1+\delta }{2}
\ee
The parameter \( \delta  \) can be both 0 or 1. On the lattice it was determined by the total number 
of roots, which now has become infinite. Restrictions on it will be appear later.
The vacuum state, or Hamiltonian ground state, corresponds to the choice \( \delta =0 \). 

With a procedure analogous to the one sketched above, it is possible to produce integral expressions 
for energy and momentum. Starting from (\ref{autovalori}), one has to isolate an extensive term, 
proportional to \( N \), to be subtracted. The remaining finite part of energy and momentum takes the 
form 
\bea
\nonumber
 E-E_{bulk} & =&{\cal M}\left(\sum ^{N_{H}}_{j=1}\cosh h_{j}-2\sum ^{N_{S}}_{j=1}\cosh y_{j}-\sum 
^{M_{C}}_{j=1}\cosh c_{j}+ \sum _{j=1}^{M_{W}}(\cosh w_{j})_{II}-\right.\\
 & &\displaystyle \left. -\int ^{\infty 
}_{-\infty }\frac{dx}{2\pi }2\,\im\left[ \sinh (x+i\eta )\log (1+(-1)^{\delta 
}e^{iZ(x+i\eta )})\right] \right) \label{energy}\\
P & =& {\cal M}\left(\sum ^{N_{H}}_{j=1}\sinh h_{j}-2
\sum^{N_{S}}_{j=1}\sinh y_{j}-\sum ^{M_{C}}_{j=1}\sinh c_{j}+
\sum _{j=1}^{M_{W}}(\sinh w_{j})_{II}-\right. \nonumber  \\
 & &\displaystyle \left. -\int ^{\infty}_{-\infty }\frac{dx}{2\pi }2\, \im\left[ \cosh (x+i\eta )
\log (1+(-1)^{\delta }e^{iZ(x+i\eta )})\right]\right) \label{momentum}
\eea
Therefore, energy and momentum can be evaluated once the counting function \( Z(\vartheta ) \) and the 
source positions \( \vartheta _{j} \) have been obtained from (\ref{nlie-cont}, \ref{quantum2}).
All these equations are exact, no approximation has been introduced in deriving them.

In practice, these equations can be treated analytically in certain limits ($l\rightarrow 0$ or 
$l\rightarrow \infty$) and numerically for intermediate values, where we actually lack a closed formula 
for them. Numerical computations are done without any approximation other than the technical ones 
introduced by the computer truncations.
In calculations, one starts from an initial guess for the counting function,
 \( Z^{(0)}(\vartheta ) \), uses it in (\ref{quantum2}) to get the roots/holes positions 
then evaluates a new  \( Z^{(1)}(\vartheta ) \) from (\ref{nlie-cont}) and so on, up to the required 
precision. 
This iterative procedure is conceptually very simple and inclined to good convergence, as one can 
easily estimate. Indeed, the following term appears within integration 
\be\label{Zapprox}
e^{iZ(x+i\eta )}\approx  e^{iZ-\eta\, l\,\cosh x}
\ee
The presence of a negative $\cosh$ at the exponent makes the support of the integral compact and 
the integral itself subdominant with respect to all the other terms, speeding up the convergence of 
the iteration, especially for large $ l $.

So, at least for the cases where holes only are considered, and no complex roots or no special roots, 
finding numerical solutions can be quite easy. The whole procedure takes few seconds of
computing on a typical Linux/Intel platform without resorting to any supercomputer or other 
technically advanced tool.

When complex roots are present, things are much more complicated and the computation time increases 
dramatically. Also, convergence at small $l$ can be problematic
because those real roots or holes that are closer to the origin can become ``special''. \label{lespecial}
In practice, 
they emanate one or two ``supplementary'' sources in consequence of a local change of sign of $Z'(x)$.
They have not been extensively treated. A similar phenomenon has been discussed in \cite{FPR2}, in the 
frame of thermodynamic Bethe ansatz.

The limit procedure described here is mathematically consistent, but the question is if from the 
physical point of view it describes a consistent quantum theory and allows for a meaningful 
physical interpretation.

The first indication comes from the emergence, in \cite{ddv87}, of the fermionic massive Thirring 
fields from the six-vertex diagonal alternating lattice of section~\ref{s:lightcone}.
This indicates that the procedure points toward a sine-Gordon/massive Thirring model. 

Before going on, an important remark must be made about the allowed values for
the XXZ spin \( S_z \). From (\ref{spinz}), only nonnegative integer values can be given to \( S_z \).

As shown in \cite{noiPL2}, one is led to include also the half-integer choice for \( S_z \),
in order to describe the totality of the spectrum. This choice seems not justified on the light-cone
lattice of section~\ref{s:lightcone} because it would require adding one column of points to the 
lattice, thus spoiling periodicity. Most probably the way to introduce it would be by inserting a twist 
in the seam or some other nontrivial boundary condition. 
In any case, half-integer values for $S_z$ are necessary and seem fully consistent with the rest of 
the model to describe odd numbers of particles.

At this point the following physical scenario appears.

\begin{enumerate}
\item The physical vacuum, or ground state, of the continuum theory comes from the antiferromagnetic 
state of the lattice so is characterized by the absence of sources, holes or complex roots.
\item All the sources are excitations above this vacuum.
\item This theory describes at least the sine-Gordon and the massive Thirring model on a cylinder; the
circumference describes a finite space of size \( L \); the infinite direction is time
\( S_z \) is the topological charge and can take nonnegative integer or half-integer values.
\item This theory describes also states that are not in sine-Gordon or in massive Thirring.
\end{enumerate}

As already observed, the real roots have disappeared from the counting equation in the continuum  
limit. They actually become a countable set and are taken into account by the integral term, both in 
the NLIE and in the energy-momentum expressions. They can be interpreted as a sort of Dirac sea on 
which holes and complex roots build particle excitations. Of course, 
the presence of holes or complex roots distorts the Dirac sea too, through the source term 
\( g(\vartheta |\vartheta _{j}) \). 

Observe that it has been indicated that only nonnegative values of \( S_z \) are required to describe 
the whole Hilbert space of the theory. Indeed the lattice theory is assumed charge-conjugation 
invariant so negative values of $S_z$, namely states with negative topological charge, have the same 
energy and momentum as their charge conjugate states. 

The assumption that all the sine-Gordon and all massive Thirring states can be described by the 
NLIE is absolutely not trivial and still misses a mathematical proof even if all the analyses 
done so far are consistent with this assumption.

\subsection{The infrared limit of the NLIE and the particle scattering\label{s:IR}}

The first task in order to understand the physics underlying the NLIE is to 
characterize the scattering of the model, by reconstructing the S-matrix. As we started from an 
integrable model, we assume that the continuum one remains integrable. We will find that this 
hypothesis is extremely reasonable because it is related to the structure of the source term 
$g(\theta|\theta_i)$. 
Indeed, the function \( \chi  \) can be written as 
\be
\chi (\vartheta )=-i\log S(\vartheta )
\ee
where \( S(\vartheta ) \) is the soliton-soliton scattering amplitude
in sine-Gordon theory \ref{matriceS}, if the parameters are fixed as in (\ref{pp}).
This means that the exponentiation of the source (\ref{source}) term is the product of several
sine-Gordon two-particle scattering amplitudes, as it appears in the factorization theorem. 

One has to remember that the theory has been constructed on a cylinder therefore the connection with the 
factorization theorem can emerge only in the limit where the circumference becomes infinite.
In this limit, the cylinder becomes indistinguishable from a plane. Here the only external parameter 
is the adimensional ``size'' $l=\mathcal{M} L$. It will be pushed to infinity $l\rightarrow \infty$. 
This can be interpreted as a very large volume or a very large mass, thus explaining the name of 
infrared limit (IR).

In this limit, the integral terms in (\ref{nlie-cont}) and in (\ref{energy}, \ref{momentum}) 
vanish exponentially fast, so they can be dropped and one remains with the momentum and the source term. 
Indeed, one can estimate that 
\be\label{stima}
\log(1+(-1)^{\delta}e^{iZ(x+i\eta )})\approx  \log(1+(-1)^{\delta}e^{iZ-\eta\, l\,\cosh x})\approx
(-1)^{\delta}e^{iZ-\eta\, l\,\cosh x}
\ee
The presence of a negative $\cosh$ at the exponent produces an exponentially fast decay for large $l$
in the integral terms.

Consider first a state with \( N_{H} \) holes only and XXZ spin \( S_z=N_{H}/2 \).
\begin{equation}
\label{IR-holes}
Z(\vartheta )=l\sinh \vartheta +\sum _{j=1}^{N_{H}}\chi (\vartheta 
-h_{j})\quad ,\quad Z(h_{j})=2\pi I_{j}
\end{equation}
This equation is the quantization of momenta in a box, for a system of particles. Indeed, by 
exponentiation one has
\be
\exp(i\,l\sinh h_k) \prod_{j=1}^{N_{H}} S(h_k-h_j)=\pm 1
\ee
where the sign depends on the parity of the quantum numbers.
This leads to interpret holes as solitons with rapidities \( h_{j} \). This is further evidenced by 
considering the energy and momentum expressions 
\be
E={\cal M}\sum _{j=1}^{N_{H}}\cosh h_{j}\,\qquad P={\cal M}\sum _{j=1}^{N_{H}}\sinh h_{j}
\ee
which is the energy of \( N_{H} \) free particles of mass \( {\cal M} \).
The identification with the particular element \( S(\theta) \) of the S-matrix
forces to give to these solitons a topological charge $Q=+1$ each, which is consistent
with the interpretation that \( Q=2S_z\). An analogous interpretation is possible in terms of pure 
antisolitons, reflecting the charge conjugation invariance of the theory. 

When considering two holes and a complex pair, the source terms can be arranged,
thanks to some identities satisfied by the functions \( \chi  \), in the form\[
Z(\vartheta _{i})=l\sinh \left( \vartheta _{i}\right) -i\log S_{+}(\vartheta 
_{i}-\vartheta _{j})=2\pi I_{j}\, \, \, ,\, \, \, i\, ,\, j=1,\, 2\]
where \[
S_{+}(\vartheta )=\frac{\sinh \left( \frac{\vartheta +i\pi }{2p}\right) 
}{\sinh \left( \frac{\vartheta -i\pi }{2p}\right) }S(\vartheta )\]
which is the scattering amplitude of a soliton on an antisoliton in the 
parity-even channel. The quantum numbers \( I_{+},I_{-} \) of the two complex roots are
constrained to be \( I_{\pm }=\mp \frac{1}{2} \) for consistency of the IR
limit. This state has \( S_z=0 \), with topological charge $Q=0$.

There is an analogous parity-odd channel in singe-Gordon \cite{zam1979}, with an $S_{-}$ amplitude.
It is realized by the state with two holes and a selfconjugate root.
In the same way, it has been possible to treat more complex cases, with different combinations of 
roots \cite{noiNP2}. See also \cite{tani} for details of the calculation.
In the attractive regime one has also to consider the breather particles that appear as 
soliton-antisoliton bound states. It turns out that the breathers are represented by self-conjugate 
roots (1st breather) or by arrays of wide roots (higher breathers). 

Thus, the whole scattering theory of sine-Gordon can be reconstructed in the IR limit,
thanks to the structure of the source term. It is now difficult to argue that the NLIE
does not describe sine-Gordon.

\subsection{UV limit and vertex operators\label{s:UV}}

It is interesting to study the opposite limit \( l\rightarrow 0 \), where one expects to see a conformal
field theory; indeed, in this limit the masses vanish and scale invariance appears. This reproduces 
the UV limit of sine-Gordon/massive Thirring, namely the \( c=1 \) conformal field theory described in 
section~\ref{sect:CFT} and in the Figure~\ref{4sectors.eps}.

The UV calculations are usually more difficult to perform than the IR ones, as they 
require to split the NLIE into two independent ``left'' and  ``right'' parts, called 
\emph{kink equations}. They correspond to the left and right movers of a two-dimensional conformal field theory. A similar manipulation is done on the energy and momentum expressions, that can finally 
be expressed in a closed form, thanks to a lemma presented in \cite{ddv95}. For the details,  
the reader is invited to consult the thesis \cite{tesidott} where all the calculations are shown in 
detail. In the present text, I give only the main results and the physical insight they imply.

A first important result is that the \( c=1 \) CFT quantum number \( m \) of the vertex operators
(\ref{vertex_operators}), which is identified with the UV limit of the topological charge,
can be related unambiguously to the XXZ spin by \( \pm m=2S_z \). Of course,
the \( \pm  \) reflects the charge conjugation invariance of the theory. Then,
by examining the states already ``visited'' at the IR limit, one can establish a bridge
between particle states and vertex operators of \( c=1 \) theory. 
\begin{enumerate}
\item The \textbf{vacuum} state has no sources, namely no holes or complex roots: only the sea of real 
roots is present.
There are two possible choices: \( \delta =0 \) or \( 1 \). The result of the UV calculation 
gives\[
\begin{array}{lll}
\rm {for}\, \delta =0:\quad  & \Delta ^{\pm }=0 & \, \, \mathrm{i}.e.\, \, 
\mathbb I\\
\rm {for}\, \delta =1:\quad  & \Delta ^{\pm }=\frac{1}{8R^{2}} & \, \, 
\mathrm{i}.e.\, \, V_{(\pm 1/2,0)}
\end{array}\]
i.e. the physical vacuum is the one with \( \delta =0 \). The other state 
belongs to the sector IV that does not describes a local CFT, as in Figure~\ref{4sectors.eps}.
\item The \textbf{two-soliton} state, described by two holes, with the smallest quantum numbers, gives
\begin{enumerate}
\item for \( \delta =0 \) and \( I_{1}=-I_{2}=\frac{1}{2} \) 
\( \Longrightarrow  \)
\( \Delta ^{\pm }=\frac{R^{2}}{2}\, \, \mathrm{i}.e.\, \, V_{(0,2)} \). 
\item for \( \delta =1 \) and \( I_{1}=-I_{2}=1 \) \( \Longrightarrow  \)
a \( V_{(\pm 1/2,2)} \) descendent, not in UV sG spectrum, as it also belongs
to sector IV.
\end{enumerate}
\item The \textbf{symmetric soliton-antisoliton} state (two holes and a self-conjugate root),
\( \delta =1 \), \( I_{1}=-I_{2}=1 \) and $ I_{c}^{\pm }=0 $
\( \Longrightarrow  \) \( \Delta ^{\pm }=\frac{1}{2R^{2}}\, \, \mathrm{i}.e.\, 
\, V_{(\pm 1,0)} \)
\item The \textbf{antisymmetric soliton-antisoliton} state (two holes and a complex pair) \\
\textcolor{black}{~~\( \delta =0 \), \( 
I_{1}=I_{c}^{-}=-I_{2}=-I_{c}^{+}=\frac{1}{2} \)
\( \Longrightarrow  \) \( \Delta ^{\pm }=\frac{1}{2R^{2}}\, \, \mathrm{i}.e.\, 
\, V_{(\pm 1,0)} \)}\\
It is obvious that these last two give two linearly independent combinations of the operators 
\( V_{(\pm 1,0)} \), one with even, the other with odd parity.
\item The \textbf{one hole} state with \( I=0 \), \( \delta =1 \) \( \Longrightarrow  \) 
\( \Delta ^{\pm }=\frac{1}{8R^{2}} \) i.e. the vertex operator \( V_{(0,1)} \), belongs to sector II.
For \( \delta =0 \) there are two minimal rapidity states with \( I=\pm \frac{1}{2} \). They are 
identified with the operators \( V_{(\pm 1/2,\pm 1)} \). As these states belong to sector
III, they are of fermionic nature and actually one identifies them with the
components of the Thirring fermion.
\end{enumerate}
These examples, taken all together, suggest the following choice of \( \delta \) to discriminate between
sine-Gordon and massive Thirring states
\begin{equation}
\label{regola_d'oro}
\begin{array}{cc}
2S_z+\delta +M_{SC}\in 2\mathbb Z & \quad \text{for\, Sine-Gordon}\\
\delta +M_{SC}\in 2\mathbb Z & \quad \text{for\, massive\, Thirring}
\end{array}
\end{equation}
where \( M_{SC} \) is the number of selfconjugate roots. This selects the sectors I and II for 
sine-Gordon states, and the sectors I and III for the Thirring ones, as in section \ref{sect:CFT}. 
The NLIE describes also the sector IV, that does not contain local operators. 
The correct interpretation of Coleman equivalence of Sine-Gordon and Thirring models is that 
even topological charge sectors are identical, the difference of the two models shows up only in 
the odd topological charge sectors, for which the content of Thirring must be fermionic while that of 
Sine-Gordon must be bosonic.

To conclude these remarks, I briefly add a comment about the special objects that were introduced in the 
classification of roots but not really used later. I need to recall their
definition: they are real roots or holes \( y_{i} \) having \( Z'(y_{i})<0 \).
Now, the function \( Z \) is globally monotonically increasing. Indeed its
asymptotic values for \( \vartheta \rightarrow \pm \infty  \) are dominated by the
term \( l\sinh \vartheta  \) which is obviously monotonically increasing. Also,
for \( l \) large, this term dominates. Therefore at IR the function \( Z \)
is surely monotonic and no special objects can appear. However, these global asymptotic estimations 
can fail at small $l$ and finite $\vartheta$. In that case the derivative $Z'(x)$ can become locally
negative. Thus, a real root or hole with negative derivative becomes ``special'' (and splits in three 
objects).
At the critical value \( l_{crit} \) of the parameter \( l \), at which the derivative become negative
moving from IR towards UV, the convergence of the iterative procedure breaks down, thus revealing
that some singularity has been encountered.  
For the scaling function to be consistently continued after this singularity, one needs
to modify the NLIE adding exactly the contributions that have been called 
special objects. A more careful analysis \cite{tesidott} reveals that these singularities
are produced by the logarithm in the convolution term going off its fundamental determination: 
``special'' objects are an artifact of the description by a counting function, they do not 
exist in the Bethe equations.
A treatment of these objects can be found in \cite{tesidott}. See also the discussion at 
page~\pageref{lespecial}. I don't know of successful numerical calculations in presence of 
special objects. In \cite{FPR2} a similar case occurred in the TBA formalism, in presence of boundary 
interactions, and was treated numerically because it was localised in some asymptotic region.

\section{Discussion}
In this chapter I have introduced the formalism of Destri-de Vega to study the sine-Gordon model 
on a cylinder with finite size space and infinite time. The presentation proposed here 
has mainly the purpose to show that this method is effective in treating finite size effects 
in quantum field theory 
and creating a bridge from a massive field theory on Minkowski space-time (visible in the IR limit) 
to a conformal field theory. 

As typical in treating integrable models, different systems meet on the way: lattice systems, 
scattering theory and conformal field theory all participate to the scenario described by the 
NLIE.  

The formalism was introduced by Kl\"umper et al. in \cite{klumper} and by Destri and de Vega in 
\cite{ddv95}. After, a number of other people participated to its development. Particularly, my 
involvement characterized my PhD years from 1996 to 1999, with the Bologna group. 

I directly contributed to four papers, \cite{noi PL1}, \cite{noiNP}, \cite{noiPL2}, \cite{noiNP2} and
I wrote my PhD thesis on this subject \cite{tesidott}. 
The main steps of my contribution are
\begin{itemize}
\item The whole formulation was revisited and corrected.
\item The study of the IR and UV limits was done systematically.
\item The spectrum of the continuum theory was carefully described, using both UV, IR and numerical calculations.
also adding the odd particle sector.
\item Many cases were studied numerically, to gain a complete control of the whole region that separates the IR and the UV.
\item The results were compared with perturbative calculations done with the truncated conformal 
space approach, 
giving a confirmation of the methods.
\item The introduction of a twist allowed to describe the restrictions of sine-Gordon, namely the 
perturbation of conformal minimal models by the thermal operator. These are massive theories
that are described by the same sine-Gordon NLIE after the introduction of an appropriate twist. 
\end{itemize}

Later, other groups profited of this NLIE to investigate a variety of models. An inexpected  
development will be presented in chapter~\ref{c:hubbard} and investigates integrability-related 
problems in gauge theory (especially $\mathcal{N}=4$ SYM).

Are there other things to be done? Even if the degree of difficulty is very high, the gain would be 
great, if one could succeed in the description of the eigenvectors in the continuum theory, or of
some correlation function.
Their  knowledge is important because they enter into the evaluation of many physical 
quantities (diffusion amplitudes, magnetic susceptibility) whose values can be compared with experiments.

\chapter{Thermodynamic Bethe Ansatz}
The first treatment of Bethe equations in their thermodynamic limit was done by Yang and Yang 
in \cite{yy1966}, for a system of nonrelativistic bosons interacting by a repulsive delta-function, 
on a line. The Bethe equations were obtained from the model Hamiltonian by the coordinate Bethe 
ansatz calculation, one of the several implementations of the Bethe ansatz methods. 
Then, the objective of that treatment was to evaluate the thermodynamic limit of the 
Bethe equations. The authors succeeded and were able to get the partition function of this 
interacting gas. They also showed the partition function analyticity in the 
temperature and the chemical potential, indicating the absence of phase transitions.

This method was generalized by Al.B. Zamolodchikov \cite{zam89} for relativistic particles 
interacting with a factorized scattering matrix. His goal was to create a contact between a 
factorized scattering theory of a given S matrix to its ultraviolet limit, typically
a conformal field theory in two dimensions. The author reasonably assumes that, at a finite
temperature, the equilibrium states of the particles in a box are described by some 
Bethe-like wave functions (asymptotic wave function), 
precisely as in the Yang and Yang approach. A quantization condition inspired to Bethe equations 
(\ref{betheTQ}, \ref{bethe}) is thus imposed.
Indeed, in the Bethe equations structure we have recognized a momentum term and an interaction term,
see near (\ref{quasipart}). Zamolodchikov assumes 
that the interaction term is given by a factor of S matrix amplitudes and the momentum is the
usual relativistic one. Notice, however, that 
the true Bethe wave functions obtained within Bethe ansatz calculations involve ``quasiparticles'', as
described near (\ref{quasipart}), while the scattering theory involves physical particles. 
For example, the Bethe roots can be complex while physical particle rapidities are always real.
The assumption works and the Zamolodchikov ``thermodynamic Bethe ansatz'' has been applied
to a variety of models. It allows to study the theory obtained by 
perturbing a conformal field theory with a relevant operator, under the condition that the
perturbation maintains integrability. 

Later, Pearce and Kl\"umper in \cite{pearceklumper1991} introduced another approach to 
``calculate analytically the finite-size corrections and scaling dimensions of 
critical lattice models'' (quoted from \cite{pearceklumper1991}). 
These authors do not make use of Bethe equations; instead, they start from the 
transfer matrix of an integrable lattice model and, knowing the arrangement of the zeros of its
eigenvalues, are able to solve certain identities satisfied by the transfer matrix itself. 
Then, one can evaluate the continuum limit of lattice models.

In the following, I will mainly talk about the last approach. 
All of them have triggered several further investigations. 
Indeed, the TBA serves as an interface between conformally invariant theories and massive or massless
integrable theories, in particular when these massive or massless theories are obtained as 
deformations of the conformal ones.

\section{Lattice TBA}
I start by defining a family of models on a square lattice of $N$ horizontal cells (faces)
and with many rows (their number will not be used) using the following diagrammatic 
representation for the double row transfer matrix \cite{BPO1996}
\be\label{transferm}
\D(N,u,\xi)_{\mbox{\footnotesize\boldmath$\sigma\,\sigma'$}}
=\!\!\sum_{\tau_{0},\dots,\tau_{N-1}}
\raisebox{-17.1mm}[19.8mm][15mm]{
\begin{tikzpicture}[scale=1.38]
\draw (1,0) -- (7,0) ;
\draw (1,-1) -- (7,-1) ;
\draw (1,1) -- (7,1) ;
\foreach \i in {1,...,7}{
  \draw (\i,-1) -- (\i,1) ;
}
\foreach \i in {1.5,...,3.5,6.5}{
  \draw (\i,-0.5) node {$u$};
  \draw (\i,0.5) node {$\lambda-u$};
}
\foreach \i in {0,...,3}{
  \draw (\i,-0.2)+(1.2,0) node {$\tau_{\i}$};
}
\foreach \i in {1,...,3}{
  \draw (\i,-1.2)+(1,0) node {$\sigma_{\i}$};
}
\foreach \i in {1,...,3}{
  \draw (\i,1.2)+(1,0) node {$\sigma_{\i}^{\prime}$};
}
\foreach \i in {1,...,3}{
  \draw (\i,1.2)+(1,0) node {$\sigma_{\i}^{\prime}$};
}
\draw (6,-1.2) node {$\sigma_{N-1}$};
\draw (5.7,-.2) node {$\tau_{N-1}$};\draw (6.8,-.2) node {$\tau_{N}$};
\draw (6,1.2) node {$\sigma_{N-1}^{\prime}$};
\draw (0.2,-1)--(1,0)--(0.2,1)--cycle [style=dashed];
\draw (0.2,-1)--(1,-1) [style=dotted];
\draw (0.2,1)--(1,1) [style=dotted];
\draw (7,-1.2)node{$1$};\draw (7.2,0)node{$2$};\draw (7,1.2)node{$1$};
\draw(0.5,0) node{$\begin{matrix}\lambda\!-\!u\\[-1mm]\xi\end{matrix}$};
\draw (0.2,-1.2)node{$r$};\draw (0.2,1.2)node{$r$};
\end{tikzpicture}
}
\ee
I use a double row transfer matrix \cite{skly} because, in presence of boundary interactions, 
it is needed to
ensure integrability. In fact, it is a transfer matrix that acts from the row $i$ to the $i+2$ while
in a more standard single row transfer matrix it would act from $i$ to $i+1$.

The diagrammatic representation is a mean to simplify the notation and to reduce the use of indices; 
it is also quite effective to realize a sort of ``graphical algebra'', useful to verify 
a number of properties \cite{BPO1996}. An entry of the transfer matrix,
$\D(N,u,\xilatt)_{\mbox{\footnotesize\boldmath$\sigma\,\sigma'$}}$, 
is obtained by multiplying the Boltzmann weights of  
each cell and summing on the indices indicated, namely summing on the central row of sites. 
The dashed triangle is a Boltzmann weight associated to
the three corresponding border sites; it actually introduces an interaction that is fully localized 
on the border of the lattice. As the actual lattice is just the one formed by the square cells,  
the triangle has been represented with dashed lines to indicate that is not a lattice cell.
The Boltzmann weights associated 
to one face are
\be\label{face}
W\!\left(\,\begin{array}{@{}cc|@{~}}d&c\\ a&b\end{array}\,u\right)=
\raisebox{-9mm}[10mm][9mm]{
\begin{tikzpicture}
\draw (0,0) --(1,0)--(1,1)--(0,1)--cycle;
\draw (0.5,0.5) node{$u$};
\draw (-0.2,-0.2) node{$a$};
\draw (1.2,-0.2) node{$b$};
\draw (1.2,1.2) node{$c$};
\draw (-0.2,1.2) node{$d$};
\end{tikzpicture}
}
\ =\ \frac{\sin(\lambda-u)}{\sin\lambda}\,\delta_{a,c}+
\frac{\sin u}{\sin\lambda}\,\sqrt{\frac{\sin (a\lambda) \sin (c\lambda)}{\sin (b\lambda) \sin 
(d\lambda)}}\,\delta_{b,d}
\ee
and those associated to the boundary interactions are
\be
B^{r,1}\!\left(\,r\pm1 \, \begin{array}{c|} r \\r \end{array}\;u, \, \xi \right)=
\raisebox{-16mm}[10mm][14mm]
{\begin{tikzpicture}
\draw (0,0)--(0.8,1)--(0,2)--cycle[style=dashed];
\draw(0.3,1) node{$\begin{matrix}u\\[-1mm]\xi\end{matrix}$};
\draw(1.3,1)node{$r\pm 1$};
\draw(0,-0.2)node{$r$};\draw(0,2.2)node{$r$};
\end{tikzpicture}
} \label{triangle}
=\sqrt{\frac{\sin(r\pm1)\lambda}{\sin r \lambda}} \quad
\frac{\sin(\xi \pm u) \sin(r\lambda+\xi \mp u)}{\sin^2 \lambda} 
\ee
where the integers associated to every vertex are called heights and must satisfy the adjacency rule 
of the Dynkin diagram $A_L$ namely adjacent sites must have height difference of 1 and 
heights are from 1 to $L$. I also use a crossing parameter
\be
\lambda=\frac{\pi}{L+1}
\ee
There are more general forms for these weights and also periodic boundary conditions 
are possible. Here we have chosen those that are more indicated to our development. Indeed, 
they are critical weights namely they describe the system at its
critical temperature. Moreover, the right boundary weight is kept fixed.

In these lattice models the interaction is characterized by the four sites around a face so
nearest neighbors and next-to-nearest neighbors interact.  On the contrary, in the six-vertex and 
XXZ models the interaction was between nearest neighbors only. The phase diagram has been 
studied in \cite{ABF} by using corner transfer matrix techniques. It is common to call these models
from the authors of the paper, ABF models. More precisely, I treat here the $A_L$ models, from 
the adjacency rules that are used. 

These models have a number of useful properties. Indeed, from direct inspection of
(\ref{face}) and (\ref{triangle}) it appears that the transfer matrix is an entire function
of the spectral parameter $u$, that means that its entries are free of poles or other
singularities in the whole complex plane. 

Transfer matrices at different spectral parameter commute, thus insuring integrability
\be
[\D(N,u,\xi),\D(N,u',\xi)]=0
\ee
Indeed, with a standard construction, one can expand in $u$ and generate integrals of motion.

In spite of a different formulation based on faces instead of vertices, the present $A_L$ models 
are very deeply related to the six-vertex one.  
The original T-Q relations of Baxter and the six-vertex Bethe ansatz 
can be modified to hold for the present case. 

The double row transfer matrix satisfy a T-Q functional equation that is fully similar to 
the one that holds in the periodic case (\ref{TQperiod}), with small modifications to account for
the boundary
\be
s(2u)\,\DD\left(u+\frac{\lambda}{2}\right)\Q(u)=s(2u+\lambda)\ f\!\left(u+\frac{\lambda}{2}\right)
\Q(u-\lambda)+s(2u-\lambda)\ f\!\left(u-\frac{\lambda}{2}\right)\Q(u+\lambda)\label{TQbordo}
\ee
where I have used
\be
f(u)=\left(\frac{\sin u}{\sin \lambda}\right)^{2N}\,,\qquad s(u)=\frac{\sin u}{\sin \lambda}
\ee
The double row transfer matrix is the one in (\ref{transferm}) but, when not strictly necessary, I omit
the dependence by $N$ and $\xi$.

These expressions hold for a fixed border on the left and the right of (\ref{transferm}), namely when 
$r=1$ and $\tau_0=r+1=2$. 
In this case the boundary coupling $\xi $ disappears. Of course, they can be modified 
to include other cases. The goal here is not to reach the largest generality but to show how
the methods work and how the TBA appears in lattice systems. For this reason, I follow the 
approach of \cite{FPW} in order to introduce the fusion and the TBA hierarchy. 

The new operator $\Q(u)$ is a family of matrices that commute each other and commute with the 
transfer matrix $[\Q(u),\Q(v)]=[\Q(u),\DD(v)]=0$. This implies that the same functional equation 
holds true for the eigenvalues $D(u)$ and $Q(u)$. The eigenvalues of $Q$ are given by
\be
Q(u)=\prod_{j=1}^n \frac{\sin(u-u_j) \sin(u+u_j)}{\sin^2 \lambda}\label{BaxQ2}
\ee
where $u_j$ are the Bethe roots. The Bethe ansatz equations result by imposing 
that the transfer matrix is an entire function. Indeed, when $Q(u)=0$, being $D(u)$ entire,
the right hand side must vanish. This forces  the following Bethe equations
\be\label{TQBethe}
\frac{\sin(2u+\lambda)}{\sin(2u-\lambda)}\ 
\left(\frac{\sin\left(u+\frac{\lambda}{2}\right)}{\sin\left(u-\frac{\lambda}{2}\right)}\right)^{2N}=-
\frac{Q(u+\lambda)}{Q(u-\lambda)}
\ee
It is convenient to shift the transfer matrix 
\be
\tilde{\D}(u)=\D(u+\frac{\lambda}{2})
\ee
With this convention, the transfer matrix is Hermitian in $\re(u)=0$. Given the position
of its zeros, that will be presented later, the relevant analyticity strip is
\be
-\lambda<\re (u)<\lambda
\ee
Using a standard notation, I let $\DD(u)=\DD_0^1$ 
\be
\DD_k^{q}=\DD^{q}(u+k\lambda),\quad \Q_k=\Q(u+k\lambda),\quad 
s_k(u)={s(2u+k\lambda)},\quad
f_k(u)=(-1)^{N}{s(u+k\lambda)}^{2N}
\ee
For ``historical'' reasons, this notation is customary here: the upper index is not an exponent
but just an index.
The T-Q relation implies that the eigenvalues $D(u)$ are determined by the eigenvalues $Q(u)$ in the 
compact form
\be
\tilde{D}(u)=\tD_0=\frac{s_1f_{1/2}Q_{-1}+s_{-1}f_{-1/2}Q_1}{s_0 Q_0}
\ee

\subsection{Fusion hierarchy and TBA hierarchy}
\label{sec:functional}
From the T-Q relation (\ref{TQbordo}) one can construct a hierarchy of models namely
a set of transfer matrices $\D^q$ recursively defined one after the other.
The process, called fusion, was introduced in \cite{kul81} as a method of obtaining new solutions to
the Yang-Baxter equation, by combining R~matrices of a known solution. 
Fusion for the ABF models has been done in \cite{date1986} and studied in detail in \cite{bazres1989}, 
in the case of periodic boundary conditions. The construction holds for the open boundary case 
for which the fusion has been done in \cite{BPO1996}.   
For what follows, the presence of the hierarchy has very important consequences because it 
imposes a very particular organization of the zeros of the transfer matrix eigenvalues in 
the complex plane of the spectral parameter $u$. At the end, this will lead to ``solve'' the eigenvalue
problem. 

From \cite{BPO1996}, the fusion hierarchy for the transfer matrices is
\be
s_{q-2}s_{2q-1}\DD_0^{q}\DD_q^{1}=s_{q-3}s_{2q}f_q\DD_0^{q-1}
		+s_{q-1}s_{2q-2}f_{q-1}\DD_0^{q+1},\qquad q=1,2,\ldots,L-1\label{FusionHier}
\ee
Notice that the level $q+1$ is extracted from the right hand side, by knowing the levels
$q,q-1,1$.
This equation is obtained for the eigenvalues, by extracting the transfer matrix eigenvalue from 
(\ref{TQbordo}) and multiplying by itself after a shift. This procedure is repeated. 
The second term on the right hand side is the higher fusion level $q+1$ and is fixed by 
the levels $q,\ q-1,\ 1$. This expresses the idea of hierarchy\footnote{It is helpful to read all 
these expressions by just ignoring all the coefficients, namely the factors $s_q$ and $f_q$.}. 
The initial conditions of the recurrence are
\be\label{initial}
\DD_0^{-1}=0,\qquad \DD_0^{0}=f_{-1}\I,
\ee
and there is a closure condition: the fusion process is upper limited by $L$ that is the largest 
value of the ``heights'' located on the corners of a lattice face 
\be
\D_0^{L}=0\label{closure}
\ee
This fusion makes sense because the obtained higher fusion level transfer matrices still have  
the remarkable properties of the original $q=1$ transfer matrix. Indeed, 
they are entire functions and they commute each other. This last property implies that  
they all have the same set of eigenvectors.
Then, they can be interpreted as transfer matrices of new lattice models. 

Starting with the fusion hierarchy (\ref{FusionHier}), one can use induction to derive the
T-system of functional equations~\cite{KP92,BPO1996}
\be\label{funct}
s_{q-2}s_{q}\DD_0^{q}\DD_1^{q}=
s_{-2}s_{2q}f_{-1}f_q\I+s_{q-1}^2\DD_0^{q+1}\DD_1^{q-1},\qquad q=1,2,\ldots,L-1
\ee 
For $q=1$, the rightmost term vanishes because of the initial conditions and what remains is just 
an inversion identity. If we further define
\be\label{deft}
\dd_0^{q}=\frac{s_{q-1}^2\DD_1^{q-1}\DD_0^{q+1}}{s_{-2}s_{2q}f_{-1}f_q},\qquad q=1,2,\ldots, L-2
\ee
then the inversion identity hierarchy (T-system) can be put in the form of a 
Y-system~\cite{KP92,BPO1996}
\be
\dd_0^{q}\dd_1^{q}=\big(\I+\dd_0^{q+1}\big)\big(\I+\dd_1^{q-1}\big)\label{Ysystem}
\ee
with closure
\be
\dd_0^{0}=\dd_0^{L-1}=0
\ee
For later convenience, I define the shifted transfer matrices
\be\label{shifted}
\tilde{\D}^{q}(u)=\D^{q}\Big(u+\frac{2-q}{2}\,\lambda\Big)\,,\qquad
\tilde{\dd}^{q}(u)=\dd^{q}\Big(u+\frac{1-q}{2}\,\lambda\Big)
\ee
so that the $Y$-system takes now the form 
\be \label{ysyst2}
\tilde{\dd}^q\big(u-\frac{\lambda}2\big)\ 
\tilde{\dd}^q\big(u+\frac{\lambda}2\big)=
\big(1+\tilde{\dd}^{q-1}(u)\big)\big(1+\tilde{\dd}^{q+1}(u)\big)
\ee
The transfer matrices are entire functions of $u$ and periodic
\be
\D^{q}(u)=\D^{q}(u+\pi)\,,\qquad \tilde{\dd}^{q}(u)=\tilde{\dd}^{q}(u+\pi)
\ee
They also are real if $u$ is real and satisfy a ``crossing symmetry''
\be
\D^{q}(u)=\D^{q}\big((2-q)\lambda-u\big),\qquad
\dd^{q}(u)=\dd^{q}\big((1-q)\lambda-u\big)
\ee
The advantage of the shifting these matrices in that the shifted matrices have the same 
analyticity strip
\be\label{phystrip}
-\lambda< \re(u)<\lambda
\ee
that means that within a periodicity strip $(\re(u),\re(u)+\pi)$ we will use only 
the analyticity strip (\ref{phystrip}) to solve for the eigenvalues. 
This choice is related to the position of the zeros and will be presented later. 

Lastly, the asymptotic values $\tilde{d}^q(+ i\infty)$ were computed in \cite{KP92}
\be\label{asym1}
\td^q(+i\infty)=
\frac{\sin [q\theta]\,\sin[(q+2)\theta]}{\sin^2\theta}
\;=\;\frac{\sin^2 (q+1)\theta}{\sin^2\theta}-1,
\qquad \theta=\frac{s\pi}{L+1}
\ee

Functional equations as the T-system and Y-system hold for the periodic boundary case, the open case, 
and off-criticality. They are a true feature of the this type of lattice models.
The same hierarchy holds for the six-vertex model but it is unlimited so there is no truncation.

Notice that my goal is to find the eigenvalues of the nonfused double row transfer matrix $\D(u)$.
The T-Q relation (\ref{TQbordo}) offers us one way: solve the Bethe equations (\ref{TQBethe}), 
put the Bethe roots in (\ref{BaxQ2}) and use (\ref{TQbordo}) to find the eigenvalues $T(u)$.
The fusion hierarchy offers another path: solve the whole family of TBA functional equations 
(\ref{ysyst2}) at all orders of fusion; this gives $\td^1(u)$ that appears as the rightmost term in 
(\ref{funct}) when $q=1$ then this equation can be inverted to find $D^1(u)=D(u)$, as wanted.
The two paths, albeit mathematically equivalent, offer different levels of difficulty. 
Solving or at least classifying all solutions of the Bethe equations can be an hard task, given that
Bethe roots are complex numbers. 
It turns out that it is simpler, in this case, to follow the second approach. This will require 
to control the zeros of the fused transfer matrices.

\subsection{Functional equations: Y-system, TBA}
In the previous section I have presented several systems of functional equations for the 
transfer matrices. I try now to motivate their relevance in relation to the Zamolodchikov
approach to thermodynamic Bethe ansatz.

The Y-system in the form (\ref{ysyst2}) can also be written as 
\be\label{ysyst3}
\tilde{\dd}^q\big(u-\frac{\lambda}2\big)\ 
\tilde{\dd}^q\big(u+\frac{\lambda}2\big)=\prod_{r}
\big(1+\tilde{\dd}^{r}(u)\big)^{A_{qr}}
\ee
where $A_{qr}=\delta_{q,r-1}+\delta_{q,r+1}$ is the incidence matrix of the Dynkin diagram $A_{L-2}$.
This equation has now the form of the one obtained by Zamolodchikov in \cite{zam91} for the RSOS
scattering theories. These theories are obtained by perturbing a unitary conformal field theory
of the minimal series (\ref{minimal}) by the operator $\phi_{1,3}$. This operator is relevant
and it preserves integrability namely the perturbed theory is still integrable.
There are two possible directions of perturbation, according to the sign of the coupling. 
One gives rise to a massive model $A_m^{(-)}$ whose scattering matrix factorizes according to the 
factorization theorem discussed in the Introduction. 
The other direction of the perturbation gives a massless theory $A_m^{(+)}$.
The name of ``RSOS theories'' for these models come from the fact that their two particle scattering 
matrix, apart overall factors, is the R-matrix of the ABF models of \cite{ABF}, that are also called 
RSOS.

For these theories in the massive regime, Zamolodchikov wrote the thermodynamic Bethe ansatz 
equations in \cite{zam91} for the ground state and showed that they are particular solutions of 
a Y-system with the same structure of (\ref{ysyst3}). The approach of Zamolodchikov does not describe 
excitations above the ground state but it is reasonable to imagine that the Y-system, 
more that the thermodynamic Bethe ansatz, describes the symmetries of the model and all its
excited states. Y-systems emerge in most (or possibly all) the thermodynamic Bethe ansatz equations 
that have been derived: see \cite{zamplb91} but also one of 135 citations to it, for example
\cite{Serban:2010sr}, that is a recent review on the integrability in AdS/CFT, containing many recent
citations to Y-systems. 

Now, it is curious to see if the Y-system can emerge, in field theory, in a direct way that avoids
the tortuous method of \cite{zam91}. Indeed, in the series of papers opened with \cite{blz1},
the authors formulated a continuum version of the transfer matrix, of the T-Q relation and of the 
subsequent functional equations for the minimal series $\mathcal{M}_{2,2n+3}$. This latter is
a family of non-unitary conformal field theories with central charge $c=1-3\frac{(2n+1)^2}{2n+3}.$ 
The formulation has been also modified in \cite{blz_exc} to describe the perturbation by the operator
$\phi_{1,3}$. So, the Y-systems can be obtained constructively from the conformal field 
theory operators, at the critical point and off criticality.

\subsection{Zeros of the eigenvalues of the transfer matrix\label{s:zeros}}
The conclusion of section~\ref{sec:functional} was an indication to solve the Y-system (\ref{ysyst2})
to obtain the eigenvalues of the transfer matrix. Before engaging in this calculation, one 
has to know the analytic properties of the transfer matrices, especially in relation to zeros
(and poles, if any). 

The eigenvalues $\tD^q(u)$ of the transfer matrices have many zeros in the complex plane. 
They can be easily counted from (\ref{transferm}), considering that each face contributes
with a trigonometric term $\sin(a+u)$  (with some $a$) and the triangular face with two such
terms. This means that $\tD^q(u)$ is a polynomial in $z=\exp (iu)$ and $\bar{z}=\exp(-iu)$
of degree $2N+2$ so we have to expect $4N+4$ zeros in a periodicity strip.
This counting can be slightly different in presence of other boundary conditions.

Many analytical and numerical observations, and also the ``string hypothesis'' formulated in
\cite{bazres1989} for the Bethe roots, indicate that the zeros have a peculiar structure within the 
strip indicated in (\ref{phystrip}) as analyticity strip. Indeed, zeros can be on the middle axis 
of the strip, such that their real part vanishes
\be\label{onestring}
\re (u)=0
\ee
(these are called 1-strings) or on the border of the strip. In that case they appear as pairs
\be\label{twostring}
\re(u_1)=-\re(u_2)=\lambda\,,\qquad \im(u_1)=\im(u_2)
\ee
and are called 2-strings, in the usual Bethe ansatz meaning of ``string''.
This pattern holds for all transfer matrices, with $q=1,2,\ldots,L-2$. 
The matrix $\D^{L-1}$ is proportional to the identity so its strip is empty.
Moreover, these are always simple zeros, because Bethe roots are mutually exclusive.

Numerical examples showing this peculiar behaviour of 1- and 2-strings have been presented in 
a number of papers, starting from the one \cite{klpe91} 
in which the methods were established. This pattern holds in presence of boundary interactions
\cite{OPW}, \cite{FPR} and with off-critical transfer matrices \cite{PCA}, with massive or 
massless perturbations.
In \cite{FPW} there is a whole ``art gallery'' of images of the zeros of the transfer matrix eigenvalues
for the critical $A_5$ model with fixed boundaries. 
\begin{figure}[t]
\hfill\includegraphics[width=0.3\linewidth]{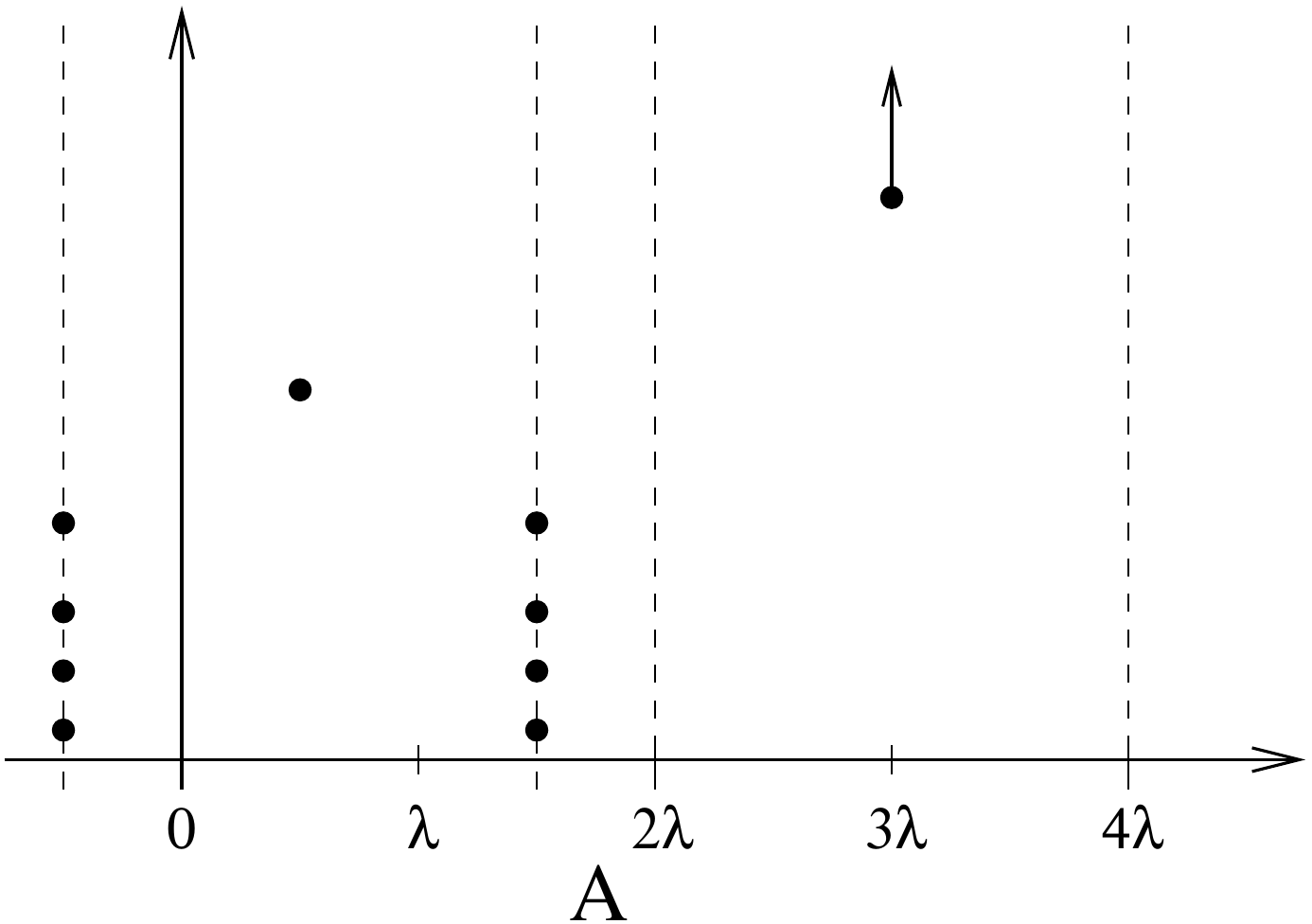}
\hfill\includegraphics[width=0.3\linewidth]{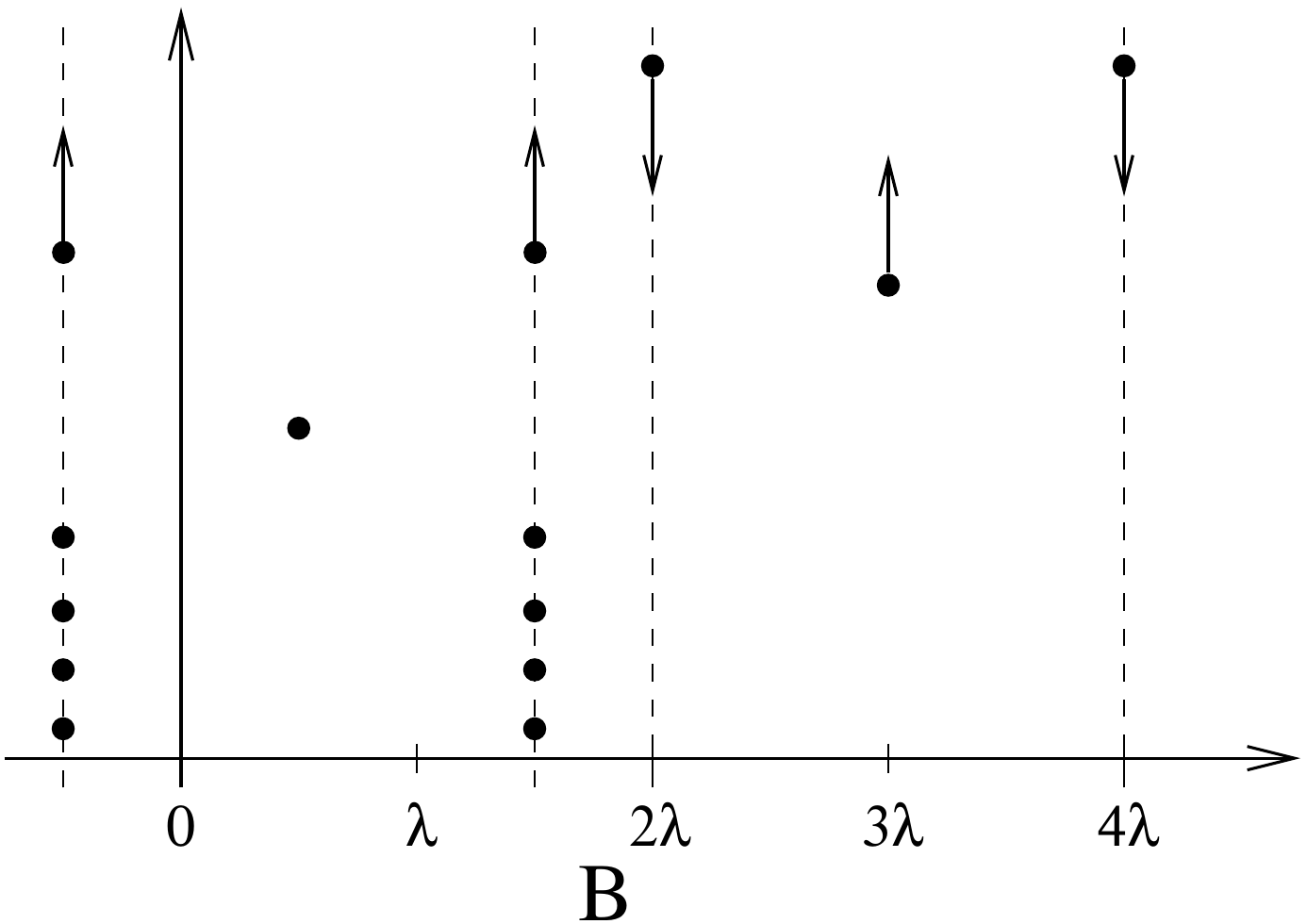}
\hfill\includegraphics[width=0.3\linewidth]{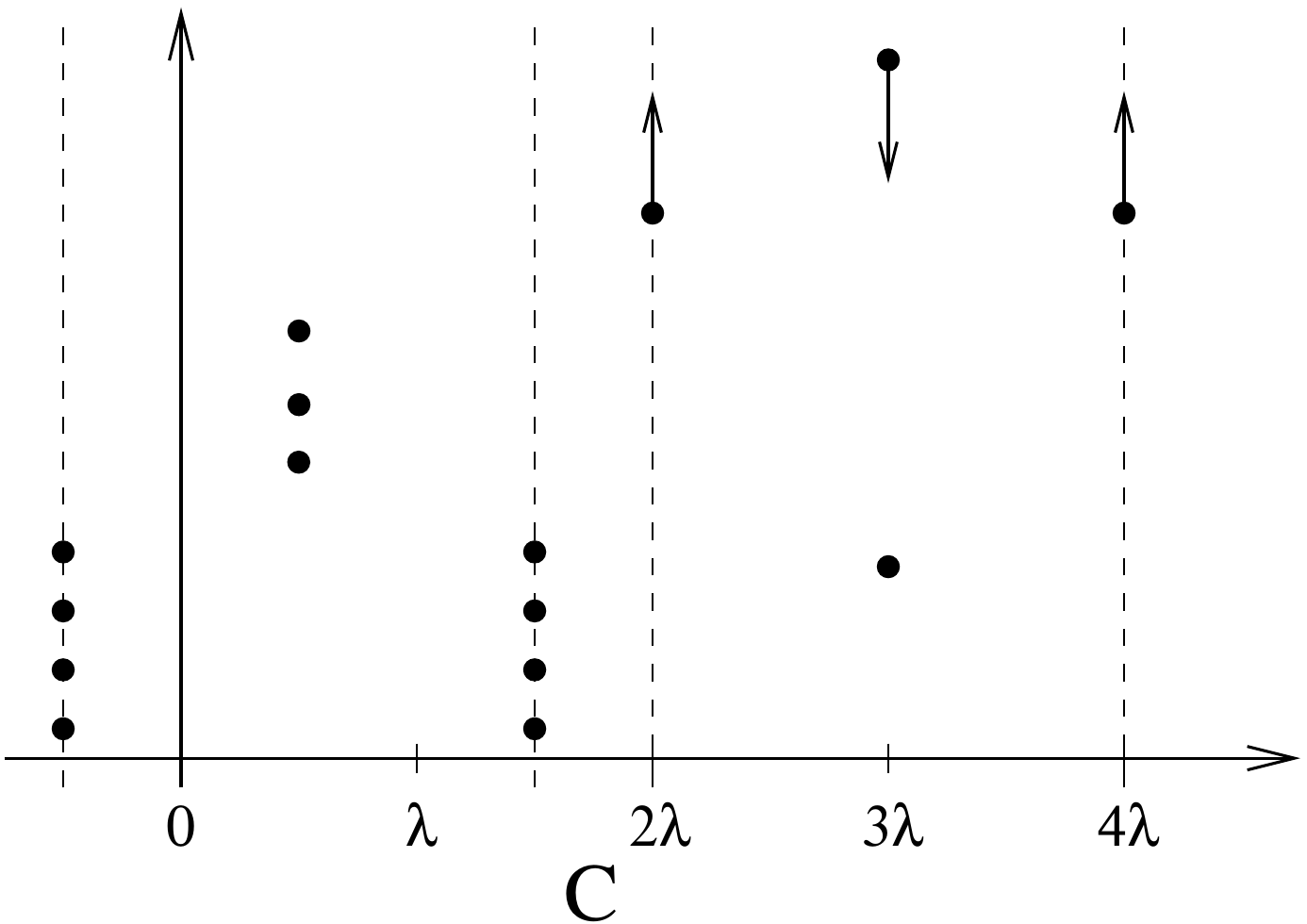}
\caption{\label{ABCmech} Three states for the tricritical Ising model are shown.
The arrows indicate three different mechanisms of displacement of the zeros triggered by the change 
of the boundary coupling constant from one to another conformal boundary condition 
${\cal B}_{(2,1)} \mapsto {\cal B}_{(1,2)}$. Note that this is the reverse of the physical flow which
actually goes from the UV point $(1,2)$ to the infrared point $(2,1)$.
The states are
A, $(0|0)\,\mapsto\,(0)_{+}$;  
B, $(1|0)\,\mapsto\,(0)_{-}$;
C, $(0\,0\,0|1)\,\mapsto\,(0\,0\,0\,|0\,0)_{-}$. In this image, taken from \cite{FPR}, the shift
(\ref{shifted}) is not used. Strip 1 is on the left, strip 2 on the right.}
\end{figure}
In Figure~\ref{ABCmech} there are three examples of states from the tricritical 
Ising model $A_4$. As it has just two levels of fusion, $q=1,2$, it has two analyticity strips.
The transfer matrices were not shifted so strips are not centered as in (\ref{phystrip}).  
In the figure, arrows indicate the dynamics of the zeros, namely their displacement after 
tuning a boundary coupling. 

In some case, often because of the normalization factors in (\ref{FusionHier}), some zeros at 
intermediate positions occur, like
\be
\re(u)=\pm \frac{\lambda}{2}
\ee
I will not consider these cases here. Actually, they do not modify the general ideas that follow, 
especially because they do not have dynamics: they appear in a fixed number, they 
show up in all states and have a fixed position. 

In addition to (\ref{onestring}, \ref{twostring}), 
several numerical investigations show that the relative order of the zeros in the 
$L-2$ fundamental strips (\ref{phystrip}) is sufficient to identify a common eigenstate of the 
transfer matrices $\D^{1}$ to $\D^{L-2}$. Indeed, the state is fixed by the order of appearance of 
1- and 2-strings, from the real axis to the asymptotic region. As an example, in Figure~\ref{ABCmech}
the state A is given by: 
\begin{itemize}
\item strip 1: four appearances of 2-strings behind one 1-string
\item strip 2: one appearance of 1-string
\end{itemize}
This means that such an eigenstate is uniquely characterized by a topological  
information and that the corresponding geometrical data, namely actual positions of zeros, can be
inferred by other means. The relative order between different strips has no relevance. 
Eigenstates/eigenvalues are thus classified by combinatorial rules.

This phenomenon is at the origin of the nomenclature of the states 
by a set of \textit{non-negative quantum numbers} $\{I_k^{(q)}\}$; considering the upper half-plane, 
in a strip the lower 1-string carries number 1 and so on counting upward; given the $k$-th
1-string of strip $q$, $I_k^{(q)}$ is the number of 2-strings above it. This definition allows an 
easy reconstruction of the sequence of 1- and 2-strings in the strip (see also Figure~\ref{ABCmech}).

Indicating with $m_q,\,n_q$ the number of 1-strings, 2-strings respectively in strip $q$ 
(above the real axis), one clearly has
\be
n_q\geq I^{(q)}_1\geq I^{(q)}_2 \geq \ldots \geq I^{(q)}_{m_q} \geq 0
\ee

It is important to mention that, even if the general scheme proposed here is only true 
``asymptotically''  (large $N$), deviations are very small and already with a lattice of 
six faces the general picture appears. Deviations consist in the displacement of the zeros by 
their asymptotic position (\ref{onestring} or \ref{twostring}). This phenomenon is also observed
in the Bethe ansatz framework: the ``string hypothesis'' is asymptotic and there are deviations 
in a finite size lattice. 

The number of zeros is expected to grow with the lattice size, according to the counting done at 
the beginning of this section. Notice first that $N$ growths in steps of 2, because of adjacency 
rules. Given a certain set of zeros with their relative order, the addition of an even number of 
new columns to 
the lattice has the only effect to add 2-strings to strip 1, near the real axis. 
In other words, the real axis of the first strip is a source of 2-strings, when $N$ increases.
The definition of the quantum numbers has been chosen consistently with this: the quantum numbers
of a state do not depend by $N$. 
The addition of new 2-strings in the first strip pushes the other zeros (in all strips) 
far and far from the real axis. This trend is very important for the scaling limit that will
be used to get contact with the conformal theory.
Indicating a 1-string position by $\frac{i\ y_k^{(q)}}{L+1}$, the vanishing of the transfer matrix 
eigenvalue is
\be
\td^q\left(\frac{i\ y_k^{(q)}}{L+1}\right)=0\,,\qquad y_k^{(q)}>0
\ee
The leading behaviour with $N$ is
\be\label{leadingzero}
y_k^{(q)}=y_k^{(q)}(N) \sim \hat{y}_k^{(q)}+\log N
\ee
where $\hat{y}_k^{(q)}$ is the asymptotic position.

Even if the relative orders and positions of different strips do not matter, there are constraints 
between the composition of the different strips. They originate in the Y-system and will be
presented later.

\subsection{Solving the Y-system}
Aiming at finding the eigenvalues of the transfer matrix, the functional system (\ref{ysyst2}) has 
been ``solved'' in \cite{klpe91} by taking Fourier transform of the logarithmic derivative. This 
method requires to
control the zeros and poles within the analyticity strip (\ref{phystrip}). After, the method 
was adapted to several other cases, as in \cite{OPW} for fixed boundaries, in \cite{FPR}, \cite{FPR2} 
and \cite{io} for interacting boundaries, in \cite{PCA} for off-critical lattices and finally in 
\cite{FPW} for the whole series of $A_L$ models and for the XXX model. 
Therefore, there is no need to repeat all the long calculations here. 

The final result will be written after taking a scaling limit or continuum limit
\be\label{scaling}
\hat{d}^q(x)=\lim_{N\rightarrow \infty} \td^q\left(N,\frac{i \,x }{L+1}\right)
\ee
where the matrices of (\ref{shifted}) are used. The lattice size has been indicated and a new spectral 
parameter $x$ has been introduced, rotated and dilatated by respect to $u$\footnote{Notice that in 
presence of boundary or bulk interactions, the corresponding parameters usually need to be scaled 
with $N$.}. 
The size $N$ appears in the limit (\ref{scaling}) first explicitly, counting the number of faces, 
second implicitly, by the ``scaling'' positions of the zeros (\ref{leadingzero}). 

There are two important reasons to require the continuum limit: first, the positions of the zeros
in (\ref{s:zeros}) are ``asymptotically'' correct therefore exact results can be obtained in the 
limit only, second because one is usually interested in comparing with conformal field theory data.
Conformal field theory is defined on a continuous space, not on the lattice.

From \cite{FPW}, the transfer matrix eigenvalues are 
\be
-\frac12 \log D(N,u)=-N \log[\kappa_{\text{bulk}}(u)]-\frac12 \log[\kappa_{\text{bound}}(u)]-
\frac12 \log\tD_{\text{finite}}\big(N,u-\frac{\lambda}2\big)
\ee
where the function $\kappa_{\text{bulk}}$ contains the main bulk contribution and is independent of $N$, 
the function $\kappa_{\text{bound}}$ contains a boundary free energy and also is independent of $N$.
These functions are given in \cite{NepomechiePearce}; they will not be used here.
$\tD_{\text{finite}}$ is finite for large $N$ but still depends by it and will be given later. 
The factor $\frac12$ has a physical meaning of normalizing to a row-to-row transfer matrix.
This produces the following lattice partition function\footnote{The lattice has $M$ rows and $N$ 
columns with periodic boundary conditions in the vertical direction and open boundary conditions 
in the horizontal one, according to (\ref{transferm}).} 
\bea\label{partition}
&&Z(N,M,u)=\mbox{Tr }\big(\D(u)^{M/2}\big)  \\ &&
=\exp \Big[ -NM  \log[\kappa_{\text{bulk}}(u)] -\frac12 M \log[\kappa_{\text{bound}}(u)] \Big] 
\mbox{ Tr }\big[\tilde{\D}_{\text{finite}}\big(N,u-\frac{\lambda}2\big)^{M/2}\big] \nonumber
\eea
The solution of the T-system leads to 
\bea\label{solY}
& &-\frac12 \log \tilde{D}_{\text{finite}}(N,0)  \\  \nonumber
& &= -\int_{-\infty}^{\infty}dy\ \frac{\log(1+\tilde{d}^{1}\big(N,\frac{i\ y}{L+1}\big))}{4\pi\cosh y}
-\sum_{k=1}^{m_1}\log\tanh\frac{y_k^{(1)}(N)}2 + \mbox{higher order corrections}\\
& &=\frac{\pi}{N} E+\mbox{higher order corrections in }\frac{1}{N} \nonumber
\eea
with $E$ independent by $N$ and given by
\be\label{sc_energ}
E=-\frac{1}{\pi^2}\int_{-\infty}^{\infty}dy\ e^{-y}\ \log\big(1+\hd^{1}(y)\big)+
2\sum_{k=1}^{m_1}\frac{1}{\pi}\ e^{-\hy_k^{(1)}}
\ee
Of course, $E$ depends by the configuration of the zeros in the various strips.
I can now express the last factor of the partition function at the isotropic point $u=\frac{\lambda}{2}$
as
\be
\chi(\hat{q})=\mbox{ Tr }\big[\tilde{\D}_{\text{finite}}(N,0)^{M/2}\big] =
\sum_{\text{configurations}} \exp \big(-\frac{\pi\ M}{N} E\big)=
\sum_{\text{configurations}} \hat{q}^E
\ee
with
\be
\hat{q}=e^{-\frac{\pi\ M}{N}}
\ee
This $\hat{q}$ is a geometrical parameter that measures the ratio between the number of row and 
columns in the lattice. 
When the system is at criticality, the function $\chi$ is the conformal partition function.
Soon I will make the connection between the ``energy'' $E$ and conformal energies of the two
dimensional conformal theory. 

In (\ref{sc_energ}), two ingredient are still missing: the transfer matrix $\hd^1$ and the zeros
$\hy^{(1)}_k$. Thanks to the limit (\ref{scaling}), I can give them an expression. This is not possible 
before the limit, namely on the lattice, because the positions indicated in section \ref{s:zeros} are 
only asymptotically correct.

The Y-system provides the missing ingredients by the thermodynamic Bethe ansatz equations
\bea
\log \hd^{q}(x)&=&-4\ \delta_{1,q}\ e^{-x}+ 
\log\prod_{j=1}^{L-2}\prod_{k=1}^{m_j} \Big[\tanh\frac{x-\hy_k^{(j)}}{2}\Big]^{A_{q,j}}
\nonumber\\
&+&\sum_{j=1}^{L-2} A_{q,j}\,\int_{-\infty}^{\infty}dy\ \frac{\log(1+\hd^{j}(x))}
{2\pi\cosh (x-y)}\,,
~\qquad q=1,\ldots,L-2 \label{sc_tba}
\eea
and the quantization conditions 
\bea\label{sc_psi}
\hat{\Psi}^{q}(x)&=&4\delta_{1,q}\, e^{-x}+i\sum_{r=1}^{L-2}{A_{q,r}}\sum_{k=1}^{m_r} 
\log\tanh\big(\frac{x-\hy_k^{(r)}}{2}-i\frac{\pi}{4}\big)  \\
&-&\sum_{r=1}^{L-2} A_{q,r} \pv dy\,
\frac{\log(1+\hd^{r}(y))}{2\pi\sinh(x-y)}\nonumber\\[3mm]
\hat{\Psi}^{q}(\hy_k^{(q)})&=&\pi\,n_k^{(q)}=\pi[ 1+2(I_k^{(q)}+m_q-k)]
\eea
where a new family of quantum numbers has been introduced $n_k^{(q)}$. They are odd integers.
These two sets of equations close the problem. Indeed by simultaneously solving the TBA
equations and the quantization conditions one has the transfer matrices in the center of the
strips (\ref{phystrip}) and the position of the zeros. Those with label $q=1$ enter in the energy 
expression (\ref{sc_energ}) namely in the original transfer matrix eigenvalues. 
Equations (\ref{sc_tba}) and (\ref{sc_psi}) are now exact and describe the whole spectrum of the
transfer matrix. 

The Y-system controls the structure of the equation, namely the appearance of the integral with a
difference-like integration kernel and the hyperbolic cosine. 
The equations still show the full $A_{L-2}$ structure of the Y-system (\ref{ysyst3}), in which 
a strip is coupled with the two neighboring ones.
The specific form of the driving term $-4\ \delta_{1,q}\ e^{-x}$ comes from the initial condition
for $\D^0$ in (\ref{initial}) and indicates how the degrees of freedom with fusion index 1 
participate to the energy. 
The hyperbolic tangent for the zeros and the convolution kernel $1/\cosh (x-y)$ are specific of a 
critical lattice and they become elliptic functions in the off-critical case. They are
mainly fixed by the form of the left side of the Y-system.
In presence of boundary interactions, one also adds a specific term as in \cite{FPR2}, \cite{io}. 

The positivity condition
\be\label{positivita}
1+\hd^{j}(x)>0
\ee
holds on the whole real axis therefore the integral terms are always real. The zeros terms can 
produce an imaginary part that corresponds to taking the logarithm of a negative value of 
$\hd^q(x)$. 

The equations (\ref{sc_tba}) are legitimately called thermodynamic Bethe ansatz because they 
correspond to the equations of Zamolodchikov \cite{zam91} for the RSOS scattering theories. 
His equations are obtained for the massive case with periodic boundary conditions while here
the model is lives on a open strip and the bulk is critical. 
More importantly, the equations of Zamolodchikov only hold for the ground state while the present 
lattice approach provides all the states. 
Notice that the Zamolodchikov equations are usually expresses in terms of the pseudo-energy 
$\epsilon_q(x)=-\log \hd^q(x)$.

In the general case, there is no explicit solution of this functional equations. For this reason, 
sometimes numerical solutions were used. It is true that, near a critical regime, it is possible
to obtain a closed form for the energy.
Indeed, after long calculations, one has a very short form for the conformal energy
\be\label{energy3}
E=-\frac{c_L}{24}+ \frac14 \m^{T}  C \m 
+\sum_{q=1}^{L-2}\sum_{k=1}^{m_q} I_{k}^{(q)}
\ee
where the central charge appears as function of $L$ 
\be
c_L=1-\frac{6}{L(L+1)}
\ee
and $\m^T=(m_1,m_2,\ldots,m_{L-2})$ is a vector with the number of 1-strings. In this case, a 
completely explicit expression for the energies has been obtained. Notice that the solution
of the TBA system is still unknown. One arrives at the expression for the energy by manipulation, 
not by actually solving the TBA or quantization conditions. 

In noncritical cases, one can arrive at equations like (\ref{energy3}) only in the UV and IR 
limits already described in section~\ref{s:IR} and \ref{s:UV}.

\section{NLIE versus TBA}
Two formalisms have been introduced. One, directly derived from the Bethe ansatz equations, 
leads to a Kl\"umper-Batchelor-Pearce-Destri-de Vega equation. The other, derived from functional 
equations for the transfer matrix, leads to the (full spectrum) thermodynamic Bethe ansatz 
equations. Both produce one or a system of exact nonlinear integral equations of Freedholm type
that allow the evaluation of energy, momenta and other observables.
Both equations have the structure
\bea
f(x)&=&\nu(x)+ \chi(x,\theta_i)+K\star \log(1+\exp(\alpha f))\nonumber \\
\nu(x)&=&\mbox{driving term }\label{struttura}\\
\chi(x,\theta_i)&=& \mbox{sources, to be fixed by quantization conditions} \nonumber
\eea
with quantization conditions given by
\be\label{zero}
f(\hat{\theta}+\theta_i)=I_i\,,\qquad I_i\in\mathbb{Z} 
\ee
($\hat{\theta}$ is a fixed shift that appears in the TBA case but not in the DdV one).
The function $f$ itself can carry an index, thus participating to a system of coupled equations.
The two approaches complement each other, precisely as the $\T$ and $\Q$ operators 
complement each other in Baxter T-Q relation (\ref{TQperiod}). Indeed, the main message is that
the $\Q$ operator is behind the Kl\"umper-Batchelor-Pearce-Destri-de Vega approach while the $\T$ 
operator gives rise to the TBA equations. 

In spite of the many analogies, the two approaches offer different paths to evaluate the main 
observables. 
The choice of using one or the other is mainly dictated by the available initial information:
Bethe roots/holes or zeros of the transfer matrix eigenvalues.
Notice that the Kl\"umper-Pearce-Destri-de Vega equation is exact in both the finite lattice and 
the continuum limit while the TBA equations exist only in the continuum limit. 
Indeed, the first approach does not assume a ``string hypothesis'' for the zeros/holes
while the second one uses the notion that zeros have a fixed real part. This is true in the 
large size limit and is very much like a ``string hypothesis'' for the transfer matrices. Actually, 
it is even more than an hypothesis as it has been widely tested. 

Another major difference is in the number of equations. For the sine-Gordon case, one Destri-de Vega
equation describes the spectrum. A corresponding TBA set would couple an infinite number of
equations. This is better understood if one defines a counting function (see (\ref{def.Zn})) by
\be
\mathcal{F}=\exp(i Z(u))=\frac{Q(u+\lambda)f(u-\frac{\lambda}2)\sin(u-\lambda)}
{Q(u-\lambda)f(u+\frac{\lambda}2)\sin(u+\lambda)}\ \frac{\kappa(u)}{\kappa(\lambda-u)}
\ee
such that Bethe equations (\ref{TQBethe}) reduce to 
\be
\mathcal{F}(u_j)=-1
\ee
The function $\kappa(u)$ is the solution of 
\bea
\kappa(u)\kappa(u+\lambda)&=&f_1\ f_1\ s_{-2}\ s_2\nonumber \\
\kappa(u)&=&\kappa(\lambda-u)
\eea 
and is given in \cite{NepomechiePearce}. This function has mainly a normalization role and is not 
very important to the present purposes. 
Notice that in the definition of the counting function the rightmost factor is actually 1.
The first case of the T-system (\ref{funct}) can be written as
\be
1+d^1(u)=\frac{s_{-1}s_1}{f_1\ f_1\ s_{-2}\ s_2}D^1_0\ D^1_1
\ee
therefore, using the T-Q relation (\ref{TQbordo}), one gets
\be\label{ddvtba}
1+d^1(u)=\left[1+\mathcal{F}(u+\textstyle\frac{\lambda}2)\right]
\left[1+\frac{1}{\mathcal{F}(u-\frac{\lambda}2)}\right]
\ee
Now it is obvious that the TBA approach has to solve for the whole hierarchy, here 
represented by the left hand side, while the counting function of the DdV equation (right hand side)
is just related to the first fusion level. 
This last equation is very important because it makes the bridge between the two formalisms.
Here is has been written for the $A_L$ models but, taking its limit $\lambda\rightarrow 0$ and rescaling
$u$ as done in the second part of \cite{FPW}, one adapts it to the XXX model. In that case the 
hierarchy has no truncation, that means that $q$ is not upper 
bounded. 
The six-vertex untruncated hierarchy holds for the sine-Gordon model; this makes clear that the TBA 
equations for sine-Gordon form an infinite system, as previously indicated.
I have derived (\ref{ddvtba}) following the paper \cite{janosarpad}. 

The Zamolodchikov scattering formulation of TBA equations \cite{zam91} was based on a dressed Bethe 
ansatz, namely a Bethe ansatz based on physical particles and not on quasi-particles; 
that formulation was not able to treat excited states. An interesting but extremely lengthy 
method to describe excitations was formulated by Dorey and Tateo \cite{doreytateo}. 
It is based on analytic continuations of the ground state TBA equations by the adimensional
parameter $r=MR$, that is the product of the mass of the fundamental particle and of the 
space size. This parameter enters the TBA equations as
\be
\nu(x)=r\cosh x
\ee
and represents the momentum of a particle with (real) rapidity $x$. Analytic continuation 
along a path that encloses the singularities of the equation and returns to the real axis produces 
equations for the excited states: 
pictorially, moving around branch point singularities let one change the Riemann sheet and access a new 
excitation level. 
The difficulty of this approach is in finding systematic classifications of these singularities. 
Clearly, the lattice formulation given here has no such difficulty and all excitations 
are easily described. 

The work of Bazhanov, Lukyanov and Zamolodchikov \cite{blz1}, \cite{blz_exc} was motivated by the 
need of describing excitations in a more systematic way by constructing a Y-system \textit{ab-initio} 
for a quantum field theory formulation of transfer matrices and $\Q$ operators. 
They could get the Y-system. The methods shown in this chapter are an efficient way to 
solve the Y-system, for the ground state and all the excitations. 

There is another important difference, in the role of the function $f(x)$. Indeed, in the TBA 
case $\alpha=-1$, the function $\exp (- f)$
indicates how the energy is distributed among the degrees of freedom and is real, as it
is a transfer matrix eigenvalue.
In the Kl\"umper-Pearce-Destri-de Vega case, $f$ is a counting function namely it controls the 
density of Bethe roots (indicated with $\rho$)
\be
\frac{dZ(u)}{du}\sim 2\pi \frac{I_{j+1}-I_j}{u_{j+1}-u_j}\sim 2\pi \rho(u)
\ee
and is especially related to the momenta of the particles. $\alpha=i$ so $\exp(i f)$ is a complex 
function.

\section{Integrals of motion}
The DdV equation and the TBA equations allow the evaluation of energy, momenta and other 
integrals of motion.
For the TBA equations, the equations for high integrals of motion have been obtained in 
\cite{fevgrinza} thus providing explicit 
expressions in the case of the tricritical Ising model with boundary perturbations
\be\label{integrals}
C_n\  I_{2n-1}(\xi)
=\frac{2}{2n-1}\sum_{k=1}^{m_1}e^{-(2n-1)y_k^{(1)}} +(-1)^{n}\int_{-\infty}^{\infty}\frac{dy}{\pi}
\log(1+\hd^1(y,\xi))\ e^{-(2n-1)y}
\ee
The constant is taken from \cite{blz1}
\be
C_n=2^{2-n}\ 3^{1 - 2 n}\ 5^{1 - n}\ \frac{(10 n-7)!!}{n!\, (4 n-2)!} \ \pi
\ee
The case $n=1$ gives the energy $E=I_1(\xi)$ as in (\ref{sc_energ}). 
The TBA equations and quantization conditions are as in (\ref{sc_tba}) and (\ref{sc_psi}) with 
$L=4$ and with the addition, on the right hand side of $\log \hd^q{(x)}$, of the boundary 
interaction term $\log g_q(x,\xi)$. 
In particular, for the boundary flow $(1,2)\rightarrow (1,1)$ the choice is
\be
g_1(x,\xi)=\tanh\frac{x+\xi}{2}\,,\qquad g_2(x,\xi)=1
\ee 
with $\xi=-\infty$ corresponding to the boundary condition (1,2), namely an unstable UV point, 
and $\xi=+\infty$ to the (1,1), a stable IR point. This perturbative flow is triggered by the 
boundary operator $\phi_{1,3}$, namely by an operator that acts on the border of the strip.
Notice also that $\hat{t}_1$ and $\hat{g}_1$ of \cite{fevgrinza} correspond to 
$\hd^2$ and $\hat{g}_2$ of the present paper; analogously $\hat{t}_2$ and $\hat{g}_2$ correspond to
$\hd^1$ and $\hat{g}_1$.
These equations are strategically important to fix the correspondence with basis vectors in the conformal
field theory. Indeed, the difficulty with TBA equations is that they provide expressions for the
energy but no indication of the states. In conformal field theory, given the high amount of symmetry,
typically many states have the same energy. For example, in a standard cylinder quantization, 
the vacuum sector of the tricritical Ising model has level degeneracies
\be
1,\ 0,\ 1,\ 1,\ 2,\ 2,\ 4,\ \ldots
\ee
How can one match lattice states and conformal field theory states?
The first few conformal field theory conserved charges are given in \cite{sasaki1987}, obtained 
after quantization of classical integrals of motion of the modified KdV and sine-Gordon models. 
This derivation is very important because it is naturally related to $\phi_{1,3}$ perturbations and 
to the structure of the Y-system (\ref{ysyst2}). The integrals of motion are
\bea
\I_1&=&L_0 -\frac{c}{24}  \nonumber \\
\I_3&=&2\sum_{n=1}^{\infty}L_{-n}L_n +L_0^2-\frac{2+c}{12}L_0 +\frac{c(5c+22)}{2880}\label{cftintegrals}\\
\I_5&=&\sum_{m,n,p}:\!L_n L_m L_p\!:\delta_{0,m+n+p}+\frac{3}{2}\sum_{n=1}^{\infty}L_{1-2n}L_{2n-1} 
+\sum_{n=1}^{\infty}\left(\frac{11+c}{6}n^2-\frac{c}{4}-1 \right)L_{-n}L_n \nonumber  \\
&&-\frac{4+c}{8} L_0^2 +\frac{(2+c)(20+3c)}{576}L_0 - \frac{c(3c+14)(7c+68)}{290304}\nonumber
\eea
Expressions for the following cases become quickly very complicated.
In terms of the generators of the Virasoro algebra, the space of states is built by linear 
superpositions of the states (\ref{levels}).
A conceptually simple (but technically very difficult) problem of linear algebra is to find 
common eigenstates of the integrals of motion on the Virasoro basis (\ref{levels}). For the first 
few levels, an explicit expression has been evaluated in \cite{fevgrinza}, together with the
corresponding eigenvalues, providing a list of eigenvalues $I_n^{\text{CFT}}$. At higher energy, 
the eigenvalues are given by solving algebraic equations of degree equal to the degeneracy. 

Using the eigenvalues (\ref{integrals}), one has another list $I_n^{\text{TBA}}$.
Matching the two lists creates a one-to-one dictionary that in \cite{fevgrinza} was appropriately called
\textit{lattice-conformal dictionary}.  The wording is inspired from \cite{melzer}. 
I'm not aware of closed expressions for the integrals (\ref{integrals}), like the energy expression
(\ref{energy3}), even if I believe they should exist. 
For this reason, numerical evaluations have been used. Notice that, even if $I_n^{\text{TBA}}$ and
$I_n^{\text{CFT}}$ are evaluated numerically, the matching is exact because the spectrum is discrete, 
as one can appreciate looking at the values given for the vacuum sector in table~\ref{t:lista}. 

\openin1=dati_I3_11cft.tex
\openin2=dati_I3_11tba.tex
\openin3=dati_I5_11cft.tex
\openin4=dati_I5_11tba.tex
\newcommand\ra{\read1 to \datoa \datoa}
\newcommand\rb{\read2 to \datob \datob}
\newcommand\rc{\read3 to \datoc \datoc}
\newcommand\rd{\read4 to \datod \datod}
\renewcommand{\v}{|0\rangle}
\begin{sidewaystable}
\caption{Comparison of the eigenvalues of $\I_3$ and $\I_5$ from conformal field theory and from TBA in the vacuum sector of the tricritical Ising model. The left column contains the level degeneracy (l.d.) as indicated in the conformal character: $d q^{l}$ \label{t:lista}}
$$
\begin{array}{@{~~}r@{\hspace{5mm}}r@{~\longleftrightarrow~}l@{~~~~~~}l@{~~~~~~}l@{~~~~~~}l@{~~~~~~}l}
\hline \rule{0mm}{6mm}
\text{l.d.} & \multicolumn{2}{l}{\text{lattice-conformal dictionary}}
 &\ra&\rb&\rc&\rd\\[2mm] \hline
\rule{0mm}{8mm} 1 & (~) &\v  &\ra&\rb&\!\!\!\!\rc&\!\!\!\!\rd \\[8mm]
1q^2 & (00) & L_{-2}\v &\ra&\rb&\rc&\rd \\[8mm]
1q^3 & (10) & L_{-3}\v &\ra&\rb&\rc&\rd \\[8mm]
2q^4 & (20) & 3(\frac{4+\sqrt{151}}{5} L_{-4}+2\, L_{-2}^2)\v &\ra&\rb&\rc&\rd \\[3mm]
     & (11) & 3(\frac{4-\sqrt{151}}{5} L_{-4}+2\, L_{-2}^2)\v &\ra&\rb&\rc&\rd\\[8mm]
2q^5 & (30) &(\frac{7+\sqrt{1345}}{2} L_{-5}+20\, L_{-3}L_{-2})\v &\ra&\rb&\rc&\rd \\[3mm]
     & (21) &(\frac{7-\sqrt{1345}}{2} L_{-5}+20\, L_{-3}L_{-2})\v &\ra&\rb&\rc&\rd \\[8mm]
4q^6 & (40) & (11.124748\, L_{-6}+9.6451291\, L_{-4}L_{-2}  &\ra&\rb&\rc&\rd \\ 
 & \multicolumn{2}{l}{\hspace{27mm}+4.4320186\, L_{-3}^2+ L_{-2}^3)\v } &&&& \\[4mm]
 & (31) &(-4.9655743\, L_{-6} + 2.3354391\, L_{-4}L_{-2}   &\ra&\rb&\rc&\rd \\
 & \multicolumn{2}{l}{\hspace{27mm}+0.71473858\, L_{-3}^2+ L_{-2})\v} &&&& \\[4mm]
 & (22) &(0.66457527\, L_{-6} - 1.2909210\, L_{-4}L_{-2} &\ra&\rb&\rc&\rd\\
 & \multicolumn{2}{l}{\hspace{27mm}-1.2605013\, L_{-3}^2+L_{-2}^3)\v} &&&& \\[4mm]
 & (0000|00) &(-1.6612491\, L_{-6} - 4.0646472\, L_{-4}L_{-2} &\ra&\rb&\rc&\rd \\
 & \multicolumn{2}{l}{\hspace{27mm}+ 1.4118691\, L_{-3}^2 + L_{-2}^3)\v} &&&&\\[3mm]  \hline
\end{array}
$$
\end{sidewaystable}
\closein1\closein2\closein3\closein4

One can wonder about the fate of such integrals of motion when a 
relevant perturbation is switched on. In the present case, where the $\phi_{1,3}$ 
boundary perturbation is concerned, the TBA formulation is preserved because, as already discussed,
this perturbation generates flows for which the Y-system and the functional equations still hold. 
This is well known on the lattice side.
It is also know that at special points of the sine-Gordon coupling, one describes the $\phi_{1,3}$ 
perturbations of the minimal models (\ref{minimal}) therefore it is natural to expect 
that the quantities (\ref{cftintegrals}) are compatible with such a perturbation, being derived from
the sine-Gordon integrals of motion.
Among the possible families of involutive integrals of motion allowed in the CFT, they  
are those whose ranks (or Lorentz spins, namely the indices $n$) are predicted to be preserved, 
by the counting argument in \cite{Zam-adv}. The perturbed operators are given in \cite{blz_exc}.
Numerical investigations of $I_3(\xi)$ were done in \cite{fevgrinza}.

The expressions (\ref{integrals}) actually work for all the other models described by (\ref{sc_tba}).
I would like to mention that very similar expressions hold in the DdV formalism \cite{marcodavide}. 
Also these authors do not find a closed form for (\ref{integrals}), except at a free fermion point. 
In the Ising case $A_3$ a closed form is known in terms of poly-logarithms \cite{nigro}, but
this is a free fermion!

\section{Numerical considerations}
The form of equations (\ref{struttura}) is particularly suited to be solved by iteration. Indeed, 
starting from an initial guess 
\be
f_0=\nu(x)
\ee
one iterates by
\be\label{itera}
 f_{k+1}(x)=\nu(x)+ \chi(x,\theta_i)+K\star \log(1+\exp(\alpha f_k))
\ee
up to the required precision. This looks very easy and, sometimes, it is so. The difficulties come 
when there are sources to fix, in particular when they are outside the real axis, in which case
one experiences also a big increase of the computational time.  
The difficulty is due to the mixing of the ``functional'' problem, namely finding a function $f$,
with the ``sources'' problem, namely finding the sources. Then, one has to iterate at the same time
on $f$ and on $\theta_k$. 

The numerical approach has been used in many occasions. In \cite{FPR2}, there is a wide discussion
in relation to the TBA case. See also the paper \cite{FFGR2} for the DdV case. 
Here I would like to discuss about the question of convergence. 
Is the iteration (\ref{itera}) converging? Is it converging to the good solution, if 
there are multiple solutions?

The {\em contraction mapping theorem} states that if a mapping $\mathcal{M}: V \rightarrow V$ on a
complete metric space $V$ is a contraction then there exists a unique fixed point 
$f_0=\mathcal{M}(f_0)$ and
all the sequences obtained under iteration starting from an arbitrary initial point $f\in V$
converge to the fixed point. 
In practice, the derivative $|\mathcal{M}'(f)|$ measures the strength of the contraction and the 
rate of convergence. The case $|\mathcal{M}'(f)|<1$ is a contraction while $|\mathcal{M}'(f)|>1$
is a dilatation. Values close to 0 converge quickly, values close to 1 converge slowly. 

If one could show that the mapping $f_{k+1}=\mathcal{M}(f_k)$ is a contraction, then the answer to the 
previous questions would be affirmative. Unfortunately, the mapping is not easy to evaluate.
Restricting to the TBA case, one can do some steps forward. By varying $f$ in (\ref{itera})
one has
\be\label{itera2}
 \delta f_{k+1}(x)=\int \frac{1}{2\pi\cosh(x-y)}\  \frac{-\exp(-f_k)}{1+\exp(-f_k)} \delta f_k(y)\ dy
\ee
By the integral  
$$
\int_{-\infty}^{\infty} \frac{1}{\cosh t}\ dt=\pi  \,,
$$
the first fraction sums up to $\frac12$ therefore ``in average'' is a contraction.
If $f_k$ is real, as it happens in the ground state, the last fraction has absolute value lower 
than 1. This suggests that the 
integration acts globally as a contraction with factor smaller than $\frac12$ therefore iteration is 
convergent to the unique solution. 

Numerical calculations have shown that, in absence of sources, the convergence of the iteration 
equations is usually fast. In the $L=4$ case, namely the tricritical Ising model, 30 iterations are 
sufficient to reach 9 significant digits, that confirms the estimate of a contraction factor of $1/2$
\be
2^{-30}\sim 10^{-9}
\ee
In presence of sources, the intuitive evaluation breaks down because one has to iterate on the sources 
position. This is related to the fact that $f_k$ can acquire an imaginary part multiple of $\pi$, 
although $1+\exp(-f)>0 $ as in (\ref{positivita}).
On numerical calculations, one immediately observes the need to iterate longer, up to
hundred times when several sources are considered. Moreover, in \cite{FPR2} it was pointed out that
certain algorithms to fix sources do not converge. The problem does not appear for the function 
$f$ itself. When algorithms for sources do not converge, they appear to be dilatations, in which case
the iteration takes away from the fixed point and the solution must be found by other means.
This fact is curious and could become more serious in other models with different kernel or sources 
structure, up to the point of preventing the unicity of the solution. 

In the models considered here, an incomplete possible argument goes as follows. In (\ref{sc_psi}) 
one neglects 
the integrals, that often appear to be much smaller that the source terms. Then, an equation with 
$q>1$ looks like 
\be
\exp(-i\ \pi\ n_k^{(q)})=
\prod_{j=1}^{m_{q-1}} \tanh\Big(\frac{\hy_k^{(q)}-\hy_j^{(q-1)}}{2}-i\frac{\pi}{4}\Big)
\prod_{h=1}^{m_{q+1}} \tanh\Big(\frac{\hy_k^{(q)}-\hy_h^{(q+1)}}{2}-i\frac{\pi}{4}\Big)
\ee
and $\hy_k^{(q)}$ can be extracted for example by 
\begin{gather}
\tanh \Big(\frac{\hy_k^{(q)}-\hy_1^{(q-1)}}{2}-i\frac{\pi}{4}\Big)=Y(\hy_k^{(q)})\\
Y(\hy_k^{(q)})\mathop{=}^{\text{def}}\exp(-i\ \pi\ n_k^{(q)})
\prod_{j=2}^{m_{q-1}} \coth\Big(\frac{\hy_k^{(q)}-\hy_j^{(q-1)}}{2}-i\frac{\pi}{4}\Big)
\prod_{h=1}^{m_{q+1}} \coth\Big(\frac{\hy_k^{(q)}-\hy_h^{(q+1)}}{2}-i\frac{\pi}{4}\Big)
\end{gather}
Notice that $Y()$ is a function of modulus one. Inverting the $\tanh$ leads to the iterative form
\be\label{iter}
\hy_k^{(q)}=\hy_1^{(q-1)}+i\frac{\pi}{2}+\log\frac{1+Y(\hy_k^{(q)})}{1-Y(\hy_k^{(q)})}=\mathcal{Y}_k^{(q)}(\hy_k^{(q)})
\ee
The derivative is
\be
\frac{d\mathcal{Y}_k^{(q)}}{d\hy_k^{(q)}}=\frac{2Y'(\hy_k^{(q)})}{1-(Y(\hy_k^{(q)}))^2}
\ee
If this derivative is larger that 1, (\ref{iter}) becomes a dilatation. Evaluating it is not easy.
If the sources can be arranged in such a way that $Y(\hy_k^{(q)})$ is sufficiently close to $\pm 1$, 
the denominator can be made close to zero. 
This probably means that the derivative can be quite large and be a dilatation, but this
analysis is not conclusive.
Several numerical investigations have shown that there are sources arrangements where iterative algorithms
as (\ref{iter}) do not converge at all. In these cases, other methods are needed. For example, one first
estimates the interval in which the source is expected then locates it by the bisection method. 
Of course, this uses much more computer time than an iterative method.

In \cite{FPR2} a similar situation was described in relation to a boundary function $g(x,\xi)$. 
The treatment given here is complementary to that one and, in some sense, is more general because 
it does hold even when boundary parameters are absent. 

In the frame of the DdV equation, in \cite{noiNP} and \cite{noiNP2} the case of ``special holes'' 
was left unsolved: in that case, the functional equation itself seems to fail to converge, or
better, it converges to a nonsense, possibly for the reasons indicated here.  

The lesson of this analysis is that 
iterations can give rise to unexpected problems and this could have consequences about the unicity 
of the solution. On the other hand, in my numerical TBA calculations, I never observed problems 
of unicity.

\section{Discussion}
I have introduced the thermodynamic Bethe ansatz method, sketched its lattice derivation
and discussed the relevance of the Y-systems to summarize the symmetries of the model.
I have presented the integrals of motion. The methods of DdV and TBA have been compared and numerical
considerations have been shown. 

Personally, I have worked on TBA equations for a fair amount of time (2000 to 2006). My contribution
has been important.
\begin{itemize}
\item I have derived and treated TBA equations for the boundary flows of the tricritical Ising model.
\item I have established the lattice-conformal dictionary.
\item I extended the TBA to all $A_L$ models and also to the XXX model.
\item I analyzed how the zeros move in consequence of boundary flows.
\item I derived the conformal characters (conformal partition functions) from TBA.
\item I have done extensive high precision numerical calculations with TBA equations, in 
presence of many sources and of several coupled equations.
\item I worked on the physical combinatorics of quasi-particles. They form a lattice gaz whose 
partition function is the conformal character.
\end{itemize}

The first development that I propose is to make systematic the lattice-conformal dictionary. 
This is probably related to the realization of a lattice Virasoro algebra. It will lead to a 
better understanding of the space of states.

The second is related to the quasi-particles and the physical combinatorics. I 
have already experimented, in \cite{fp}, algebraic formalisms to express these quasi-particles. 
The formulation was primitive but other authors worked on it and proposed more effective 
formalisms, see \cite{mathieu} and the papers that followed it. 
I think the clarification of an algebraic formalism for the quasi-particles would help 
to work on the space of states of the minimal models and their perturbations.

\chapter{Hubbard model and integrability in $\mathcal{N}=4$ SYM\label{c:hubbard}}

\newcommand{\cc}[1]{c_{#1}^{\phantom{\dagger}}}
\newcommand{\cd}[1]{c_{#1}^{\dagger}}
\newcommand{\nn}[1]{n^{\phantom{\dagger}}_{#1}}
\newcommand{\su}{^{\uparrow}}
\newcommand{\giu}{^{\downarrow}}
\newcommand{\cV}{\mathcal{V}}
\newcommand{\cW}{\mathcal{W}}
\newcommand{\wt}[1]{\widetilde{#1}}
\newcommand{\MM}{\mbox{${\mathbb M}$}}
\newcommand{\pir}{\raisebox{-.12ex}{$\stackrel{\circ}{\pi}$}{}}
\newcommand{\Wr}{\raisebox{.12ex}{$\stackrel{\circ}{W}$}{}}
The Hubbard model was introduced in order to investigate strongly correlated
electrons in matter \cite{Hubbard,Gutzwiller} and since, it has been widely studied,
essentially due to its connection with condensed matter physics. 
It has been used to describe the Mott metal-insulator transition \cite{Mott,Hubbard3}, high critical 
temperature $T_c$ superconductivity \cite{Anderson,Affleck}, band magnetism \cite{Lieb}
and chemical properties of aromatic molecules \cite{heilieb}.
The literature on the Hubbard model being rather large, I refer to the books \cite{Monto,EFGKK} and 
references therein. Exact results have been mostly obtained in the case of the
one-dimensional model, which enters the framework of our study. In
particular, the 1D model Hamiltonian eigenvalues have been obtained by means of the coordinate 
Bethe Ansatz by Lieb and Wu \cite{LiebWu}. 

One of the main motivations for the present study of the Hubbard model and
its generalisations is the fact that it has unexpectedly appeared in the context of $N=4$ super 
Yang-Mills theory. 
This is a superconformal gauge theory in four dimensions, conjectured to be dual to 
a string theory in a $\mbox{AdS}_5\times\mbox{S}^5$ background, a ten dimensional space. 

Indeed, it was noticed in \cite{Rej:2005qt} that the Hubbard model at half-filling,
when treated perturbatively in the coupling, reproduces the long-ranged integrable spin chain 
of \cite{Beisert:2004hm} as an effective theory. It thus provides a localisation of the long-ranged 
spin chain model and gives a potential solution to the problem of describing interactions which are 
longer than the length of the spin chain. The Hamiltonian of this chain was conjectured in 
\cite{Beisert:2004hm} to be an all-order description of the dilatation operator of $N=4$ super 
Yang-Mills in the $SU(2)$ sector. That is, the energies of the spin chain where conjectured
to be proportional to the anomalous dimensions of the gauge theory operators in this sector. 
After, it was shown that starting with the fourth
loop terms, the Hubbard model is incomplete in describing the dilatation operator 
\cite{bes2006}, certain highly nontrivial phase factors being required.
But this wasn't the end of the story! 
The full factorized scattering matrix of the gauge theory has been studied and Beisert has shown the 
relation of this 
S-matrix with the Shastry R-matrix of the Hubbard model \cite{beis2006}. This means that the integrable
structure of the Hubbard model enters in the conjectured integrable structure of the SYM theory.

In this chapter, I will present two different approaches to the Hubbard model, one based on the 
Kl\"umper-Batchelor-Pearce-Destri-de Vega method \cite{FFGR, FFGR2} and one based on 
R-matrices \cite{DFFR}. 
The first one has led to the evaluation of energies for the antiferromagnetic state. 
It allows also to control the order of the limits of high coupling and high lattice size. 
The large size of the model is easily treated at all values of the coupling. 
This is important as in the SYM frame it corresponds to very long monomials of local operators, totally 
inaccessible with ordinary diagrammatic techniques.
For the second approach, in 2005-2006 I thought there may be the possibility that some integrable extension 
of the Hubbard model could be put in relation to other subsectors of the $N=4$ super Yang-Mills theory
given that the Hubbard model itself was observed in relation to the sector  $SU(2)$.
Here I will discuss a general approach to construct a number of supersymmetric
Hubbard models. Each of these models can be treated perturbatively and thus
gives rise to an integrable long-ranged spin chain in the high coupling limit.

Other symmetric or supersymmetric generalizations of the Hubbard model have been
constructed, see e.g. \cite{EKS}. These approaches mainly concern high
$T_c$ superconductivity models. They essentially use the $gl(1|2)$ or $gl(2|2)$ superalgebras, 
which appear as the symmetry algebras of the Hamiltonian of the model. The approach 
I have adopted in  \cite{DFFR}
however is different, being based on transfer matrices and quantum inverse scattering framework. 
It ensures the integrability of the model and allows one to obtain local Hubbard-like
Hamiltonians for general $gl(N|M)$ superalgebras. After a Jordan-Wigner transformation, these 
Hamiltonians appear to describe one or more families of charged and chargeless fermions.

\section{The Hubbard model\label{sect:Hubbard}}
The Hubbard model, introduced in \cite{Hubbard,Gutzwiller}, describes hopping electrons on a lattice, 
with an ultralocal repulsive potential that implements a screened Coulomb repulsion, with $U>0$. 
The 1-dimensional Hamiltonian is given by
\be\label{oldHubb}
H=-t \sum _{i=1}^L \sum _{\rho=\uparrow , \downarrow}
\left(e^{i\phi} \cd{\rho,i}\cc{\rho,i+1}+e^{-i\phi}\cd{\rho,i+1}\cc{\rho,i}\right) 
+U \sum_{i=1}^L \big(1-2\nn{\uparrow,i}\big)\big(1-2\nn{\downarrow,i}\big)
\ee
where $\cc{},\cd{} $ are usual fermionic operators, $i$ indicates the lattice site and $\rho$ is the 
``spin orientation''. I will always use periodic boundary conditions. 

The physical idea behind this Hamiltonian is that the metallic positive ions create the  
crystalline structure. Each ion puts up to two electrons in the conductive band. Ions are much 
heavier than electrons so for most investigations they can be considered as static thus the lattice
has no dynamics.
Electrons in the conductive band repel each other (of course!) but they also experience major 
screening effects. 
Indeed, an electron feel the repulsion of the other electrons but also the strong, periodic, 
attraction by the ions. 
This makes the Coulomb repulsion short-ranged. In the Hubbard model, the electronic repulsion is
modeled with an ultralocal term: electrons interact only if they are on the same site. 
Pauli exclusion then implies 
that they interact if they have opposite spin only. Pauli exclusion implies also that 
a maximum of $2L$ electrons can be accommodate in the lattice, in which case it is ``fully 
filled''. I will often use the ``half filled'' case that contains precisely $L$ electrons. 
The phase $\phi$ in the Hamiltonian represents a uniform magnetic field. For many purposes, one 
can put it to zero. In the approach of \cite{FFGR2} this phase was introduced to fit with 
the Hubbard model used in \cite{Rej:2005qt}.

There are some features that can be explored without too much calculative effort. If $U=0$, the Hamiltonian 
describes free fermions (electrons). The first term in (\ref{oldHubb}) describes hopping between 
nearest neighbor sites in such a way that
electrons can freely move around, yielding a conductor. On the other hand, when 
$U$ becomes very large, it appears that the total energy is lower if one can make negative the 
contribution from the potential term 
$\big(1-2\nn{\uparrow,i}\big)\big(1-2\nn{\downarrow,i}\big)$, namely if on each site there is just one 
electron. At half filling and large $U$, the ground state has precisely this form with one electron 
per 
site, no empty sites and no doubly occupied sites. At zero temperature and large $U$ this state is
fully frozen because overturning a spin would require an amount of energy of $U$ to create a
state with a doubly occupied site. This ``frozen'' state describes a Mott insulator namely a system
whose conductive band is not empty but the Coulomb interaction forbids any electronic 
displacement. At positive temperature, the ground state is always conductive because thermal excitations
can provide the amount of energy needed to create vacances and double occupances.

The large $U$ regime is the spin chain limit. Indeed, the Hubbard model looks very close to an 
Heisenberg XXX model: one (quantum) spin per site, up or down. 
Notice that if the lattice is not half filled, there is conduction whatever is the value of $U$. 

The underling algebraic structure leads to superalgebras. In a first instance, I consider a single
fermion 
\be\label{fermi}
\{\cc{},\cd{}\}=\I \,,\qquad n=\cd{} \cc{}
\ee
where $\I$ is the identity operator, $n$ is the number operator and $\{,\}$ is the anticommutator.
The operators $\cc{},\cd{},\I,n$ form a realization of a $gl(1|1)$ superalgebra. 
One way to see this is to write down the whole set of ``commutation rules''
\be
[n,c]=-c\,,\qquad [n,\cd{}]=\cd{}\,,\qquad [X,\I]=0 \mbox{ ~for~ } X\in\{n,c,\cd{}\} 
\ee
These and (\ref{fermi}) can be realized by the two dimensional matrices of $gl(1|1)$ 
$$
E_{12}=\begin{pmatrix} 0 & 1\\ 0 & 0\end{pmatrix}=c \,,\quad 
E_{21}=\begin{pmatrix} 0 & 0\\ 1 & 0\end{pmatrix}=\cd{} \,,\quad
E_{11}=\begin{pmatrix} 1 & 0\\ 0 & 0\end{pmatrix}=n\,,\quad
E_{22}=\begin{pmatrix} 0 & 0\\ 0 & 1\end{pmatrix}=\I-n\,. 
$$
On each site of the Hubbard lattice there are two ``spin polarizations'' so on each site
there is a $gl(1|1)\oplus gl(1|1)$ superalgebra and, on the whole lattice, the fermionic structure 
\be
\{\cc{\rho,i},\cd{\rho',j}\}=\delta_{\rho,\rho'}\delta_{i,j} \, \qquad
\{\cc{\rho,i},\cc{\rho',j}\}=\{\cd{\rho,i},\cd{\rho',j}\}=0
\ee
is $L$-times the tensor product of the one site structure. 
I can easily represent the fermionic structure by a graded tensor product of the matrices
\be
E_{12;\rho,i}=\cc{\rho,i} \,,\quad  
E_{21;\rho,i}=\cd{\rho,i} \,,\quad
E_{22;\rho,i}=\nn{\rho,i}=\cd{\rho,i}\cc{\rho,i} \,,\quad
E_{11;\rho,i}=1-\nn{\rho,i}=\cc{\rho,i}\cd{\rho,i} \label{JW}
\ee
When it occurs, the second pair of labels $\rho,i$ indicates the spin polarization $\rho$ 
and the site $i$.
The matrices $E_{12}\,,E_{21}$ are taken of fermionic character (they satisfy anticommutation relations whatever their spin and space labels are) and $E_{11}\,,E_{22}$ are taken of bosonic character (they always enter commutation relations whatever their spin and space labels are).
The relation (\ref{JW}) is a graded Jordan-Wigner transformation\footnote{The ordinary
Jordan-Wigner transformation is 
$\displaystyle\cd{\uparrow,i}=\sigma^{-}_{\uparrow,i} \prod _{k>i} \sigma^{z}_{\uparrow,i}$
for the up polarization; an additional term occurs for the down polarisation.} 
and respects periodic boundary conditions\footnote{The standard one violates periodicity.}.
I now rewrite the Hubbard Hamiltonian in the matrix language
\be\label{hubb}
H = -t \sum _{i=1}^L \sum _{\rho=\uparrow , \downarrow}
\left( E_{21;\rho,i}\ E_{12;\rho,i+1}+E_{21;\rho,i+1}\ E_{12;\rho,i}\right) 
+U \sum_{i=1}^L  
\big(E_{11;\uparrow,i}-E_{22;\uparrow,i}\big)\big(E_{11;\downarrow,i}-E_{22;\downarrow,i}\big) 
\ee
and I split it into the sum of the two polarizations 
\begin{gather}
H= H_{\text{XX}}\su +H_{\text{XX}}\giu +U \sum_{i=1}^L 
\big(E_{11;\uparrow,i}-E_{22;\uparrow,i}\big)\big(E_{11;\downarrow,i}-E_{22;\downarrow,i}\big)\;; \label{hubbspin}\\
H_{\text{XX}}^{\rho} = -t \sum _{i=1}^L 
\left( E_{21;\rho,i}\ E_{12;\rho,i+1}+E_{21;\rho,i+1}\ E_{12;\rho,i}\right)\;. \nonumber
\end{gather}
Taking one polarization of the kinetic term one easily sees
\begin{gather}
E_{21;\rho,i}\ E_{12;\rho,i+1}+E_{21;\rho,i+1}\ E_{12;\rho,i}=
\frac12 \Big[ E_{x;\rho,i} \ E_{x;\rho,i+1}+ E_{y;\rho,i}\  E_{y;\rho,i+1} \Big]\\[3mm]
E_{x;\rho,i}=\begin{pmatrix} 0 & 1\\ 1 & 0\end{pmatrix}_{\rho,i} \,,\quad 
E_{y;\rho,i}=\begin{pmatrix} 0 & -i\\ i & 0\end{pmatrix}_{\rho,i} \notag
\end{gather}
the appearance of two (graded) XX spin chain Hamiltonians\footnotemark,
one for each polarisation, within the Hubbard model. 
\footnotetext{At this point it should be clear that the difference between graded and non
graded cases appears when boundary effects are observed; the thermodynamic limit usually ignores such terms, being sensitive to bulk contributions only.} 

It turns out that the breaking of (\ref{hubbspin}) into the Hamiltonian of two XX models plus a 
potential term allows one to generalise this model to higher algebraic structures by maintaining 
integrability\footnote{The flux $\phi$ does not affect integrability properties.}. 

Exact investigations on the Hubbard model required many years of work.
A first hint of integrability came from the coordinate Bethe Ansatz solution obtained by Lieb and 
Wu \cite{LiebWu} in 1968 but a full understanding of it by an R-matrix satisfying a Yang-Baxter
equation came much later. An R-matrix was first constructed by Shastry \cite{shastry,JWshas} 
and Olmedilla et al.~\cite{Akutsu}, by coupling the R-matrices of two independent $XX$ models, 
through a term depending on the coupling constant $U$ of the Hubbard potential. 
The proof of the Yang-Baxter relation for the R-matrix was given by Shiroishi and Wadati \cite{shiro2}
in 1995. 

The construction of the R-matrix was then generalised to the $gl(N)$ case by Maassarani et al.,
first for the XX model \cite{maasa} and then for the $gl(N)$ Hubbard model
\cite{maasa2,maasa3}. Later, I will use this approach to generalize to $gl(N|M)$ models.

The Lieb-Wu equations \cite {LiebWu, Rej:2005qt} for the Hubbard model are, 
in the half-filling case,
\begin{eqnarray}
e^{i\hat{k}_jL}&=&\prod _{l=1}^M \frac {u_l - \frac {2t}{U}\sin
  (\hat{k}_j+\phi)-\frac {i}{2}} {u_l - \frac {2t}{U}\sin
  (\hat{k}_j+\phi)+\frac {i}{2}} \nonumber \\
\prod _{j=1}^L \frac {u_l - \frac {2t}{U}\sin
  (\hat{k}_j+\phi)+\frac {i}{2}} {u_l - \frac {2t}{U}\sin
  (\hat{k}_j+\phi)-\frac {i}{2}}&=& \mathop{\prod _ {m=1}}_{m \not=l}^M \frac
  {u_l-u_m+i}{u_l-u_m-i} \, , \label {lw}
\end{eqnarray}
where $M$ is the number of down spins; here they are modified to include the phase. The spectrum 
of the Hamiltonian is then given in terms of the momenta $\hat{k}_j$ by the 
{\it dispersion relation}
\begin{equation}\label{energia}
E=-2 t \sum _{j=1}^{L} \cos (\hat{k}_j+\phi) \,.
\end{equation}
Starting from these ``Bethe equations'', I will present two coupled nonlinear integral equations 
for the antiferromagnetic state of the model. These equations are derived in \cite{FFGR2}
in the same framework of the Kl\"umper-Batchelor-Pearce-Destri-de Vega approach of the chapter~\ref{c:nlie}.

For reason of completeness, it is important to point out that the thermodynamics (infinite length 
$L$, but finite temperature) of the Hubbard model has been studied in \cite {KB,JKS} by means of three 
nonlinear integral equations. This approach was based on the equivalence of the quantum 
one-dimensional Hubbard model with the classical two-dimensional Shastry model. The work 
presented here was oriented to gauge theory understanding. The objective was to obtain energies of 
the Hubbard model at zero temperature but at any value of the lattice size $L$ therefore the 
approach of \cite {KB,JKS} was not appropriate. 

There are some features of (\ref{lw}) that deserve some attention. In the large coupling limit 
(large $U$), the second set of equations decouples form the first one and coincides with the 
XXX Bethe equations (\ref{betheTQ}), once the limit $\lambda\rightarrow 0$ has been taken. 
This is consistent with the argument, given earlier, that the spin chain limit of the Hubbard model
is the XXX model.
On the opposite, if $U=0$ the second group becomes useless because the first group is enough to fix
$\hat{k}_j$ and the energy. The first group reduces to 
\be
e^{i\hat{k}_jL}=1
\ee
that is the box quantization of free particles. The momenta are all different, as usual in Bethe 
ansatz, therefore particles are fermions. Indeed, in this limit the Hamiltonian describes free 
fermions.

From this analysis, one can see that the Lieb-Wu equations describe the phenomenon of spin-charge 
separation. Indeed, the momenta $\hat{k}_j$ are as many as the electrons so they carry charge. 
Instead, the ``rapidities'' $u_{\ell}$ are as many as the down spins so they carry sping.
In the spin limit the quasiparticles described by $\hat{k}_j$ disappear from the equations 
while at the free fermion point $U=0$ it is the opposite.

\subsection{$\mathcal{N}=4$ super Yang-Mills and AdS/CFT}
This superconformal field theory in its planar limit,  namely the limit of an infinite number 
of colors, is probably an integrable theory. It seems related to the Hubbard model, as it was 
first observed in 
\cite{Rej:2005qt}. The following relation between coupling constants was proposed 
\be\label{couplings}
\frac{t}{U}=\frac{g}{\sqrt{2}}=\frac{\sqrt{\lambda}}{4\pi^2}\,,\qquad U=-\frac{1}{g^2}=
-\frac{8\pi^2}{\lambda}
\ee 
where $\lambda$ is the 't Hooft coupling of the theory and $g$ is related to the SYM coupling. 
The Hubbard lattice must be taken half-filled.
The energy (\ref{energia}) is related to the anomalous dimensions $\gamma_{\text{SYM}}$ 
of the super Yang-Mills operators in the scalar sector $SU(2)$ by
\be
\gamma_{\text{SYM}}=\frac{\lambda}{8\pi ^2}E
\ee
The lattice size $L$ is identified with 
the ``length'' of the operators in terms of the fundamental scalar fields of the theory.

This theory is believed to be dual to a type II string theory
on a $AdS_5\times S^5$ background. This and other dualities between quantum field theory and string 
theory are known as AdS/CFT dualities, after Maldacena \cite{maldacena}. The duality has a very 
nice and curious feature: it exchanges strong and weak couplings. As strong coupling calculations 
are usually difficult, the duality makes them accessible \textit{via} weak coupling calculations 
in the dual theory. 

After the important work of Minahan and Zarembo \cite{MZ}, there has been an explosion of researches 
in this domain. The AdS/CFT duality has been enriched of tools and new calculation methods 
by recognizing that there are integrable models, on both sides of the duality.

\section{Universal Hubbard models}
Following the methods of \cite{maasa} and \cite{martins}, it has been possible to generalize the 
Hubbard model to include more general symmetries than the original $SU(2)$ one.
In a first stage, the XX model is generalized to (almost) arbitrary  vector
spaces and symmetries. Secondly, two copies of the XX model are ``glued'' to form a Hubbard model. 
This is the usual construction of the R-matrix of the Hubbard model. 

I will use the standard notation in which the lower index indicates the space on which the operator 
acts. For example, to $A\in \mbox{End}(V)$, I associate the operator $A_{1}=A\otimes \I$ and
$A_{2}=\I\otimes A$ in $\mbox{End}(V)\otimes \mbox{End}(V)$. More
generally, when considering expressions in $\mbox{End}(V)^{\otimes k}$,
$A_{j}$, $j=1,\ldots,k$ will act as the identity in all spaces $\mbox{End}(V)$ except the $j^{th}$ one.

To deal with superalgebras, I will also need a $\mathcal{Z}_{2}$ grading $[.]$ on
$V$, such that $[v]=0$ will be associated to bosonic states
and $[v]=1$ to fermionic ones. 

The construction of a universal XX model is mainly based on general properties 
of projectors and permutations. The needed projectors $\pi,\wt\pi$ select a proper subspace of $V$
\begin{eqnarray}
\pi:\ V\to\ W\,,\quad 
\wt\pi=\I-\pi:\  V\to\ \wt{W} \mbox{~~~with~~~} V=W\oplus\wt{W}
\label{def:univpi}
\end{eqnarray}
In the tensor product of two vector spaces I take the possibly graded permutation
\begin{eqnarray}
P_{12}:
\begin{cases} V\otimes V \ \to\ V\otimes V\\
v_{1}\otimes v_{2}\ \to\ (-1)^{[v_{1}][v_{2}]}\, v_{2}\otimes v_{1}
\end{cases}
\end{eqnarray}
and also $\Sigma_{12}$ 
\begin{eqnarray}
\Sigma_{12} &=& 
\pi_{1}\,\wt\pi_{2}+\wt\pi_{1}\,\pi_{2} 
\label{def:univSigma}
\end{eqnarray}
It is easy to show that $\Sigma_{12}$ is also a projector in $V\otimes V$:
$\left(\Sigma_{12}\right)^2=\Sigma_{12}$.
The operator $C$ will also be used later
\begin{equation}
C = \pi-\wt\pi\,.
\label{eq:opC}
\end{equation}
It obeys $C^{2}=\I$ and is related to $\Sigma_{12}$ through the equalities
\begin{equation}
\Sigma_{12}=\frac12(1-C_{1}C_{2}) \mbox{ ~and~ }
\I\otimes\I-\Sigma_{12}=\frac12(1+C_{1}C_{2})
\label{eq:univSig-C}
\end{equation}
From the previous operators, one can construct an R-matrix acting on $V\otimes V$ and with spectral
parameter $\lambda$
\begin{equation}
R_{12}(\lambda) = \Sigma_{12}\,P_{12} + \Sigma_{12}\,\sin\lambda +
(\I\otimes\I-\Sigma_{12})\,P_{12}\,\cos\lambda
\label{def:univRXX}
\end{equation}
Several properties of the R-matrix are given in \cite{DFFR}, \cite{FFR}. The most important is
the Yang--Baxter equation 
\begin{gather}
R_{12}(\lambda_{12})\,R_{13}(\lambda_{13})\,R_{23}(\lambda_{23}) = 
R_{23}(\lambda_{23})\,R_{13}(\lambda_{13})\,R_{12}(\lambda_{12})
\qquad\notag \\[2mm]
\mbox{where~ } \lambda_{ij} = \lambda_i-\lambda_j.
\label{eq:univYBE}
\end{gather}
With a very standard construction, from the R-matrix one constructs the ($L$ sites) transfer matrix
(\ref{trasf})
\begin{equation}
t_{1\ldots L}(\lambda) = \mathop{\mbox{strace}}_{0}R_{01}(\lambda)\,R_{02}(\lambda)\ldots R_{0L}(\lambda)
\end{equation}
by taking the supertrace in the auxiliary space. So far, the calculation has been very general and no 
special properties of the space $V$ are required. Now, if $V$ has infinite dimension, it is
necessary to assume the existence of a trace or supertrace with the cyclic property. 
If $V$ has finite dimension, the trace always exists. 
The relation (\ref{eq:univYBE}) implies that the transfer matrices commute for
different values of the spectral parameter, thus granting integrability.

Since the R-matrix is regular (namely in $\lambda=0$ it is a permutation), logarithmic derivatives 
in $\lambda=0$ give local operators as in (\ref{hamilt}). The first one can be chosen as XX-Hamiltonian
\begin{gather}
H=t_{1\ldots L}(0)^{-1}\, \frac{dt_{1\ldots L}}{d\lambda}(0) =\sum_{j=1}^{L} H_{j,j+1} 
\label{eq:univXXHam}\\
\mbox{with}\quad H_{j,j+1}=P_{j,j+1}\,\Sigma_{j,j+1} \nonumber
\end{gather}
where periodic boundary conditions have been used, i.e. the site $L+1$ is identified with the first one. 

For example, the original XX model (related to the algebra $gl(2)$) is obtained without gradation 
with local vector space $V=\mathbb{C}^2$ and $2\times 2$ matrices
\be \label{example}
\pi=E_{1,1}\,, \quad \wt{\pi}=\I-\pi = E_{2,2}
\ee
Then the Hamiltonian is the XX model
$$
H=\sum_{j=1}^{L} \left( E_{12;j} E_{21;j+1}+E_{21;j} E_{12;j+1} \right)=
\sum_{j=1}^{L} \left( \sigma^{+}_{j} \sigma^{-}_{j+1}+\sigma^{-}_{j} \sigma^{+}_{j+1} \right)=
\frac12 \sum_{j=1}^{L} \left( \sigma^{x}_{j} \sigma^{x}_{j+1}+\sigma^{y}_{j} \sigma^{y}_{j+1} \right)
$$
For this reason, (\ref{eq:univXXHam}) defines generalized XX models that in \cite{FFR} were called 
\textit{universal}. With the same choice (\ref{example}) but using a grading such that the index
1 is bosonic and the index 2 is fermionic, the $gl(1|1)$ XX model has Hamiltonian
\be
H=\sum_{j=1}^{L} \left( -E_{12;j} E_{21;j+1}+E_{21;j} E_{12;j+1} \right)=
\sum_{j=1}^{L}\left(\cd{j}\cc{j+1}+\cd{j+1}\cc{j}\right)
\ee
because the matrices $E_{12}$ and $E_{21}$ are both ``fermionic''; they anticommute on different sites
so the fermionic realization (\ref{JW}) can be used. The index $\rho$ here is not necessary.
The relation between the XX model and the Hubbard model is now more clear.

``Gluing'' two possibly different universal XX models produces a generalized integrable 
Hubbard model. The $R$-matrices of two universal XX models are distinguished by the arrow
$R^{\uparrow}_{12}(\lambda)$ and $R^{\downarrow}_{12}(\lambda)$. 
The Hubbard-like $R$-matrix has two spectral parameters $\lambda_1\,,\lambda_2$ and is constructed by 
tensoring on each site an ``up'' and a ``down'' copy
\begin{equation}
R_{12}(\lambda_{1},\lambda_{2}) =
R^{\uparrow}_{12}(\lambda_{12})\,R^{\downarrow}_{12}(\lambda_{12}) +
\frac{\sin(\lambda_{12})}{\sin(\lambda'_{12})} \,\tanh(h'_{12})\,
R^{\uparrow}_{12}(\lambda'_{12})\,C^{\uparrow}_{1}\,
R^{\downarrow}_{12}(\lambda'_{12})\,C^{\downarrow}_{1}
\label{R-XXfus}
\end{equation}
where $\lambda_{12}=\lambda_{1}-\lambda_{2}$ and $\lambda'_{12}=\lambda_{1}+\lambda_{2}$. 
Moreover, $h'_{12}=h(\lambda_{1})+h(\lambda_{2})$ and the choice of the function $h(\lambda)$
is fixed within the proof of the Yang-Baxter equation.
Indeed, when the function $h(\lambda)$ is given by $\sinh(2h)=U\, \sin(2\lambda)$
for some free parameter $U$, the R-matrix (\ref{R-XXfus}) obeys the Yang-Baxter equation:
\begin{eqnarray}
R_{12}(\lambda_{1},\lambda_{2})\, 
R_{13}(\lambda_{1},\lambda_{3})\, 
R_{23}(\lambda_{2},\lambda_{3}) 
&=& 
R_{23}(\lambda_{2},\lambda_{3})\, 
R_{13}(\lambda_{1},\lambda_{3})\, 
R_{12}(\lambda_{1},\lambda_{2})\,.
\end{eqnarray}
Notice that, this time, the equation is not of difference type.
As remarked in \cite{DFFR} the proof relies only on some intermediate properties
that are not affected by the choice of the fundamental projectors (\ref{def:univpi}). 
The proof follows the steps of the original proof by Shiroishi \cite{shiro2} for the Hubbard model. 
The same proof has been used for general $gl(N)$ algebras in \cite{EFGKK}.

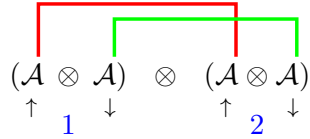
\begin{figure}[h]
\begin{center}\begin{tikzpicture}[baseline,scale=1]
\draw(1.6,0) node{$\begin{array}{ccccccc}(\mathcal{A}\!\!&\!\!\otimes\!\!&\!\!\mathcal{A})&
\otimes&(\mathcal{A}\!\!&\!\!\otimes\!\!&\!\!\mathcal{A}) \\[-1mm]
\phantom{(}\scriptstyle \uparrow\!\! &&\!\! \scriptstyle \downarrow && \phantom{)}\scriptstyle \uparrow\!\! &&\!\! 
\scriptstyle \downarrow \\[-2mm] &\textcolor{blue}{1} &&&& \hspace*{-3.5mm} \textcolor{blue}{2}\hspace*{-3.5mm}
\end{array}$};
\draw[color=red,line width=1.3pt] (0,0.6) -- (0,1.3)--(2.6,1.3)--(2.6,0.6) ;
\draw[color=green,line width=1.3pt] (1,0.6) -- (1,1.1)--(3.4,1.1)--(3.4,0.6) ;
\end{tikzpicture}\end{center}
\caption{\label{updown}This scheme shows the coupling between two universal XX models. 
The blue indices represent the sites 1 and 2 of the Hubbard model. On each site there is one 
XX up and one XX down. $\mathcal{A}$ 
represents the local vector space $V$ or the local algebra $\mbox{End}(V)$ if vectors or 
matrices are considered, respectively. }
\end{figure}

The Hubbard R-matrix is regular but non symmetric. It satisfies unitarity.
A commuting family of transfer matrices is obtained by fixing one of the two spectral parameters
\begin{equation}\label{redmonodromy}
t_{1\ldots L}(\lambda)= \mathop{\mbox{str}_{0}} 
R_{01}(\lambda,\mu)\ldots R_{0L}(\lambda,\mu) \Big|_{\mu=0}\,.
\end{equation}
Any other choice for $\mu$ is possible but, at least in view of obtaining a local Hamiltonian, 
they do not give new information. 
The `reduced' R-matrices that enter in the previous equation take a particularly simple factorised form
\begin{equation}
R_{12}(\lambda,0) = 
\,R^{\uparrow}_{12}(\lambda)\,R^{\downarrow}_{12}(\lambda)\,I^{\uparrow\downarrow}_{1}(h)
\end{equation}
where
\begin{equation}
I^{\uparrow\downarrow}_{1}(h) = \I\otimes\I+\tanh(\frac{h}{2})\,C^{\uparrow}_{1}\,C^{\downarrow}_{1}
\end{equation}
and one arrives at a Hubbard-like Hamiltonian
\begin{equation} \label{eq:HubHam}
H = \sum_{j=1}^{L}H_{j,j+1} = \sum_{j=1}^{L} \Big[
\Sigma^{\uparrow}_{j,j+1}\,P^{\uparrow}_{j,j+1}
+\Sigma^{\downarrow}_{j,j+1}\,P^{\downarrow}_{j,j+1}
+U\,C^{\uparrow}_{j}\,C^{\downarrow}_{j} \Big]
\end{equation}
where periodic boundary conditions hold. Clearly, the up and down components are put in interaction only
by the potential term $U \,C^{\uparrow}_{j}\,C^{\downarrow}_{j} $.
The tensor product of the up and down component is represented in Figure~\ref{updown}.
Operators ``up'', as $R^{\uparrow}_{12}(\lambda)$, act on the first and third spaces and are the 
identity on the others while operators ``down'', as $R^{\downarrow}_{12}(\lambda)$, act on the 
second and fourth. 
The Hubbard model itself is obtained by two graded $gl(1|1)$ models 
as in (\ref{example}). Its local (one site) vector space is 
\be
V=V^{\uparrow}\otimes V^{\downarrow}\,,\qquad V^{\rho}=\mathbb{C}^2
\ee
The universal Hamiltonian (\ref{eq:HubHam}) has the same structure of the Hubbard model. What can make
different the dynamics is the fact that the projectors $\pi,\wt\pi$ seem to introduce several types of 
particles. 

These models were introduced in relation to their symmetries. In \cite{DFFR, FFR} it has been shown
that the transfer matrix admits as symmetry (super)algebra the direct sum of the symmetry algebras 
of the XX components
\be\label{simmetria}
\mathcal{S}=\mbox{End}(W^{\uparrow})\oplus \mbox{End}(\wt{W}^{\uparrow})
\oplus \mbox{End}(W^{\downarrow})\oplus \mbox{End}(\wt{W}^{\downarrow})
\ee
(the up R-matrix commutes with the down generators and vice versa). 
The generators of the symmetry have the form of the sum of local matrices acting on a single site at 
a time
\begin{equation}
\MM=\MM^{\uparrow}+\MM^{\downarrow}\,,\qquad  \MM^{\uparrow}=\sum_{j=1}^{L}\MM^{\uparrow}_{j} \,,\qquad 
\MM^{\downarrow} =\sum_{j=1}^{L}\MM^{\downarrow}_{j}
\end{equation}
where 
\be
\MM=\MM^{\uparrow}+\MM^{\downarrow}  \qquad \mbox{and} \qquad 
\MM^\sigma\in \mbox{End}(W^{\sigma})\oplus \mbox{End}(\wt{W}^{\sigma})\,.
\ee
They commute with the monodromy/transfer matrix and with the Hamiltonian.

In this formalism, given that $W^{\uparrow}=\wt W^{\uparrow}=W^{\downarrow}=\wt W^{\downarrow}=\mathbb{C}$, 
the Hubbard model seems to have just the symmetry algebra
\be\label{simm}
gl(1)\oplus gl(1)\oplus gl(1)\oplus gl(1)
\ee
where each term is a single operator 
\be\label{simmH}
\hat{E}_{1,1;\uparrow}=\sum_{j=1}^{L} E_{1,1;\uparrow j}\,,\qquad
\hat{E}_{1,1;\downarrow}=\sum_{j=1}^{L} E_{1,1;\downarrow j}\,,\qquad
\hat{E}_{2,2;\uparrow}=\sum_{j=1}^{L} E_{2,2;\uparrow j}\,,\qquad
\hat{E}_{2,2;\downarrow}=\sum_{j=1}^{L} E_{2,2;\downarrow j}
\ee
These operators count the number of ``particles''. This is more visible after a Jordan-Wigner 
transformation (\ref{JW}): indeed the operator 
\be
\hat{E}_{2,2;\uparrow}=\sum_{j=1}^{L} n_{\uparrow, j}
\ee
counts how many up fermions are in a given state. Similarly, the operator $\hat{E}_{2,2;\downarrow}$ 
counts the number of down fermions. From the local (=on site) identity 
$ E_{1,1}+ E_{2,2}=\I$, the following sums give the lattice size
\be
\hat{E}_{1,1;\uparrow}+\hat{E}_{2,2;\uparrow}=\hat{E}_{1,1;\downarrow}+\hat{E}_{2,2;\downarrow}=L
\ee
so in the symmetry algebra (\ref{simm}) there is an amount of redundancy. 

It is well known that the Hubbard symmetry algebra is $su(2)$ and becomes $su(2)\times su(2)$ if
the number of sites is even. 
Indeed, the cases where $V^{\sigma}$ is two dimensional are special because, in addition to the list of 
generators contained in (\ref{simmetria}, \ref{simmH}), they have new generators given by
\be
S^{\pm}_j=\sigma^{\pm}_{\uparrow j}\otimes\sigma^{\mp}_{\downarrow j}\,,\qquad
\mathcal{M}^{\pm}_j=\sigma^{\pm}_{\uparrow j}\otimes\sigma^{\pm}_{\downarrow j}\,.
\ee
To be precise, the first commutes with the Hamiltonian in all cases and promotes the Hubbard symmetry 
algebra to $su(2)$, where the third generator would be 
$S^3_j=\frac12(\hat{E}_{2,2;\uparrow j}-\hat{E}_{2,2;\downarrow j})$.  
The operators $\mathcal{M}^{\pm}_j$ commutes only if $L$ is even enhancing the symmetry to 
$su(2)\oplus su(2)$. In that case, the third generator is 
$\mathcal{M}^3_j=\frac12(\hat{E}_{2,2;\uparrow j}+\hat{E}_{2,2;\downarrow j})$.

Unfortunately, this strongest symmetry doesn't extend to higher dimensional cases.
Some enlargement of the symmetry appears at large coupling in perturbative calculations 
but it does not survive at higher orders.

\subsubsection{$gl(2|2) \oplus gl(2|2) $ Hubbard Hamiltonian }
This model implements two identical copies (up and down) of an XX both with 
\be
\pi=E_{11}+E_{33}\,,\qquad \wt\pi=E_{22}+E_{44}
\ee
and with indices $3,4$ of fermionic nature. Using a graded Jordan-Wigner transformation one arrives 
at a fermionic form for the Hamiltonian 
\begin{eqnarray}\label{hamiltgl22}
H &\!\!=\!\!& \sum_{i=1}^{L} \; \Big\{ \;
\sum_{\sigma=\uparrow,\downarrow} \big( \cd{\sigma,i} \cc{\sigma,i+1}
+ \cd{\sigma,i+1} \cc{\sigma,i} \big)
\big( c_{\sigma,i}'^\dagger c_{\sigma,i+1}' + c_{\sigma,i+1}'^\dagger c_{\sigma,i}' + 1 
- n_{\sigma,i}' - n_{\sigma,i+1}' \big) \nonumber \\
&& + \, U (1-2n_{\uparrow,i})(1-2n_{\downarrow,i}) \; \Big\}
\end{eqnarray}
where the factor 
\be
\mathcal{N}'_{\sigma,i,i+1}=\big( c_{\sigma,i}'^\dagger
c_{\sigma,i+1}' + c_{\sigma,i+1}'^\dagger c_{\sigma,i}' + 1 - n_{\sigma,i}'
- n_{\sigma,i+1}' \big)
\ee
multiplies an ordinary Hubbard hopping term; only unprimed particles enter into the potential.
There are four types of fermionic particles, respectively generated by 
$\cd{\uparrow,i}\,,\cd{\downarrow,i}\,,c_{\uparrow,i}'^\dagger \,,c_{\downarrow,i}'^\dagger$
so that they define a 16 dimensional vector space on each site
\be
V_{\uparrow,i}\otimes V_{\downarrow,i}\otimes V_{\uparrow,i}' \otimes V_{\downarrow,i}'
\ee
with each $V=\mathbb{C}^2$.
The corresponding numbers of particles are conserved.

The factor $\mathcal{N}'_{\sigma,i,i+1}$ works on a $4\times 4$ one-site space; its eigenvalues 
can be easily obtained and are $\pm 1$ with two-fold multiplicity. This means that it cannot vanish, 
$\mathcal{N}'_{\sigma,i,i+1}\neq 0$.
Moreover, if no primed particles are present, $\mathcal{N}'_{\sigma,i,i+1}=1\,, ~\forall ~\sigma,i$. 
The same is true if the lattice is fully filled with primed particles
in which case $\mathcal{N}'_{\sigma,i,i+1}=-1$ therefore two of the sectors 
described by this Hamiltonian are equivalent to the ordinary Hubbard model. A Russian doll 
structure is appearing: if the projectors are well chosen, a larger model contains 
the small ones.

If there are primed particles only, the energy vanishes but not the momentum. This actually means that 
primed particles do not have a dynamics independent of the unprimed. This fact is curious and I 
am not aware of other cases in which it has been observed.
If the potential term is interpreted as a Coulomb repulsion, then unprimed particles only carry electric 
charge so primed particles are neutral. 

The compound objects formed by ~$\cd{\sigma,i}\,c_{\sigma,i}'{^\dagger}$~ are rigid: no other term in 
the Hamiltonian can destroy them. In this sense, there are four types of carriers, with the same 
charge but different behaviours:
two are the elementary objects $\cd{\sigma,i}$ in two polarisations 
$\sigma=\uparrow\,,\downarrow$, 
two are the compound objects, in two polarisations.

\definecolor{verdone}{rgb}{0.1,0.55,0.1}
\newcommand{\pve}{{\color{verdone}$\bullet$}}
\newcommand{\pr}{{\color{red}$\bullet$}}
\newcommand{\pn}{{\tiny\color{black}$\bullet$}}
\begin{figure}[h]
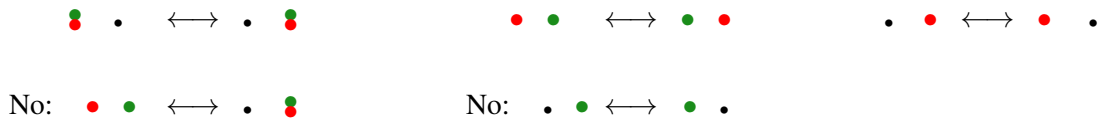

\begin{tabular}{c@{$\longleftrightarrow$\hspace{3mm}}c@{\hspace{20mm}}c@{$\longleftrightarrow$\hspace{3mm}}c
@{\hspace{20mm}}c@{\hspace{3mm}$\longleftrightarrow$\hspace{3mm}}c}
\begin{tabular}{c}\pve\\[-3.4mm] \pr \end{tabular}~~\pn & \pn ~~\begin{tabular}{c}\pve\\[-3.4mm] \pr \end{tabular} &
\pr ~~ \pve & \pve ~~ \pr & \pn ~~~~\pr & \pr ~~~~ \pn \\[6mm]
No: ~~ \pr ~~ \pve ~~~~ & \pn ~~\begin{tabular}{c}\pve \\[-3.4mm] \pr\end{tabular} &
No: ~~ \pn ~~ \pve~~ & \pve ~~ \pn 
\end{tabular}
\caption{The different elementary processes that are described in (\ref{hamiltgl22}); unprimed particles 
are charged \pr, primed particles are neutral \pve. The compound object has both the colors. 
The two lower processes cannot exist, namely the compound object cannot be created or destroyed and
the neutral particle alone is static.}
\end{figure}
The study of the two-particle scattering matrix has been done in \cite{FFR}, with a preliminary account
of the Bethe equations. There are general features that emerge. First the vacuum state is chosen as
(other choices are possible)
\be
\Omega= \mathop{(e_1^{\uparrow}\otimes e_1^{\downarrow})}_{\scriptscriptstyle 1} \otimes
\mathop{(e_1^{\uparrow}\otimes e_1^{\downarrow})}_{\scriptscriptstyle 2} \otimes \dots 
\mathop{(e_1^{\uparrow}\otimes e_1^{\downarrow})}_{\scriptscriptstyle L}
\ee
where the index behind the tensor product labels the lattice sites. All other states are considered 
excitations above it. From the projector $\pi$ one has to remove the part that projects on the vacuum
so the operator $\pir$ projects on the subspace $\Wr$ 
\be
\pir=\pi-E_{1,1}\,,\qquad \Wr=\pir\ V=\pir\ W
\ee
Particles are classified by the type, according to the various subspaces 
$\Wr^{\uparrow},\wt{W}^{\uparrow},\Wr^{\downarrow},\wt{W}^{\downarrow}$.

Within the universal XX models, all particles satisfy the exclusion principle, namely they 
cannot appear on the same site. If two particles are both from $\Wr$ or both from 
$\wt{W}$, they reflect each other; if they are one from $\Wr$, one from $\wt{W}$,
they traverse each other but still remaining on different sites. 

In the universal Hubbard models, the coupling activates a sort of electrostatic interaction 
felt by particles of opposite ``polarisation'' only. Indeed, the potential term in (\ref{eq:HubHam})
squares to the identity (\ref{eq:opC}) so on one site it has eigenvalues $\pm U$. 
Which sign occurs is dictated by the membership to $\Wr$ or $\wt{W}$ according to the rule:
with $U>0$, equal type particles (both in $\Wr$ or in $\wt{W}$) repel each other with an amplitude
-1 while different type particles one each from $\Wr\,,\wt{W}$ attract each other with an
amplitude that is just a phase.
Observe that the vacuum itself is in the repulsive case so actually the only 
"visible" effect is the attractive one. 

The most important interaction comes when a particle from $\wt{W}\su$ and one from  $\wt{W}\giu$
meet at a point. This gives rise to the usual transmission and reflection amplitudes of the 
Hubbard model, $T(p_{1},p_{2}), R(p_{1},p_{2})$. Notice that they are the same for all particles.
The two-particle S-matrix, directly taken from \cite{FFR}, is 
\begin{eqnarray*}
S_{12}(p_{1},p_{2}) &=& S^{X\uparrow}_{12}(p_{1},p_{2})
+S^{X\downarrow}_{12}(p_{1},p_{2})+S^{\updownarrow}_{12}(p_{1},p_{2})
+S^{H}_{12}(p_{1},p_{2})\\[1.2ex]
S^{X\rho}_{12}(p_{1},p_{2}) &=& 
e^{-ip_{1}}\,\pir^{\rho}\otimes \wt\pi^{\rho}
+e^{ip_{2}}\,\wt\pi^{\rho}\otimes \pir^{\rho}
-P_{12}\Big(
\pir^{\rho}\otimes \pir^{\rho} +\wt\pi^{\rho}\otimes \wt\pi^{\rho}\Big)
\,,\ \rho=\uparrow,\downarrow\qquad\\
S^\updownarrow_{12}&=& 
\pir^{\uparrow}\otimes (\pir^{\downarrow}+\wt\pi^{\downarrow})+
(\pir^{\downarrow}+\wt\pi^{\downarrow})\otimes \pir^{\uparrow}+
\pir^{\downarrow}\otimes \wt\pi^{\uparrow}
+\wt\pi^{\uparrow}\otimes \pir^{\downarrow}
\\
S^{H}_{12}(p_{1},p_{2}) &=& \Big(T(p_{1},p_{2})\,\I\otimes\I
+R(p_{1},p_{2})\,P_{12}\Big)\,
\Big(\wt\pi^{\uparrow}\otimes \wt\pi^{\downarrow}
+\wt\pi^{\downarrow}\otimes \wt\pi^{\uparrow} \Big)
\\
T(p_{1},p_{2}) &=& 
\frac{\sin(p_{1})-\sin(p_{2})}{\sin(p_{1})-\sin(p_{2})-2iU}\\
R(p_{1},p_{2}) &=& 
\frac{2iU}{\sin(p_{1})-\sin(p_{2})-2iU} \ =\ T(p_{1},p_{2}) -1
\end{eqnarray*}
In summary, the generalizations of the Hubbard model describe new aspects, mainly in relation to the
$\pi$ projector and $\Wr$ space. They were not present in Hubbard because its space of states is 
too small. 
The generalized models can describe many different fermionic particles, all living in the same lattice, some charged and some chargeless.
The core of the interactions within $\wt{W}$ remains the same 
as in Hubbard, with the same amplitudes. 

The exposition on the generalizations of the Hubbard model stops here.

\section{A system of two non-linear integral equations for the Hubbard model}
Following the methods of Chapter~\ref{c:nlie}, the system of equations for the Hubbard model 
is introduced. The full derivation is given in the original paper \cite{FFGR2}. 
The main purpose of this work was to study 
certain super Yang-Mills operators; this has conditioned some choices, as the systematic use of the 
phase $\phi$, which is  related to a global magnetic field; of course, the whole construction holds for the 
Hubbard model itself. 
Looking at the Lieb-We equations (\ref {lw}), I define the function
\begin{equation}
\Phi (x,\xi)=i \ln \frac {i\xi +x}{i\xi -x} \, , \label {Phi}
\end{equation}
with the branch cut of $\ln(z)$ along the real negative $z$-axis in such a way that 
$-\pi < \arg z <\pi$. Then, I introduce the gauge transformation which amounts to add the 
magnetic flux
\begin{equation}
k_j=\hat{k}_j+\phi \,.
\end{equation}
Using the following counting functions
\begin{eqnarray}
W(k)&=&L(k-\phi) -\sum _{l=1}^M \Phi \left (u_l-\frac {2t}{U}\sin k, \frac{1}{2} \right ) \, , 
\label {Wdef} \\
Z(u)&=&\sum _{j=1}^L \Phi  \left (u - \frac {2t}{U}\sin k_j, \frac{1}{2} \right ) -
\sum _{m=1}^M \Phi \left (u-u_m, 1 \right ) \, , \label {Zdef}
\end{eqnarray}
the Lieb-Wu equations take the form of quantisation conditions for the Bethe roots $\{k_j,u_l\}$,
\begin{eqnarray}
W(k_j)&=&\pi (M +2 I^w_j) \, , \\
Z(u_l)&=&\pi (M-L+1+2I^z_l) \, .
\end{eqnarray}
From now on, the treatment focuses on the highest energy state, consisting of the maximum number 
$M=L/2$ of real roots $u_l$ and of $L$ real roots $k_j$. For simplicity reasons, it is useful
to restrict the calculation to the case $M\in 2\mathbb{N}$ (the remaining
case $M\in 2\mathbb{N}+1$ is a simple modification of this case), which obviously implies 
$L\in 4\mathbb{N}$.

With the integral definition of the Bessel function $J_0(z)$,
\begin{equation}
J_0(z)=\int _{-\pi}^{\pi} \frac {dk}{2\pi}\, e^{i\,z\sin k}\,,
\end{equation}
and also the following shorthand notations
\begin{equation}
L_W(k)= {\mbox {Im}}\ln \left [1-e^{iW(k+i0)}\right ] \, , \quad
L_Z(x)= {\mbox {Im}}\ln \left [1+e^{iZ(x+i0)}\right ] \, .
\end{equation}
the first of two nonlinear integral equations for the counting functions is
\begin{eqnarray}
Z(u)&=&L \int _{-\infty}^{\infty} \frac {dp}{2p} \sin (pu) \frac
{J_0\left ( \frac {2tp}{U}\right )}{\cosh \frac {p}{2}}+2 \int
 _{-\infty}^{\infty} dy \ G(u-y) \ {\mbox {Im}}\ln \left
[1+e^{iZ(y+i0)}\right ]- \nonumber \\
&-&\frac {2t}{U}\int _{-\pi}^{\pi} dk \cos k \frac {1}{\cosh \left
( \pi u - \frac {2t\pi}{U}\sin  k \right ) } \
 {\mbox {Im}}\ln \left [1-e^{iW(k+i0)}\right ] \, , \label {Zeq4}
\end{eqnarray}
where $G(x)$ is the same kernel function that appears in the spin $1/2$ XXX chain and in the BDS 
Bethe Ansatz\footnote{The Beisert, Dippel, Staudacher model was a deformation of the XXX Bethe 
equations introduced to describe all loops in the $SU(2)$ sector of SYM, see \cite{FFGR}.}, as in eq. 2.24 of \cite{FFGR},
\begin{equation}
G(x)=\int _{-\infty}^{\infty} \frac {dp}{2\pi} e^{ipx} \frac
{1}{1+e^{|p|}} \, . \label {Gxxx}
\end{equation}
The first line of the NLIE for $Z$ (\ref {Zeq4}) coincides with the NLIE (eq. 3.15 of \cite{FFGR}) 
for the counting function of the highest energy state of the BDS model. The second
line of (\ref {Zeq4}) is the genuine contribution of the Hubbard model.
The second nonlinear integral equation is
\begin{eqnarray}
W(k)&=&L\left[ (k-\phi) + \int _{-\infty}^{\infty} \frac {dp}{p}
\sin \left ({\frac
  {2tp}{U}\sin k }\right ) \frac {J_0\left (\frac {2tp}{U}\right )}{1+e^{|p|}}\right]
- \nonumber \\
&-&\int _{-\infty}^{\infty}dx \, \frac {1}{\cosh \left ( \frac
{2t\pi}{U}\sin k-\pi x \right ) } \,
{\mbox {Im}}\ln \left[1+e^{iZ(x+i0)}\right ]-  \label {Weq2}\\
&-& \frac {4t}{U} \int _{-\pi}^{\pi} dh \ G \left ( \frac {2t}{U}
\sin h-\frac {2t}{U}\sin k \right ) \cos h \mbox{ Im} \ln
\left[1-e^{iW(h+i0)}\right ] \, . \nonumber
\end{eqnarray}
The two equations (\ref{Zeq4}, \ref{Weq2}) are coupled by integral
terms and are completely equivalent to the Bethe equations for the
highest energy state.

The eigenvalues of the Hamiltonian (\ref {oldHubb}) on the Bethe states are given by (\ref{energia}), 
that can now be expressed in terms of the counting functions.
I use the Bessel function
\begin{equation}
J_1(z)=\frac {1}{2\pi i}\int _{-\pi}^{\pi} dk \sin k ~ e^{iz
  \sin k} \, , \label {J1}
\end{equation}
The highest eigenvalue energy is expressed in terms of the counting functions $Z$ and $W$ as follows
\begin{eqnarray}
E&=&-2t\left\{ L \int _{-\infty}^{\infty} \frac {dp}{p} \frac
{J_0\left (\frac
  {2tp}{U}\right ) J_1\left (\frac  {2tp}{U}\right ) }{e^{|p|}+1}
+  \int _{-\infty}^{\infty} dx \left [ \int _{-\infty}^{\infty}
\frac
  {dp}{2\pi}\frac {e^{ipx}}{\cosh \frac {p}{2}}i J_1\left (\frac
  {2tp}{U}\right ) \right ]  L_Z(x) - \right. \nonumber  \\
&-& \frac{2t}{U} \left. \int _{-\pi}^{\pi} \frac {dh}{\pi} L_W(h)
\cos h \left [ \int _{-\infty}^{\infty} \frac {dp}{i}
 e^{i\frac {2tp}{U}\sin h } \frac {J_1\left (\frac
  {2tp}{U}\right ) }{e^{|p|}+1} \right ]-
\int _{-\pi}^{\pi} \frac {dh}{\pi} L_W(h)
\sin h  \right\} \nonumber \\
&\equiv& E_L+E_Z+E_{W1}+E_{W2} \,, \quad \text{and}\quad E_W\equiv
E_{W1}+E_{W2}.  \label {Eexp}
\end{eqnarray}
The first line of (\ref {Eexp}), namely $E_L+E_Z$, coincides
formally with the expression of the highest energy of the BDS
chain as given in equation (3.24) of \cite{FFGR}. However, in this case $Z$ satisfies a NLIE which
is different from that of the BDS model. On the other hand, the second line, i.e. $E_W=E_{W1}+E_{W2}$, 
is a completely new contribution.

\begin{figure}[h]\hspace*{3mm}
\includegraphics[width=0.95\linewidth]{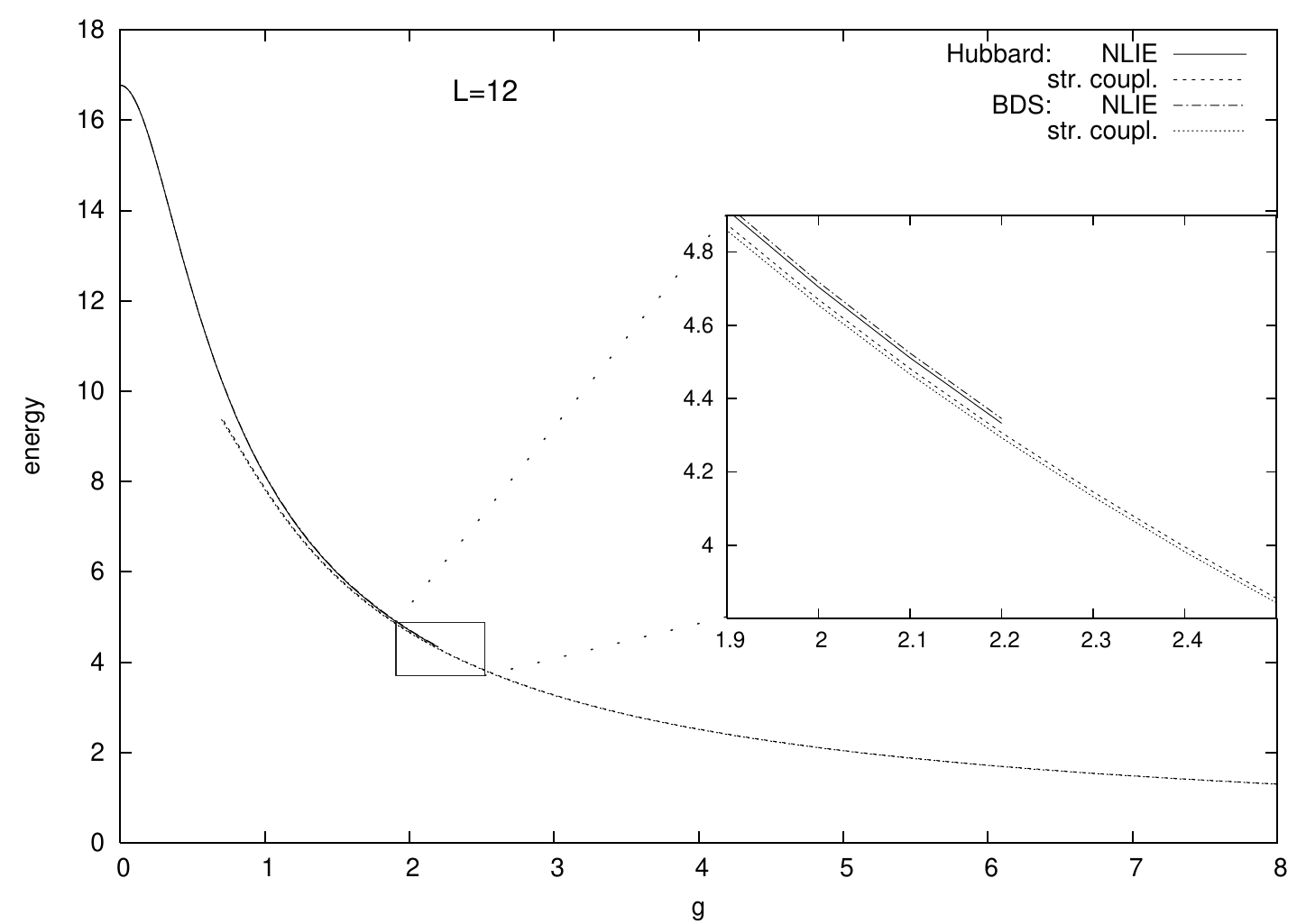}
\caption{\label{ws} The behaviour of the energies for Hubbard and 
BDS model  from small to strong coupling is plotted here
for a lattice of 12 sites. The left branches of the curves are obtained 
by solving numerically the NLIE while the right branches are plotted using the 
the strong coupling expansion from the NLIE. In the small
picture there is a zoom of the region where the branches overlap. The approximated match is due to numerical errors in the left curve.}
\end{figure}

This system of equations can be extended to include all excitations. Indeed, one has to include
appropriate sources for the real holes and for all the complex roots that can appear, precisely as 
in (\ref{struttura}). The main goal of the papers \cite{FFGR, FFGR2} was, however, to use the 
nonlinear integral equations for a careful investigation of several limits: large volume (large $L$
expansion, namely a thermodynamic limit), strong ($U\rightarrow +\infty$) and weak coupling 
($U\rightarrow 0$) and possibly to study the effect of interchanging the order of the limits. 
At the beginning of this work (2005-2006) the belief was that the Hubbard model could represent
anomalous dimensions of the $\mathcal{N}=4$ super Yang-Mills theory. 
Within this correspondence between 
Hubbard and super Yang-Mills, the study of the large volume limit corresponds to study 
very long super Yang-Mills operators. 
In Figure~\ref{ws}, the energy (\ref{Eexp}) is plotted as function of the coupling constant, for 
a lattice of 12 sites.

\section{Discussion}
I have presented two different works on the subject of the Hubbard model in relation to the 
$\mathcal{N}=4$ super Yang-Mills theory. Both of them have been developed in the years 
2005-2007 and were amongst the first attempts to use integrability techniques in the gauge theory.

The integrable generalizations of the Hubbard model were introduced in \cite{DFFR}. In \cite{FFR} 
I have started to work on the derivation of the Bethe equations. The scattering matrix 
is fully presented in that article. The full set of Bethe equations has been obtained more recently 
by colleagues of mine \cite{fomin}. 

The work on the nonlinear integral equations for the Hubbard model in \cite{FFGR2} was actually 
the continuation
of a work presented in \cite{FFGR} on the XXX model with excitations of type hole,  
on the BDS Bethe ansatz \cite{Beisert:2004hm} and on the $SO(6)$ spin chain. 
I treated these models in the frame of the integrability within the $\mathcal{N}=4$ super 
Yang-Mills theory. Today we know that these models are at best approximations of the correct
Bethe equations \cite{beis2006}. In spite of this, it was important to start working with
the methods presented here.   
The work of \cite{FFGR2} has shown to the community of SYM that the methods of nonlinear integral 
equations are effective in treating certain questions starting from Bethe equations. 
For this reason, my co-authors are still active in the field. They have treated a number of new 
cases, including models with non-compact symmetry groups, large number of holes, etc.
\cite{davidemarcopaolo}. 

After these publications, my research activities have taken a new direction, that will be presented
in the next chapters.

\renewcommand\bibname{Bibliography: Integrability}

\part{Biophysics: quantitative methods for biology}

\chapter*{Motivations}\label{introbio}
\addcontentsline{toc}{chapter}{Motivations}
The application of quantitative methods to biology is presently the object of a large
amount of theoretical studies. 
Theoretical study and modeling of biological phenomena are not a substitute to biological in vivo
investigation. Instead, they are a very ``economical'' way to formulate quantitative relations between 
relevant quantities and to make predictions with them. 

Nowadays, modeling biological phenomena corresponds to the approach that has been adopted innoumerous 
times in other domains of science, especially physics.
Weren't the three Kepler's laws a model? Of course they were.
Kepler was not aware of more fundamental and general laws to use (namely Newtonian mechanics and 
gravity), he just formulated a quantitative description of his observations.
The Bohr-Sommerfield quantization of adiabatic invariants was a model for old quantum theory.
In these examples the model preceded the theory and somehow helped formulating it. Of course, the 
model can follow the theory and somehow simplify it, as the Ising model for magnetism
simplifies a full quantum mechanical approach to the problem. 
The power of computers makes it possible to develop theoretical tools and models to elaborate and 
speculate on the vast amount of data accumulated on the genome and on the proteome.

Strongly motivated by this, in collaboration with Dr. F. Musso of the University of Burgos, 
in 2006 I introduced a model of evolution based on a population of
``Turing machines''. Each machine is actually defined by a finite number of ``states'' that
form its own code or genome. This code undergoes stochastic evolution with certain rates 
that implement different aspects of genome mutation. Then, a performance based selection process 
creates a new generation of machines with increased performance. The process is repeated for a large 
number of generations. 
The goal of this model is to explore features of the Darwinian evolution with a full control
of the parameters that participate to it. Indeed, in silico evolution and mathematical modeling 
are ideal environments to test evolutive hypotheses that are otherwise difficult to test in the real
biological environment.

In a similar spirit, in 2007 I co-founded the Gemini team in Annecy, with C.~Lesieur, biologist, 
and other collaborators from the mathematical-physical background of the ``Federation de 
recherche Modelisation, Simulation, Interactions Fondamentales''. 
The team agreed to work on questions of assembly in oligomeric proteins\footnote{Oligomeric proteins 
are those whose native state, or functional state, is the aggregate of two or more polypeptidic chains.}. 
The team biologist had already approached the subject 
with standard experimental tools but needed to use theoretical methods 
describe the problem in general terms through the analysis of 
a larger number of cases.

A mathematician, L. Vuillon, is part of the Gemini team. This is a signal that the project on the 
oligomeric proteins has a sufficiently high degree of ``synthesis'', in the Greek sense,
to be effective in biology and to require a theoretical development as a complex system.

The approach that we carry on is inductive, based on the systematic analysis of structure 
data of oligomeric proteins. I will present it later. 
The research is revealing features related to the process of the interface formation  
and to the process of the assembly of different chains. 

The inductive reasoning, which goes from specific observations to broader generalizations and theories, 
is not always applied in mathematical and theoretical physics.
Indeed, it is typical of the periods when a paradigm, or theory, is missing but
a scenario is emerging from observations, that forces to move toward a more accurate and complete 
understanding.
Deductive reasoning is common when the paradigm is established and can be applied to predict
a variety of phenomena.
Modelization is intermediate because the construction of the model often 
comes from empirical knowledge but the model allows one to deduce further effects.

I hope the next two chapters on biophysics will suggest why a theoretical physicist, like me,
is engaging in research on evolution and proteome, and how he can help in the inductive reasoning
towards a paradigm.

\chapter{Modelling Darwinian evolution\label{c:turing}}
\newcommand{\nt}{N_{\text{t}}}
\newcommand{\nc}{N_{\text{c}}}
\newcommand{\nnc}{N_{\text{nc}}}
\newcommand{\pmm}{p_{\text{m}}}
\newcommand{\pii}{p_{\text{i}}}
Darwinian evolution is the today paradigm that unifies paleontological records with modern biology. It creates a bridge between the microscopic view (genome, proteome) and 
the macroscopic features of the living organisms (phenotype). The phenotypic, or macroscopic, 
mechanism of Darwinian evolution is natural selection namely a differential scrutiny of the 
phenotypes by environment. The microscopic mechanism is genome mutations. In neodarwinism, it
is very important  to appreciate the fact that even if the genome is a physical memory that is 
transmitted from the parent(s) to the offspring, the phenotype of a single organism 
is not inherited: if a person looses a leg, its children will still 
have both of their legs.
Notice that the physical memory or genome is also under control of the selection so it is
correct to say that the genome is part of the phenotype.
The opposite is false: the phenotype is not contained in the genome, 
otherwise it would be automatic to inherit acquired characters, as in Lamarkism.

The basic functioning of the genome\footnote{Here a possible distintion between genome and DNA is not needed} is to record long sequences of four letters that later on 
can be mapped into amino acid sequences by a known mapping that biologists call 
``the genetic code''. This means that the biological
function of proteins, intrinsically three-dimensional and based on physico-chemical properties
of atomic aggregates, can be described by a discrete and finite amount of information. 
It's unavoidable to imagine that amount of information as being an algorithm that, executed 
with given rules, realizes some tasks. This analogy between genome and algorithm pervades 
the whole domain of the artificial life \cite{fogel2006} namely the tentative of realizing 
in silico organisms that exploit the main features of living organisms\footnote{More or less, one can summarize them in: existence of a 
separation interior/exterior, existence of a metabolism, response to external stimuli, 
self-identical reproduction.}.  

The need of modeling Darwinian evolution, or else the need of creating an artificial life, 
comes from several directions all related to the difficulty
to make quantitative experimental studies: 
we cannot rewind the Earth history to study past organisms ``alive'', 
the fossil record is incomplete, in vitro evolution experiments are long and expensive and 
only few of them have successfully been done. 
Moreover, the biggest difficulty is to keep apart the different causes that produce the observed
evolutionary dynamics. 
In this scenario, in silico evolution can help precisely where observation biology fails: 
in silico experiments can be done on today's calculus grids, the control on the parameters is
complete and the full record of the evolution is available. 
I cannot avoid a comparison with cosmology: we know a single universe, we cannot rewind its
history and observational data on far objects (in remote time and in space) are few. The difference
is that artificial cosmological experiments are just impossible while artificial evolution 
experiments are possible and will slowly be realized. In this sense, biological evolution
is much more affordable than cosmology.
From Maynard Smith, one reads: ``...we badly need a comparative biology. So far,
we have been able to study only one evolving system and we cannot wait for interstellar flight 
to provide us with a second. If we want to discover generalizations about evolving systems, we 
will have to look at artificial ones.''
Modeling evolution has definitely not the goal to replace observatory or experimental biology
but has the goal to help finding ``universal'' features in the evolutionary dynamics and in 
the mechanisms of mutation, selection. After all, universality is the way critical phenomena 
are studied in statistical mechanics: specific details do not participate in determining 
universal features.
This point has been particularly stressed in my recent article \cite{FM3}. 

The existence of a small genome within a much larger phenotype and the basic functioning by the 
genetic code are strategic for modelization purposes because they allow to investigate at 
least some of the features of evolution without paying too much attention to the whole organism 
and its proteome but just focusing on algorithm features.

\section{The model}
\begin{figure}
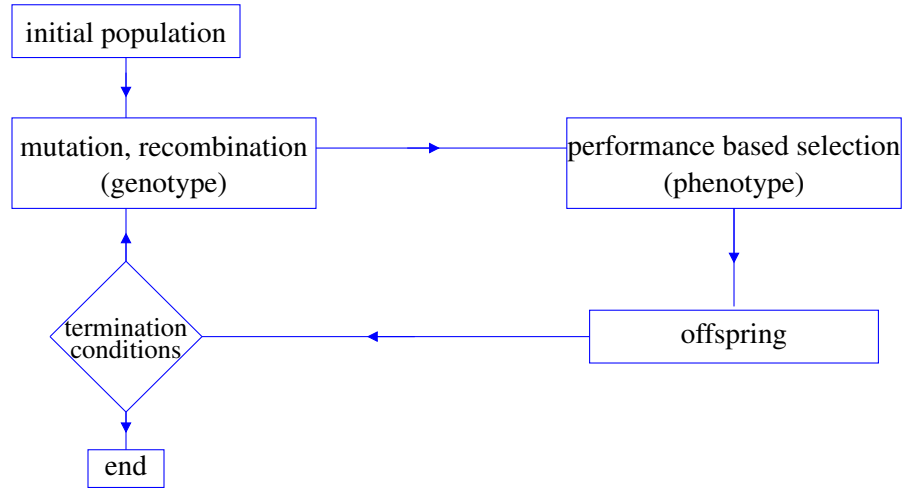

\begin{pgfpicture}{-3cm}{2.3cm}{11cm}{9cm}
\pgfputat{\pgfxy(1,8.5)}{\pgfbox[center,center]{initial population}}
\pgfputat{\pgfxy(1.5,7)}{\pgfbox[center,center]{mutation, recombination}}
\pgfputat{\pgfxy(1.5,6.5)}{\pgfbox[center,center]{(genotype)}}
\pgfputat{\pgfxy(9,7)}{\pgfbox[center,center]{performance based selection}}
\pgfputat{\pgfxy(9,6.5)}{\pgfbox[center,center]{(phenotype)}}
\pgfputat{\pgfxy(9,4.5)}{\pgfbox[center,center]{offspring}}
\pgfputat{\pgfxy(1,4.63)}{\pgfbox[center,center]{\small termination}}
\pgfputat{\pgfxy(1,4.33)}{\pgfbox[center,center]{\small conditions}}
\pgfputat{\pgfxy(1,2.8)}{\pgfbox[center,center]{end}}
{\color{blue}
\pgfrect[stroke]{\pgfxy(-0.5,8.2)}{\pgfxy(3,0.7)}
\pgfrect[stroke]{\pgfxy(-0.5,6.2)}{\pgfxy(4,1.2)}
\pgfrect[stroke]{\pgfxy(6.8,6.2)}{\pgfxy(4.4,1.2)}
\pgfrect[stroke]{\pgfxy(7.1,4.15)}{\pgfxy(3.8,0.7)}
\pgfrect[stroke]{\pgfxy(0.5,2.5)}{\pgfxy(1,0.5)}
\pgfline{\pgfxy(0,4.5)}{\pgfxy(1,5.5)}\pgfline{\pgfxy(1,5.5)}{\pgfxy(2,4.5)}
\pgfline{\pgfxy(2,4.5)}{\pgfxy(1,3.5)}\pgfline{\pgfxy(1,3.5)}{\pgfxy(0,4.5)}
\pgfline{\pgfxy(1,8.2)}{\pgfxy(1,7.4)}
\pgfline{\pgfxy(3.5,7)}{\pgfxy(6.8,7)}
\pgfline{\pgfxy(9,6.2)}{\pgfxy(9,4.9)}
\pgfline{\pgfxy(1,6.2)}{\pgfxy(1,5.5)}
\pgfline{\pgfxy(2,4.5)}{\pgfxy(7.1,4.5)}
\pgfline{\pgfxy(1,3.5)}{\pgfxy(1,3.)}
\pgfsetendarrow{\pgfarrowtriangle{3pt}}
\pgfline{\pgfxy(1,8)}{\pgfxy(1,7.7)}
\pgfline{\pgfxy(4.7,7)}{\pgfxy(5.1,7)}
\pgfline{\pgfxy(9,6.2)}{\pgfxy(9,5.5)}
\pgfline{\pgfxy(4.8,4.5)}{\pgfxy(4.2,4.5)}
\pgfline{\pgfxy(1,5.5)}{\pgfxy(1,5.85)}
\pgfline{\pgfxy(1,3.5)}{\pgfxy(1,3.25)}
}
\end{pgfpicture}
\caption{Basic flow of evolutionary computations or artificial life cycle. Notice the distinction
of genotype and phenotype.\label{insilico}}
\end{figure}

The idea is to have a population of algorithms that evolve with generations. At each generation, 
each algorithm undergoes mutation (possibly in several flavours). The
algorithm is then executed, which corresponds to the life of the organism. The output represents
the phenotype and the interaction with the environment, therefore selection acts on it. 
The selection, precisely as in the biological case, evaluates differentially two or more phenotypes
retaining the best fitted for reproduction. This creates an artificial 
life cycle, as in Figure~\ref{insilico}, that can be repeated a large number of times, 
to study the evolution of the population features.

In my Turing machines model, the algorithms are precisely the Turing machines. This choice was mainly 
motivated by the generality attained by the Turing machines language, in spite of a very simple set 
of basic instructions. This aspect is extremely important as it allows to do some theoretical 
investigation of the model. Many other 
formalizations of algorithms have actually been adopted in evolutionary computation \cite{fogel2006}.

\begin{figure}[h]
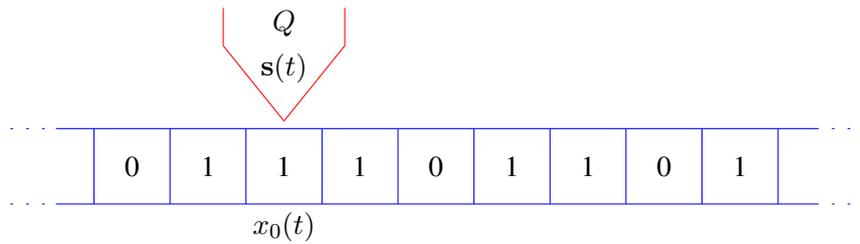

\begin{pgfpicture}{-3cm}{0.5cm}{8cm}{4.0cm}
{\color{blue}
\pgfline{\pgfxy(0.5,1)}{\pgfxy(10.5,1)}
\pgfline{\pgfxy(0.5,2)}{\pgfxy(10.5,2)}
\pgfline{\pgfxy(1,1)}{\pgfxy(1,2)}
\pgfline{\pgfxy(2,1)}{\pgfxy(2,2)}
\pgfline{\pgfxy(3,1)}{\pgfxy(3,2)}
\pgfline{\pgfxy(4,1)}{\pgfxy(4,2)}
\pgfline{\pgfxy(5,1)}{\pgfxy(5,2)}
\pgfline{\pgfxy(6,1)}{\pgfxy(6,2)}
\pgfline{\pgfxy(7,1)}{\pgfxy(7,2)}
\pgfline{\pgfxy(8,1)}{\pgfxy(8,2)}
\pgfline{\pgfxy(9,1)}{\pgfxy(9,2)}
\pgfline{\pgfxy(10,1)}{\pgfxy(10,2)}}
{\color{red}
\pgfline{\pgfxy(3.5,2.1)}{\pgfxy(4.3,3.1)}
\pgfline{\pgfxy(3.5,2.1)}{\pgfxy(2.7,3.1)}
\pgfline{\pgfxy(4.3,3.1)}{\pgfxy(4.3,3.6)}
\pgfline{\pgfxy(2.7,3.1)}{\pgfxy(2.7,3.6)}
}
{\color{blue}
\pgfsetdash{{1pt}{1ex}}{0pt}
\pgfline{\pgfxy(-0.1,1)}{\pgfxy(0.5,1)}
\pgfline{\pgfxy(-0.1,2)}{\pgfxy(0.5,2)}
\pgfline{\pgfxy(10.5,1)}{\pgfxy(11.3,1)}
\pgfline{\pgfxy(10.5,2)}{\pgfxy(11.3,2)}}
\pgfputat{\pgfxy(1.5,1.5)}{\pgfbox[center,center]{0}}
\pgfputat{\pgfxy(2.5,1.5)}{\pgfbox[center,center]{1}}
\pgfputat{\pgfxy(3.5,1.5)}{\pgfbox[center,center]{1}}
\pgfputat{\pgfxy(4.5,1.5)}{\pgfbox[center,center]{1}}
\pgfputat{\pgfxy(5.5,1.5)}{\pgfbox[center,center]{0}}
\pgfputat{\pgfxy(6.5,1.5)}{\pgfbox[center,center]{1}}
\pgfputat{\pgfxy(7.5,1.5)}{\pgfbox[center,center]{1}}
\pgfputat{\pgfxy(8.5,1.5)}{\pgfbox[center,center]{0}}
\pgfputat{\pgfxy(9.5,1.5)}{\pgfbox[center,center]{1}}
\pgfputat{\pgfxy(3.5,0.7)}{\pgfbox[center,center]{$x_0(t)$}}
\pgfputat{\pgfxy(3.5,2.8)}{\pgfbox[center,center]{$\mathbf{s}(t)$}}
\pgfputat{\pgfxy(3.5,3.4)}{\pgfbox[center,center]{$Q$}}
\end{pgfpicture}
\caption{\label{f:turing}Graphical representation of a Turing machine at time $t$, in the internal state
$\mathbf{s}(t)$, located on the $x_0(t)$ cell of an infinite tape.}
\end{figure}

Turing machines are abstract symbol-manipulating devices that implement a ``one-point'' 
discrete evolution law. Given a finite alphabet or list of symbols $\mathcal{A}$ and given  
the total number of internal ``states'' $\nt$, one defines a Turing 
machine by giving the evolution law $Q$
\be
(r,\mathbf{s})\mathop{\mapsto}^Q (r',d',\mathbf{s}')\,,\qquad r,r'\in \mathcal{A}\,,\qquad 
d'\in \{R,L\}\,,\qquad \mathbf{s},\mathbf{s}'\in \{1,2, \ldots \nt\}
\ee
It is the set of actions that the machine performs,
determined by the value read on the tape $r=r(t)$, at a position $x_0(t)$, and by the internal state 
of the machine $\mathbf{s}=\mathbf{s}(t)$. These variables depend on the execution time, so
$r'=r(t+1)$ and so on. Notice that the mapping $Q$ does not depend on the old displacement right/left 
$d(t)$ but it produces the new displacement $d'=d(t+1)$.
The mapping $Q$ is the core of the Turing machine and can be represented by the triplets
``write,displace,call'' $(w,d,c)$ acting on a tape, as in table~\ref{t:somma} and in 
figure~\ref{f:turing}:
\begin{enumerate} 
\item \textbf{write:} it writes a new symbol at position $x_0(t)$,
\item \textbf{displace:} it moves right R or left by one cell L,
\item \textbf{call:} it changes its internal state.
\end{enumerate} 
An initial configuration is given by assigning the function $T(x,0)\in\mathcal{A}\,,\ \forall\ x \in \mathbb{Z}$. 
Recursively, a new configuration $T(x,t+1)\in\mathcal{A}\,, \ \forall\ x \in \mathbb{Z}$
is computed from $T(x,t)$ with the mapping $Q$ in such a way that only a single mutated position 
$x_0(t)$ can exist at each time
$$
\begin{array}{c} 
T(x,t+1)=T(x,t) \quad \forall\ x\neq x_0(t) \\[3mm]
T(x_0(t),t+1) \mbox{ could be } \neq T(x_0(t),t)
\end{array}
$$
and such that $x_0(t+1)=x_0(t)\pm 1$. Here I will use the binary set of symbols 
$\mathcal{A}=\{0,1\}$. This choice is mainly dictated by the simplicity of coding it offers.
Representing $T(x,t)\ \forall\ x $ by a tape of cells at position $x\in\mathbb{Z}$, one has the 
familiar representation of figure~\ref{f:turing}.

\begin{table}
$$
\begin{array}{|c|c|c|c|}
\hline \raisebox{-5mm}[7mm][4mm]{\begin{pgfpicture}{0cm}{0cm}{15mm}{10mm}
\pgfline{\pgfxy(1.7,0.1)}{\pgfxy(-0.2,1.2)} \pgfputat{\pgfxy(0.4,0.4)}{\pgfbox[center,center]{read}} \pgfputat{\pgfxy(1.1,1)}{\pgfbox[center,center]{\bf state}}\end{pgfpicture}} & {\bf 1} & {\bf 2} & {\bf 3}  \\
\hline 0 & 1-\mbox{Right}-\bf 2 & 0-\mbox{Left}-\bf 3 & \_-\_-\_  \\
\hline 1 & 1-\mbox{Right}-\bf 1 & 1-\mbox{Right}-\bf 2 & 0-\_-\mbox{\bf Halt}  \\ \hline
\end{array}
$$
\caption{\label{t:somma}Table of states of a Turing machine that performs the sum of two positive 
numbers represented by ``sticks'': $\ \ldots 0\ 1\ 1\ 1\ 0\ 0\ $ represents the number three and so on.
Missing entries are irrelevant and can be fixed arbitrarily. Here and in the following, the states of the 
machine are written in bold character, to ease the reading.}
\end{table}

Considering that the general characterization of Turing machines is not needed here, for computational
reasons the tape is taken of finite length usually fixed to $L=300$  boxes. In some simulations
the tape has been ``periodized'', by identifying the last cell+1 with the first one. 
Periodic boundary conditions were also used in lattice models.   
Moreover, as it is extremely easy to generate machines that run forever, the maximum of 4000 
temporal steps is imposed. When a machine reaches it, it is stopped and its tape is taken without
further modification. 

The simulations start with a population of $npop=300$ Turing machines each with just one state of
the following form 
$$
\begin{array}{|c|c|}
\hline
 &\bf 1 \\ \hline
 0 & 0-\mbox{R}-\mbox{\bf Halt} \\ \hline
 1 & 1-\mbox{R}-\mbox{\bf Halt} \\ \hline
\end{array}
$$
and go on for $ngen=50000$ generations or more. 
At each generation every TM undergoes the following three processes, in the order:
\begin{enumerate}
\item (insertion) states-increase,
\item (point) mutation,
\item selection and reproduction.
\end{enumerate}
In the states-increase process, with a probability $\pii$, the TM passes from $\mathbf \nt$ to 
$\mathbf \nt+\mathbf 1$ states by the addition of the further state
$$
\begin{array}{|c|c|}
\hline & \mathbf \nt+\bf 1\\
\hline 0 & 0-\mbox{R}-\mbox{\bf Halt}\\
\hline 1 & 1-\mbox{R}-\mbox{\bf Halt}\\ \hline
\end{array}
$$
This state will be initially non-coding since it cannot be called by any other state. Indeed,
the Turing machine cannot call a state that does not exist. The only way this state can be activated 
is if a mutation in an already coding state changes the state call to $\mathbf \nt+\mathbf 1$. 
Notice that, when called, this particular state does not affect the tape but halts the machine.
Consequently the activation of this state is mainly harmful or neutral and it can be advantageous 
only in exceptional cases therefore the TM can benefit from the added states only if they are 
mutated before their activation.   
This form of mutation vaguely resembles DNA insertion. 

During point mutation, all the entries of each state of the TM can be randomly changed with a 
probability $\pmm$. The new entry is randomly chosen among all corresponding permitted values 
excluded the original one. The permitted values are: 
\begin{itemize}
\item 0 or 1 for the ``write'' entries;
\item Right, Left for the ``move'' entries;
\item The \textbf{Halt} state or an integer from {\bf 1} to the number of states $\mathbf \nt$ of 
the machine for the ``call'' entries.
\end{itemize}
This mechanism of mutation is reminiscent of the biological point mutation. 
Notice that the states-increase process is actually a form of mutation. 
Here it has been chosen to keep the two 
mutations separate in order to differentiate their roles. Other biological mechanisms like 
traslocation, inversion, deletion, etc.\ are not implemented.

In the selection and reproduction phase a new population is created from the actual one (old population).
The number of offspring of a TM is determined by its ``performance'' and, to a minor extent, by chance.
The performance\footnote{The word ``performance'' is preferred to ``fitness'' as this last one 
indicates two different concepts in biology and in the field of algorithms. The word ``fitness'' will 
be used in the biological sense.}  of a TM is a function 
that measures how well the output tape of the machine reproduces a given ``goal'' tape starting from a 
prescribed input tape. It is computed in the following way. The performance is initially set to zero. 
Then the output tape and the goal tape are compared cell by cell. The performance is increased by one 
for any $1$ on the output tape that has a matching $1$ on the goal tape and it is decreased by 3 for any 
$1$ on the output tape that matches a $0$ on the goal tape. 
    
As a selection process, I use what in the field of evolutionary algorithms is known as ``tournament 
selection of size 2 without replacement''. In it, two TMs are randomly extracted from the old 
population and let run on the 
input tape. At the end, a performance value is assigned to each machine on the basis of its 
output tape. The performance values are compared and the machine which scores higher creates two copies 
of itself in the new population, while the other is eliminated. This reproduction is fully asexual. 
If the performance values are equal, each TM creates a copy of itself in the new population.
The two TMs that were chosen for the tournament are eliminated from the old population and the 
process restarts until the exhaustion of the old population.

The goal tapes are chosen according to the criterion of providing two difficult and qualitatively 
different tasks for a TM. The distribution of the ``1'' on the goal tape has to be 
extremely non-regular since a periodic distribution would provide a very easy task for a TM. 

In the various simulations several goal tapes have been used. The tape ``primes'' has ``1'' on the 
cell positions corresponding to prime numbers, with $1$ included for convenience, and zeros elsewhere:
$$
\begin{array}{l}
1110101000.1010001010.0010000010.1000001000.1010001000.0010000010.1000001000.1010000010.\\
0010000010.0000001000.1010001010.0010000000.0000001000.1000001010.0000000010.1000001000.\\
0010001000.0010000010.1000000000.1010001010.0000000000.1000000000.0010001010.0010000010.\\
1000000000.1000001000.0010000010.1000001000.1010000000.0010000000.
\end{array}
$$
In the previous expression I inserted a dot every ten cells to facilitate the reading.
The second goal tape $\pi$ is given by the binary expression of the decimal part of $\pi$, 
namely $(\pi-3)_{\text{bin}}$:
$$
\begin{array}{l}
0010010000.1111110110.1010100010.0010000101.1010001100.0010001101.0011000100.1100011001.\\
1000101000.1011100000.0011011100.0001110011.0100010010.1001000000.1001001110.0000100010.\\
0010100110.0111110011.0001110100.0000001000.0010111011.1110101001.1000111011.0001001110.\\
0110110010.0010010100.0101001010.0000100001.1110011000.1110001101.
\end{array}
$$
Notice that while for prime numbers the ``1'' become progressively rarer so that the task becomes 
progressively more difficult, in the case of the digits of $\pi$ they are more or less equally 
distributed. 
Another difference is that prime numbers are always odd (with the exception of 2) therefore in the 
goal tape
two ``1'' are separated by at least one ``0''. On the contrary, the digits of $\pi$ can form clusters 
of ``1'' of arbitrary length; this feature is actually visible only in very long tapes of
thousands of cells or more and is not important here.

According to the definition, the maximal possible value for the performance is 63 for the prime 
numbers and 125 for the digits of $\pi$.

In \cite{chevalier}, the objective was introduced of gathering ``1'' on the left side of the output tape,
simulating the process of resources accumulation. The actual definition of the score is involved
so it will not be written here.

\section{Results}

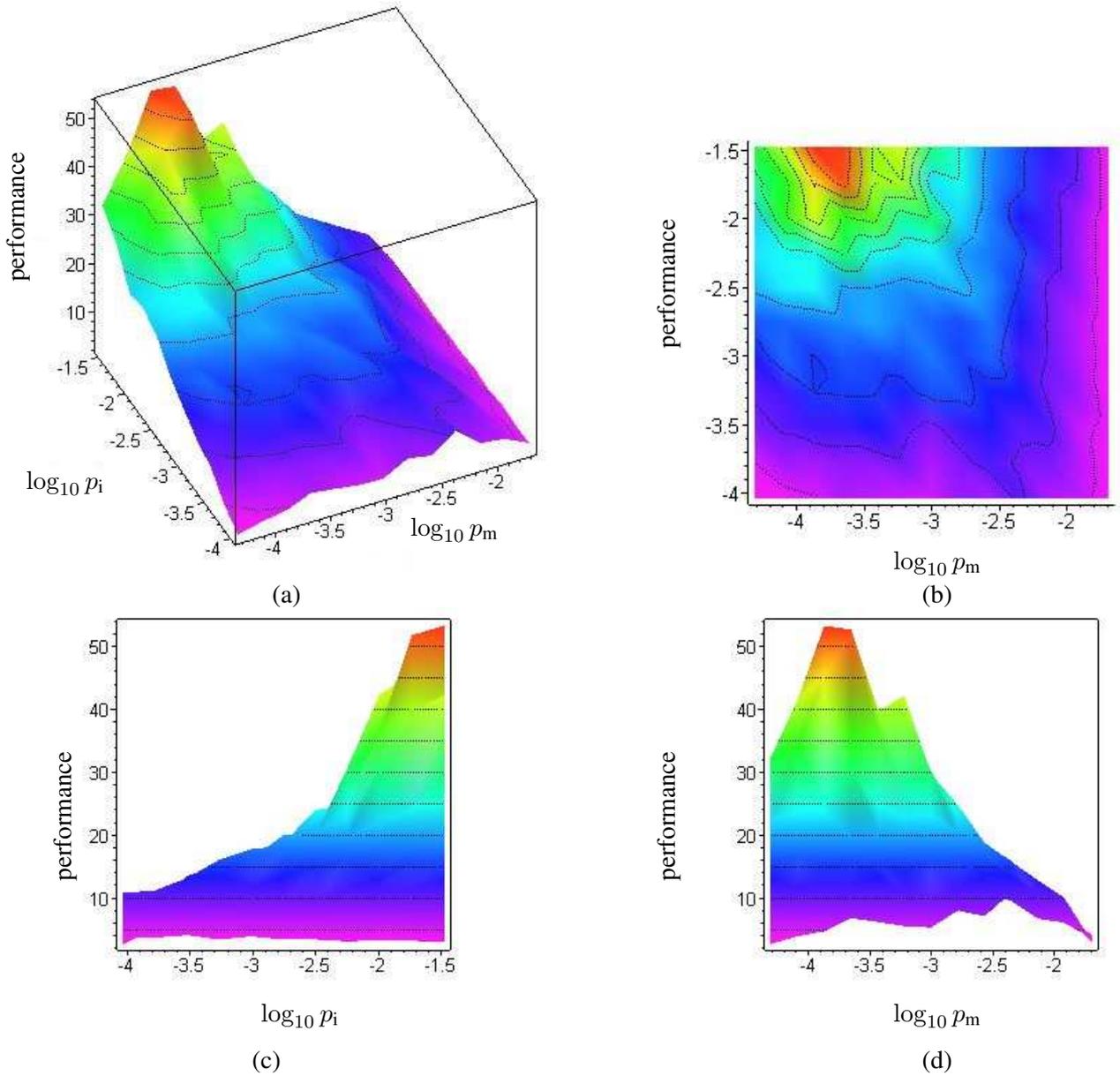
\begin{figure}[!ht]
\includegraphics[viewport=30 10 438 368,clip,scale=0.7]{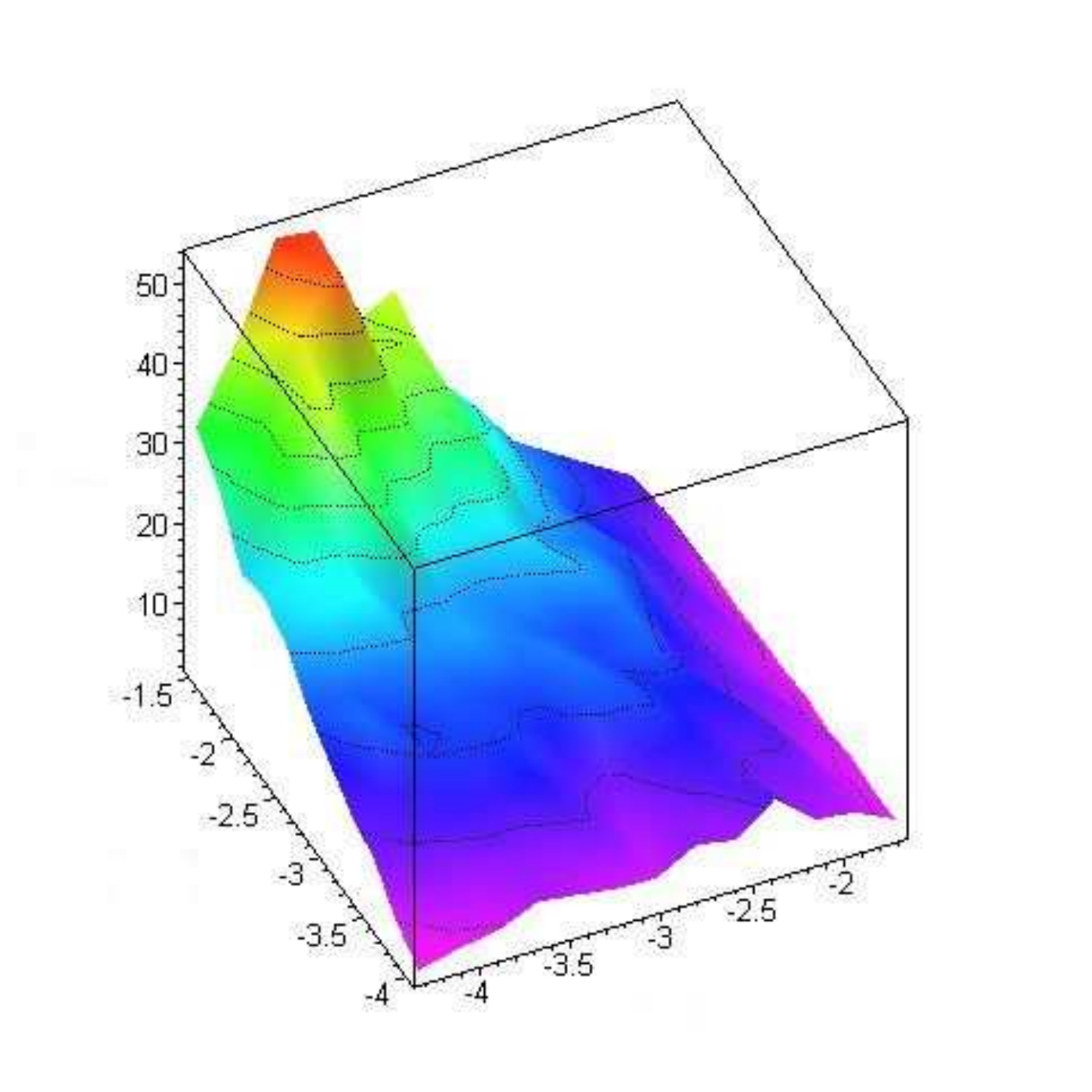}\hspace*{-4mm}
\includegraphics[viewport=66 53 318 348,clip,scale=0.77]{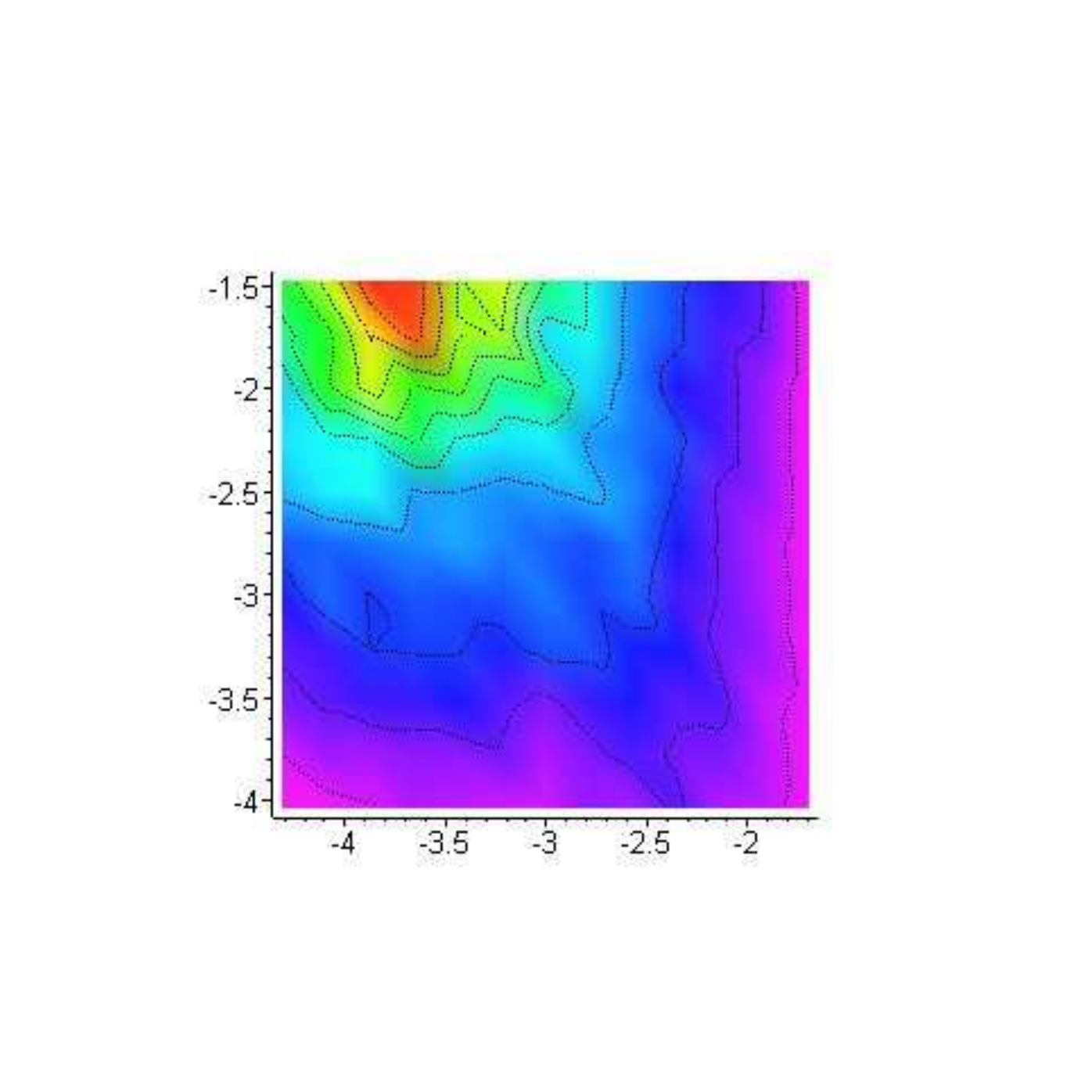}\\
\hspace*{2mm}
\includegraphics[viewport=60 60 468 318,clip,scale=0.7]{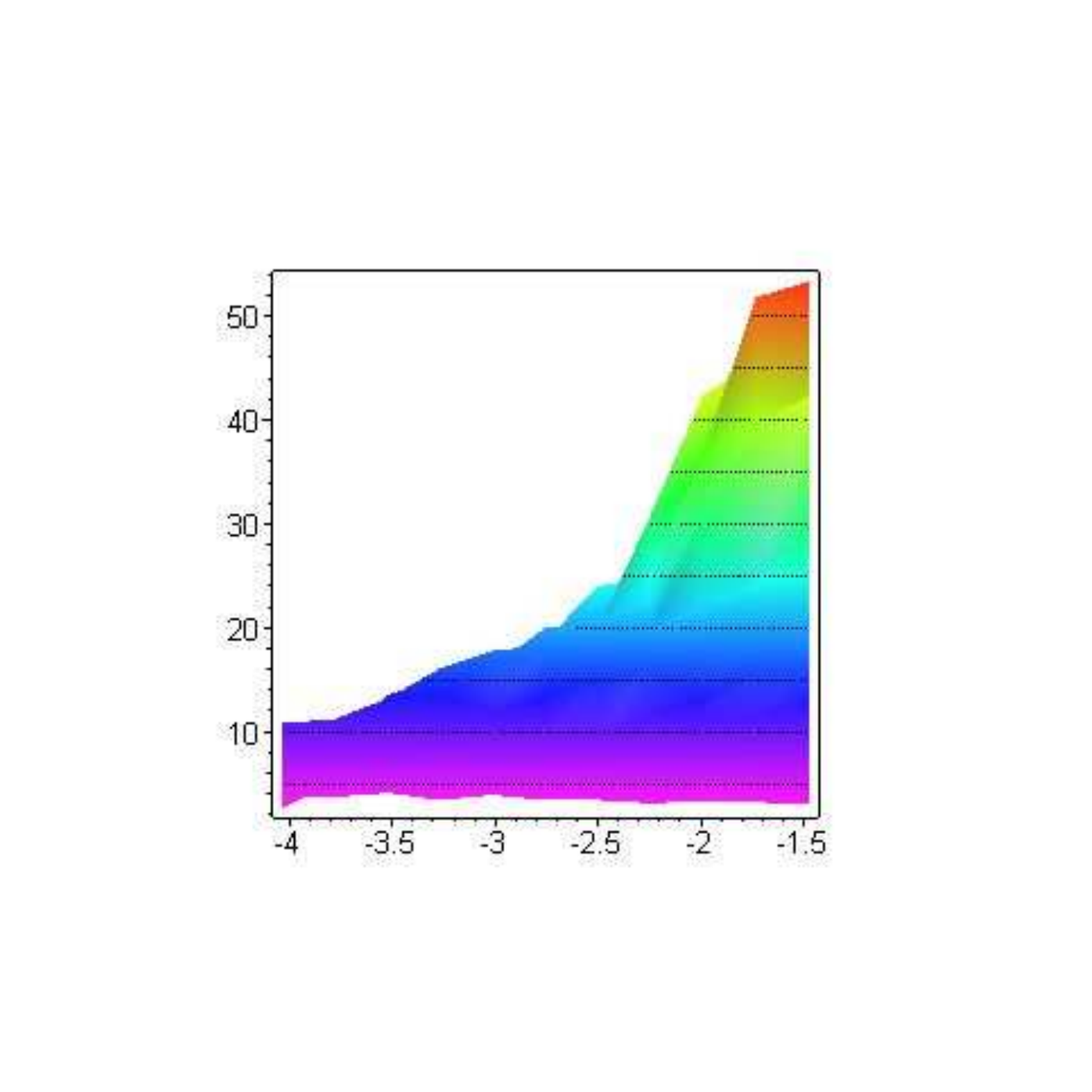}
\includegraphics[viewport=80 60 468 318,clip,scale=0.7]{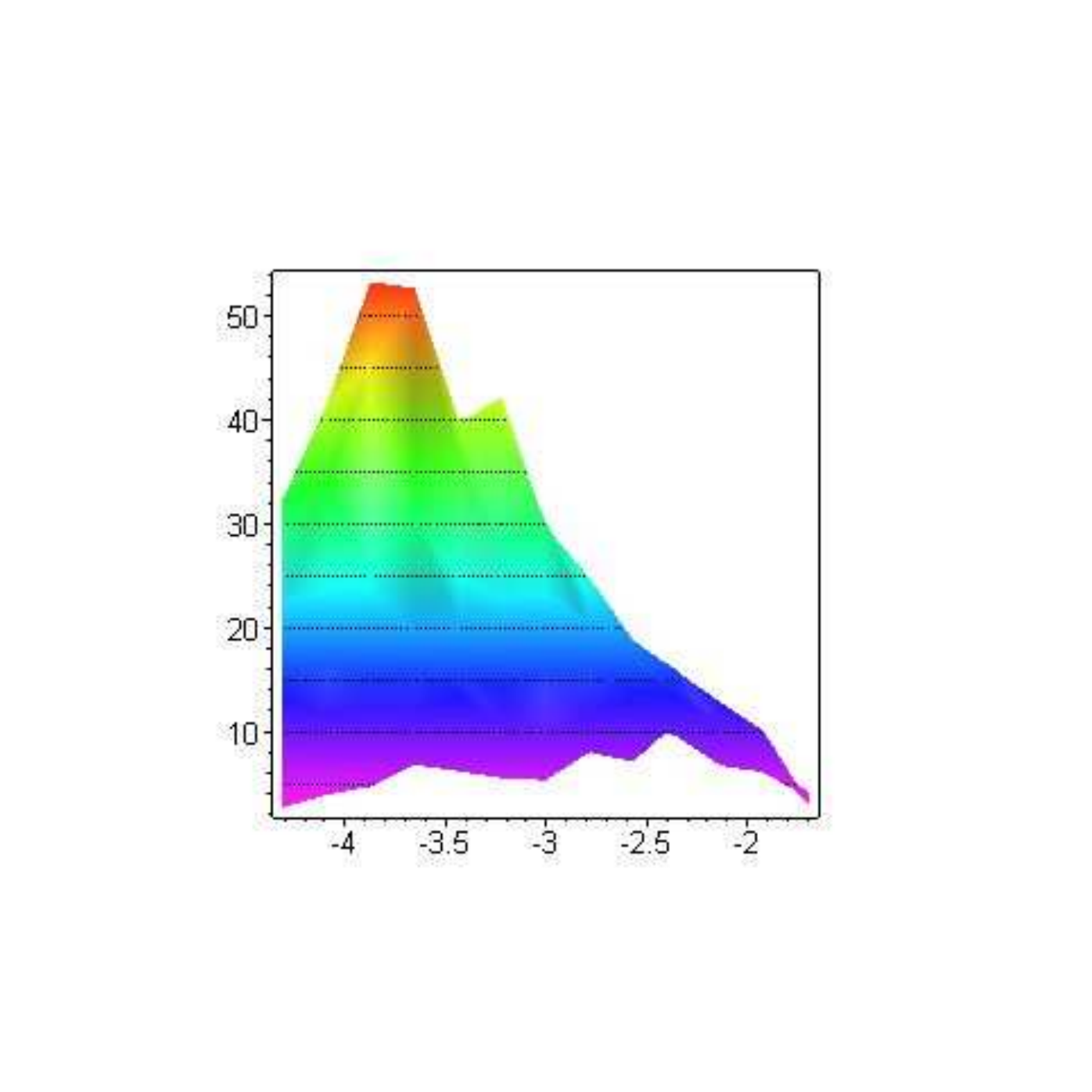}\\[2mm]
\begin{tikzpicture}
\useasboundingbox (0,0) rectangle (16,0.3);
\draw (6.3,8) node{$\log_{10} \pmm$};\draw (0.5,8.7) node {$\log_{10} \pii$}; 
\draw (13.5,0.7) node {$\log_{10} \pmm$}; \draw (4,0.7) node {$\log_{10} \pii$};
\draw (13.5,7.5) node {$\log_{10} \pmm$}; 
\begin{pgfrotateby}{\pgfdegree{90}}
\pgfputat{\pgfxy(3.7,-0.5)}{\pgfbox[center,center]{performance}}
\pgfputat{\pgfxy(12.7,0.2)}{\pgfbox[center,center]{performance}}
\pgfputat{\pgfxy(11.7,-9.5)}{\pgfbox[center,center]{performance}}
\pgfputat{\pgfxy(3.7,-9.5)}{\pgfbox[center,center]{performance}}
\end{pgfrotateby}
\draw (3.8,7) node {(a)}; \draw(13.5,7) node {(b)};\draw (3.5,0.) node {(c)};\draw (13.5,0.) node {(d)};
\end{tikzpicture}
\caption{\label{plotpi}For the goal $\pi$, in (a) it is shows the 3D plot of the best performance  
value in the population, averaged on the ten different seeds of the simulation, as function of the 
states-increase rate $\pii$ and of the mutation rate $\pmm$. The three orthogonal projections of (a) 
are also shown.}
\end{figure}

\begin{figure}[!h]
\hspace{22mm}\includegraphics[angle=90,scale=0.6]{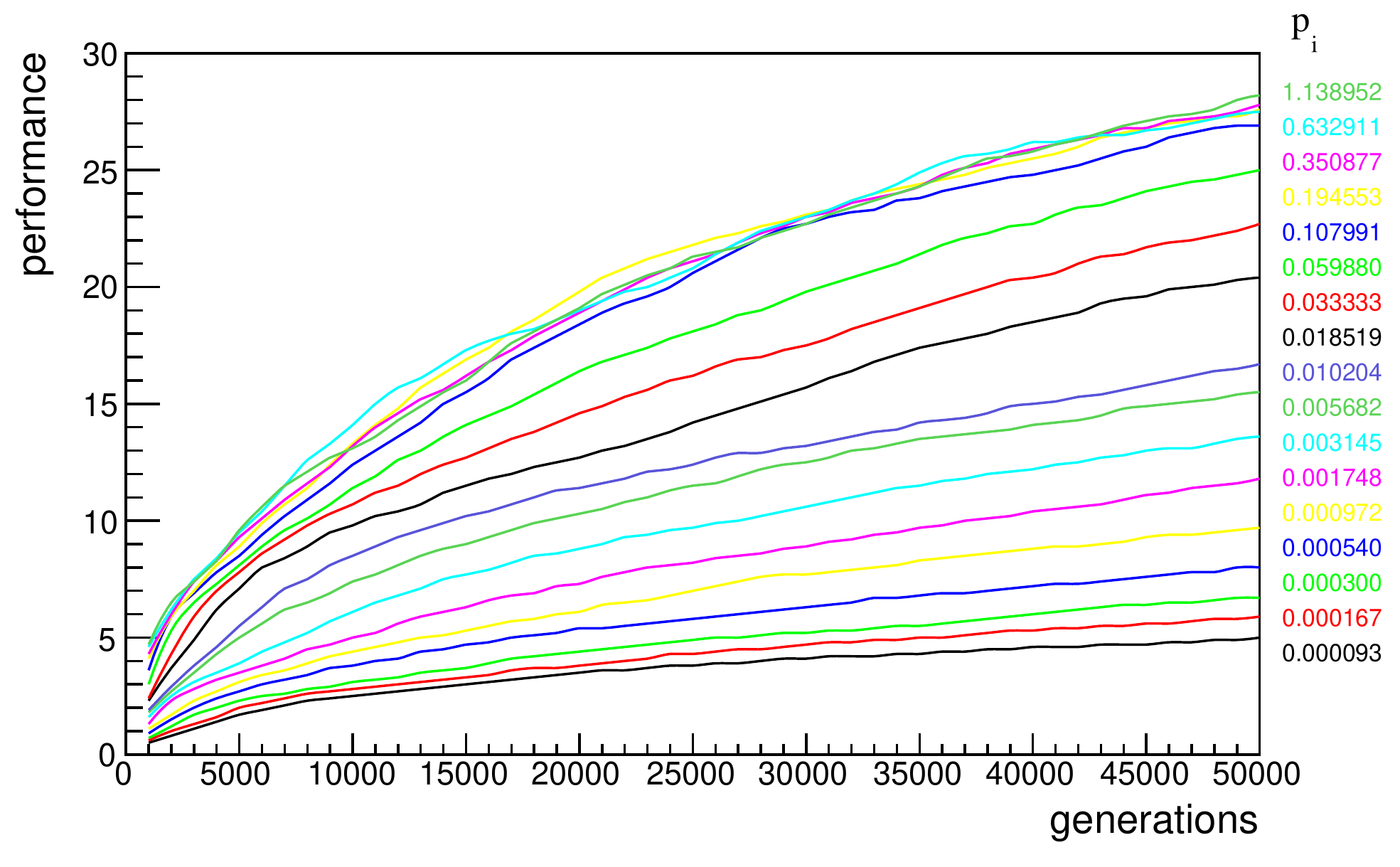}
\caption{\label{f:storia}The evolution of the performance with the generations, 
at various values of the state-increase rate. Notice the overlap of the four highest lines, possibly 
related to the presence of a plateau in the plot of the performance versus $\pii$. This interpretation
is reasonable but still uncertain and difficult to prove because of computational time.}
\end{figure}

In \cite{FM1} the dependence of the performance by the external parameters $\pmm,\pii$ was studied with 
both the output tapes indicated. 
The two series of simulations show very similar features. The most evident effect is that 
having a large amount of non-coding states speeds up evolution and allows to reach larger values of the 
population performance, as in Figure~\ref{plotpi}. 
It is important to remember that when new states are added by the states-increase process, 
they are and remain non-coding until activation by point mutation. 
Of course, the model has a bias toward the growth of the number of states, because no deletion is 
introduced and no cost for large genomes is used. This bias is on the total number of states, not
on the actual number of coding triplets. The latter is not biased, it can both increase and decrease.
This bias
\begin{enumerate}
\item \textbf{implies} that the total number of states $\nt$ cannot decrease but 
\item \textbf{does not imply}
that the performance grows faster if $\nt$ is large. For this reason, there is no need to
add deletion or metabolic costs for large genomes.  
\end{enumerate}
The total number of triplets is approximately 
\be
\nt \approx 2(1+ ngen \cdot \pii) 
\ee
If $\nc$ is the number of coding triplets, the ratio $\nc/\nt$ has been measured and it is of the order of 
few percent, often less, so approximately $\nnc=\nt-\nc\approx \nt$ is the number of non-coding triplets. 
The ratio $\nc/\nt$ is observed to decrease with the growth of the performance. This enhances the effect
of ``reservoir'' of the non-coding triplets: simulations show that the performance grows faster if $\nnc$ 
is large, Figure~\ref{plotpi} (a) and (c), Figure~\ref{f:storia}. This means that the non-coding triplets 
are used to explore new strategies. 
While the phenomenon in itself is not totally unexpected, its amplitude and persistence surprises.  
Several simulations, in part not yet published, seem to indicate that 
the phenomenon continues at higher $\pii$ with a ratio $\nc/\nt< 0.5$ namely with an enormous 
excess of non-coding versus coding triplets. 
It is important to stress that $\nt$ is positively selected, namely it is larger that the value attained
in the absence of selection (random evolution). Also, $\nnc$ is larger than in the random choice case. 
As the bias is present with and without selection, the effect on $\nt$ is not produced by the bias.  
It is a real effect, indirectly produced by selection. It is indirect because the algorithms of 
selection do not act on $\nt$.

Besides numerical investigations, the model allows one to perform some analytical evaluations. 
The spirit of the papers \cite{FM2, FM3} has been precisely to develop a mathematical description of 
the mutation-selection dynamics and complete it with numerical data. 
The mutation-selection dynamics is the set of rules that are used by the Turing machines evolving
population during simulations. They can be treated mathematically, thus showing the presence of 
an error threshold. This is a value of the mutation probability $\pmm^{\star}$ 
such that if $\pmm> \pmm^{\star}$ the highest performance class degrades faster than it is generated;
said otherwise, the occupation number of the highest performance class reduces to zero in such a 
way that the population performance decreases. Degradation is due to harmful mutations, that are
very frequent events. Generation is due to rare good mutations and mainly to the selection mechanism,
that favours the replication of high performance individuals. 
Moreover, the population evolves toward the error threshold; this means that, granted a sufficiently
large number of generations, the population will occupy all the performance classes up to the 
error threshold. 

\begin{figure}[!h]
\hspace*{22mm}\includegraphics[scale=0.6]{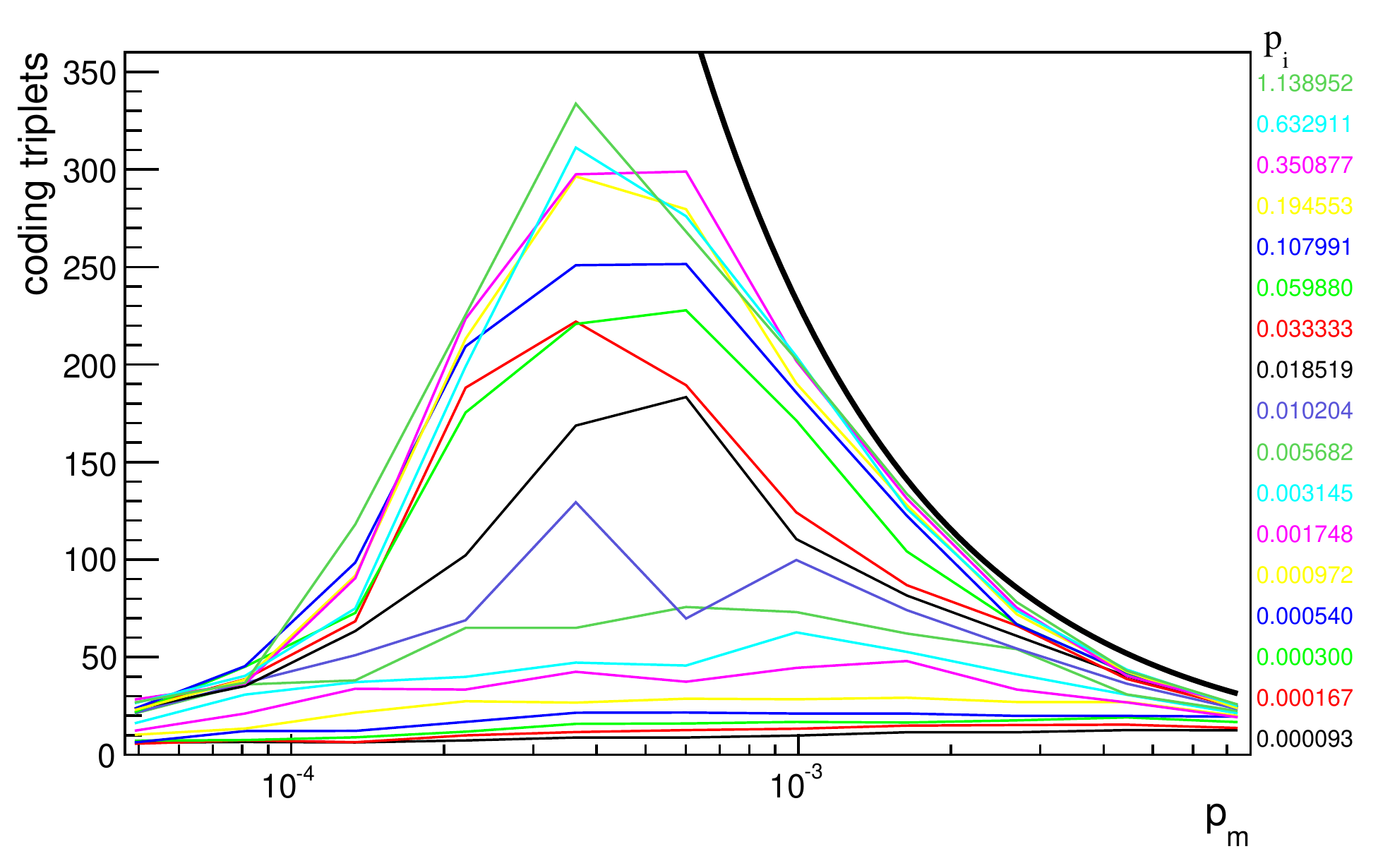}
\caption{\label{f:ncodif}The average number of coding triplets $\langle\nc\rangle$ at the end of 
simulation ($ngen=50000$) is  shown as  a function of $\pmm$, for all the values $\pii$. The value $\nc$
is taken for the best performing machine in the population and is averaged on the seeds. The black thick
line on the right represents the critical number of coding triplets $\nc^{\star}$, as function of 
$\pmm^{\star}$, extracted from (\ref{errort}).}
\end{figure}

The error threshold evaluated in \cite{FM3} is
\be\label{errort}
\pmm^{\star}=1-2^{-\frac{1}{3\nc^{\star}}}
\ee
Both of these effects are shown in Figure~\ref{f:ncodif}. The black thick line represents
the relation between the critical number of coding triplets $\nc^{\star}$ and the 
error threshold $\pmm^{\star}$. It is never crossed by the averaged population data. 
Clearly, the error threshold equation expresses quite a general feature of the systems evolving by 
random mutation and performance based selection: the effect of random mutations is to put an 
upper bound to the size of the genome. The only way to escape this fate is to reduce the effect of 
mutations by using reparation mechanisms, optimization, small coding part in a large non-coding genome...

Other effects studied with the Turing machine model include the extinction time of the machines and the 
evolutive effect of punctuated equilibria.

\section{Discussion}
The Turing machines model has been introduced by myself and F. Musso in 2007. 
From the first paper \cite{FM1} on, the model has been used for mathematical evaluations and several
numerical investigations \cite{FM2, FM3, chevalier}. One of the next challenges will be to introduce
recombination and study the evolution and maintenance of sexual reproduction, a very general 
reproductive form in nature that is still theoretically poorly understood. Work in this 
direction has been carried on during the internship of my master student \cite{chevalier}.

At the very beginning of this Turing machines program, it wasn't clear to me if the model was just 
a personal exercise or if it could be of some use. The positive answer came later, by comparing 
with other evolutionary models and also by appreciating that the TM model could be developed 
in several directions, for example the one already cited of sexual reproduction.

There are not many other evolutionary models designed to study Darwinian evolution, the most famous one
being Avida \cite{Avida} with its ancestor Tierra \cite{Tierra}. They both are elegant and complete 
platforms that create an in silico life. Organisms are programs that live on computer grids and 
compete for resources: CPU time and memory.
The simplicity is not the feature of these models, neither is the parsimony in terms of computer
resources. 

A model that is much closer to TM has been developed by some members of the Institute of complex 
systems in Lyon \cite{aevol}. Their model ``Aevol'' has both a genome and a proteome therefore 
it implements the transcription/translation mechanism. 

In common, all these models have the idea that the genome is an algorithm that is created by random 
mutations and is selected on the basis of its performance, as in Figure~\ref{insilico}, measured
by the ability of realizing some task: self-replication in the Avida platform, evaluation of 
some complex mathematical function in Aevol and Turing.
Of course, they all have an incomparably lesser degree of complexity than living organisms. 
More importantly, they do not try to describe the DNA by some ``close'' description
or the protein functions by some catalytic process. Is this a real difficulty?
Probably it would be of interest to describe a ``realistic'' genome, with four bases, the genetic code,
the mRNA and amino acid sequences. One could try to simulate the evolution of very basic enzymatic 
functions in bacteria. To do that, one should find the ``functional site'' of a protein from its
amino acid sequence, with one of the prediction tools that are known to work. 
In perspective, this project offers an interesting development but, so far, it has not been realized.
Instead, all the cited models claim that a certain algorithm can simulate a genome; they
claim that a mathematical function can represent the phenotype and work with these simplifying 
hypotheses. 

Given that, the problem of knowing if realistic results can emerge from non-realistic models is 
conceptually extremely important. I think the answer that all these authors implicitly assume is very 
deep and smart: 
no matter the details of the model, results are universal if universal hypotheses have been formulated.
``Universal'' is employed in the sense of Kadanoff and Wilson, as it is used in statistical physics and 
in the renormalization group. 
Therefore, having different models is important because the comparison of their features and
predictions leads to understand which are the universal features of evolution.

\chapter{Protein assembly}
A large number of proteins become biologically functional only after association of a  
number of amino acid chains. The ensuing structures are called oligomers. In addition to folding, the
oligomers need to assemble, which takes place through the formation of interface areas
that are mutually interacting.  
From \cite{goodsell}, one learns that 20\% of the proteins in Escherichia coli are monomeric, the rest 
oligomeric/polymeric, with a clear preference for dimers (38\%) and tetramers (21\%).
Amongst oligomers, the large majority is homooligomeric. Few are true polymers. In general, 
polymers differ from oligomers by their variable stoichiometric number, that can take values
of hundred to millions of subunits. 
In the Protein Data Bank (PDB), 20\% only of the recorded proteins are oligomers.
It has been noticed, however, that the Protein Data Bank (PDB) over-represents small monomers,   
because of the difficulties in protein crystallization, thus the value of 20\% is underestimated.
Given all these data, the importance of investigations of oligomeric proteins is apparent.

Folding and assembly are two processes that occur in oligomeric proteins after ribosomal synthesis.
It is believed that in most cases at least a partial folding is required before the assembly 
can start. The reason for this is that assembly requires the encounter of at least two parts that are 
in solution in the cell; this process is diffusion limited and can be quite lengthy. In spite of this,
it is known that sometimes a ``fly-casting'' mechanism takes place in which assembly comes very early 
and the several subunits fold together only after assembly. 
Thus, the two processes cannot be considered as separated and independent from each other. 
Moreover, in the 
first case, it is reasonable to expect at least a partial rearrangement of the structure after assembly.

Even if the microscopic description of folding and assembly undergoes the principles of 
molecular dynamics-molecular mechanics approaches, it would be of great value to obtain a more
macroscopic understanding. Some of the relevant questions are now indicated.
What does differentiate two amino acid sequences, one incapable of association, the other capable? 
Namely, given an unknown sequence, can one predict if it will give rise to a monomer or to an oligomer?
Is it possible to predict, from the sequence, which amino acids will constitute the interface? 
What will be the interface three-dimensional form? 

These questions are not so different from those of protein folding. For example, given the sequence,
it is possible to do secondary structure predictions because it has been shown that
certain groups of few amino acids have particular propensities to one or other possible
secondary structure. For example, this has lead to formulate the Chou-Fasman rule. 
These predictions, however, are not free of errors. 

In a similar fashion, it is reasonable to imagine that groups of amino acids or perhaps certain 
secondary elements have a propensity to form or not interfaces. Or else, there is a propensity for 
interfaces of a specific geometric form. 
Are there propensities for a preferred association mechanism? 
Notice that examples are known where sequences with 90\% of identity follow different association 
patterns. This means that few key amino acids can actually decide the association mechanisms
and, why not, the folding itself.
These and other questions motivate the present studies on oligomeric proteins. 

Important is to focus on the interfaces. In an oligomeric protein the interface has a high degree
of specificity and is very stable. Indeed, the mutual recognition of the two sides of the interface 
is extremely accurate. Early it was recognised that this happens if the interface 
is made of many weak ``contacts'' \cite{crane}, namely hydrogen bonds, and if the geometric and 
chemical arrangements of atoms on the 
two parts are complementary \cite{chothia}. Indeed, strong contacts, as in a ionic bond,
would be able to attract and fix 
several different molecules so they would be non specific.
Absence of complementarity would increase the interatomic distance thus reducing the strength of 
the contacts and possibly creating space for spurious molecules of water. 

The project that I will detail in the next sections focuses on the interfaces of trimeric and higher 
stoichiometry proteins. Using experimental approaches, it has been observed that few residues, 
located on the interface of a protein oligomer, are crucial for its assembly. Some of them control 
the formation of interfaces (association 
steps) while others control the stability of the oligomer (maintenance of the associated state) 
\cite{luke}. These key residues are not necessarily conserved among proteins of identical function 
or even of similar fold \cite{ngling}. 
This could mean that the few residues dedicated to protein assembly would have to be identified 
experimentally, for each particular case. Alternatively, a theoretical approach could reveal
which features characterize interfaces, by a systematic investigation of the
known three dimensional structures of protein interfaces. There are about 4000 cases deposited on the
PDB data bank, from the trimeric to the dodecameric stoichiometry.
The aim is to identify key residues involved in the different steps of the protein assembly
and possibly to derive some of the basic principles that manage protein assembly.

To this purpose, I have created a series of programs (Gemini) that sort out the protein interfaces 
and describe them 
as interaction networks (graphs). The interface structure is thus efficiently coded into 
graphs that allow to identify (or at least propose) the chemical links responsible for the 
interface's formation. At present, 3000 cases have been screened. The programs have been 
successfully tested on known protein interfaces.

\section{Gemini}
It is a series of programs and database utilities that have been created under the common name of 
Gemini to investigate properties of the interfaces of oligomers: presently, the most important are
GeminiDistances, GeminiRegions, GeminiGraph and GeminiData \cite{FL1}.

These programs come from the need to make systematic the analysis of oligomeric interfaces in 
three-dimensional protein structures. The main criterion followed has been to propose a framework of 
the amino acid interactions involved in an interface so their role in providing the interface its 
specificity and in regulating the mechanism of assembly can be addressed, for example by comparing 
protein interfaces of similar geometry.
The objective is to find all pairs of atoms (one atom per chain) located at distances small enough 
for intermolecular interaction, and to reduce this set of interaction pairs to a minimum: the 
smallest set that still describes the protein interface.

\subsection{GeminiDistances}
This program has the main goal to recognize the interface between two adjacent chains M and M+1 
in an oligomeric protein from its 3D structure. 

A first screening is done on the backbone $\alpha$ carbons of adjacent chains: all pairs of amino acids 
(one per chain), whose C$\alpha$ are separated by a distance lower than a given cut-off, 
fixed to cut1=20 \AA, are retained for the next step, the others are discarded. This has 
the unique goal 
to speed up the calculation and is legitimated by the observation that the maximal amino acid 
theoretical length is about 8 \AA. With smaller distance cuts-off (e.g. 10 \AA), some of 
the amino acids of the interfaces were missed. 

In the second screening all the atoms of the amino acids previously retained are examined and the pairs 
at distance lower than cut2=5 \AA\ are kept to form the so-called \textit{raw interface}. 
This $5$ \AA\ distance covers the range of distances that corresponds to weak chemical bonds involved 
in interfaces: Van der Waals, electrostatics, hydrogen bonds. Notice that these cuts-off can be 
freely modified. The presence of the second cut-off makes the raw interface de facto independent 
of the first one: values of cut1 of 17, 20, 25 \AA\  and higher give identical results.

The raw interface is a long list of pairs of atoms that may form chemical bonds. For example, the 
interface of the heptamer co-chaperon 10, produced by Mycobacter tuberculosis (PDB code: 1HX5), 
has 328 pairs of atoms selected in the raw interface. Because the aim of GeminiDistances is to propose a
framework with a minimum of interactions, it is necessary to add another constraint to deselect a 
maximum number of pairs. The deselection is performed by a 
\textit{symmetrization} procedure which only retains a single interaction per atom, the one involving
the closest partner, even for atoms having more than one partner on the adjacent chain. Precisely, 
for each atom of M, in the raw interface, only the closest atom on M+1 is retained, yielding a set of 
pairs $L1$. Similarly, for each atom of M+1, in the raw interface, only the closest one on M is 
retained to form a second set of pairs $L2$. The pairs common to both lists, $L1 \cap L2$,
form the interface used for the investigations of this paper, also called \textit{symmetrized interface}.
In other words, a pair of atoms $(i, j)$ is in the interface if both $i$ is the closest to $j$ 
and $j$ is the closest to $i$. 

The symmetrization makes the symmetrized interface almost cut-off independent. Indeed, values in the 
range cut2=4.5 to 6 \AA\ have been explored. In the former case, some interactions are lost and 
the raw interface forms a subset of the raw interface obtained with cut2=5 \AA. Vice versa, in the 
latter case the raw interface is bigger. After symmetrization, one observes remarkably small 
variations: in average, they do not exceed 10\% of the interface in the indicated range for cut2. 
Variations are even smaller if only amino acids and not atoms are considered.

It is important to keep in mind that the symmetrization discards many atoms at distances for which 
a chemical interaction is plausible. Therefore, the output generated by GeminiDistances may miss 
atoms, that will be called \textit{false negative}. It may also select atoms which are not 
chemically the most plausible, indicated as \textit{false positive}. But the selection of the most 
chemically plausible interactions is a more 
difficult task than the geometrical selection performed by GeminiDistances. A more chemical 
selection would be necessarily slower and might not necessarily be more accurate. Such a method may 
be better for a case-to-case study, but the symmetrization is more appropriate for comparing the 
interfaces of many oligomers. 

For example, from the 328 pairs of atoms selected for the raw interface of 1HX5, only 18 pairs 
remain after symmetrization.  In a more coarse grained interpretation, the atoms 
of the symmetrized interface are replaced by the amino acids they belong to. This \textit{amino acids 
interface} is used by the next program GeminiRegions.
 
GeminiDistances is written in C and runs in less than 0.2 s for an average size protein, 
on a normal desktop computer.

\subsection{GeminiRegions, GeminiGraph}
This program separates the amino acids interface, given by GeminiDistances, into regions, 
understood as elementary interaction networks between the amino acids of two adjacent chains. 
Many criteria can be used: the one adopted so far considers that amino acids in a region must be 
``close'' along the sequence, in addition to be close in space as considered in the construction of 
the interface itself.
Another interesting criterion that is implemented in Gemini is based on connected components 
in graph theory. According to it, atoms are grouped if there are paths that connect them by steps 
shorter than a given distance. This criterion ignores the sequential nature of the proteins.

This C++ program runs in the infinitesimal time of 2 ms per protein.

\begin{figure}\begin{center}
\includegraphics[height=5cm]{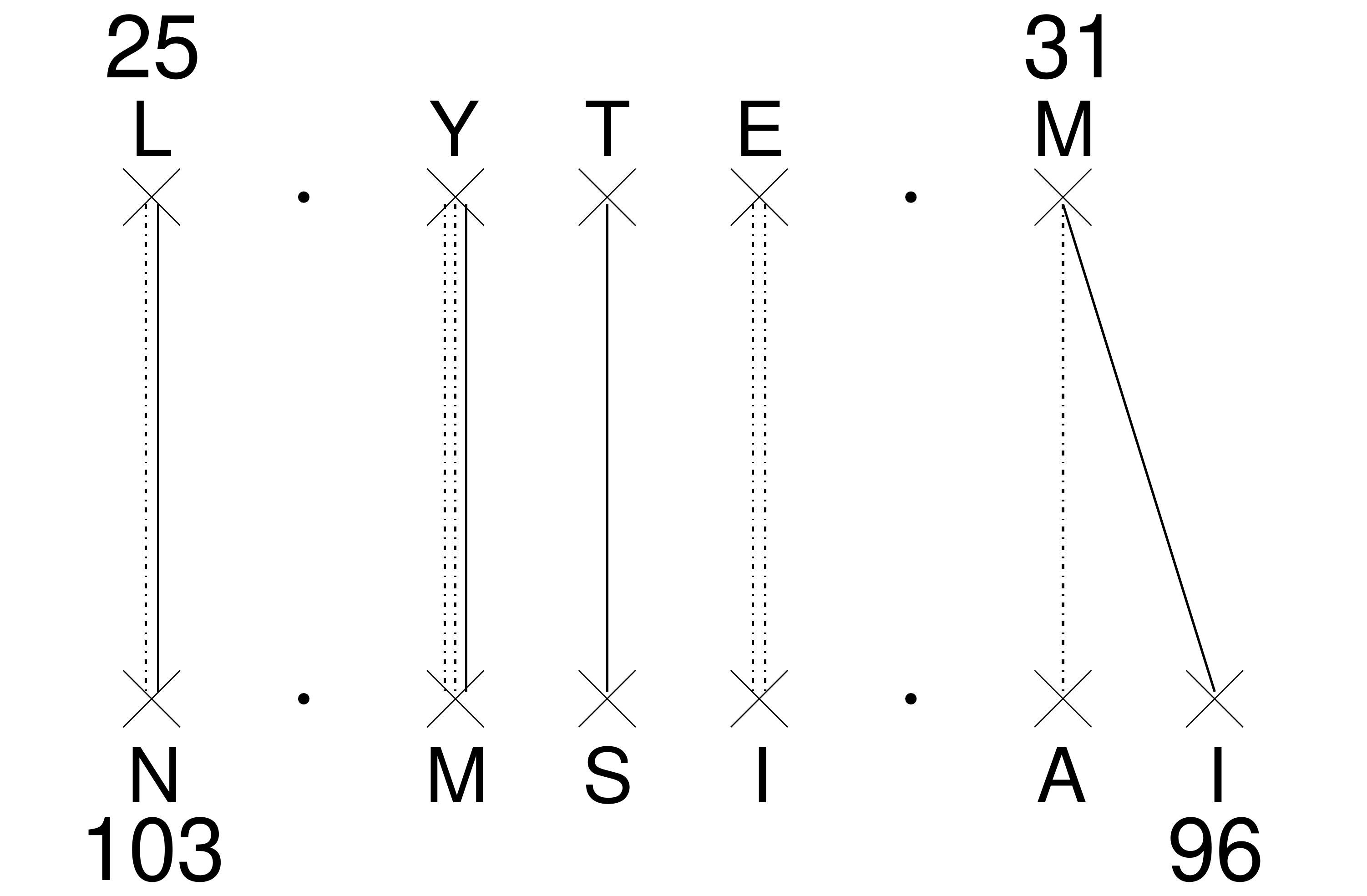}
\end{center}
\caption{\label{1efi-1-10}The largest region of the pentamer 1EFI (subunit B of the heat-labile 
enterotoxin of Escherichia coli) is represented here. The ladder structure formed by backbone-backbone 
interactions is present in most of the interfaces formed by the alignment of two parallel or 
antiparallel $\beta$ strands.}
\end{figure}

\begin{figure}[t]\begin{center}\vspace*{-7mm}
\includegraphics[height=7cm]{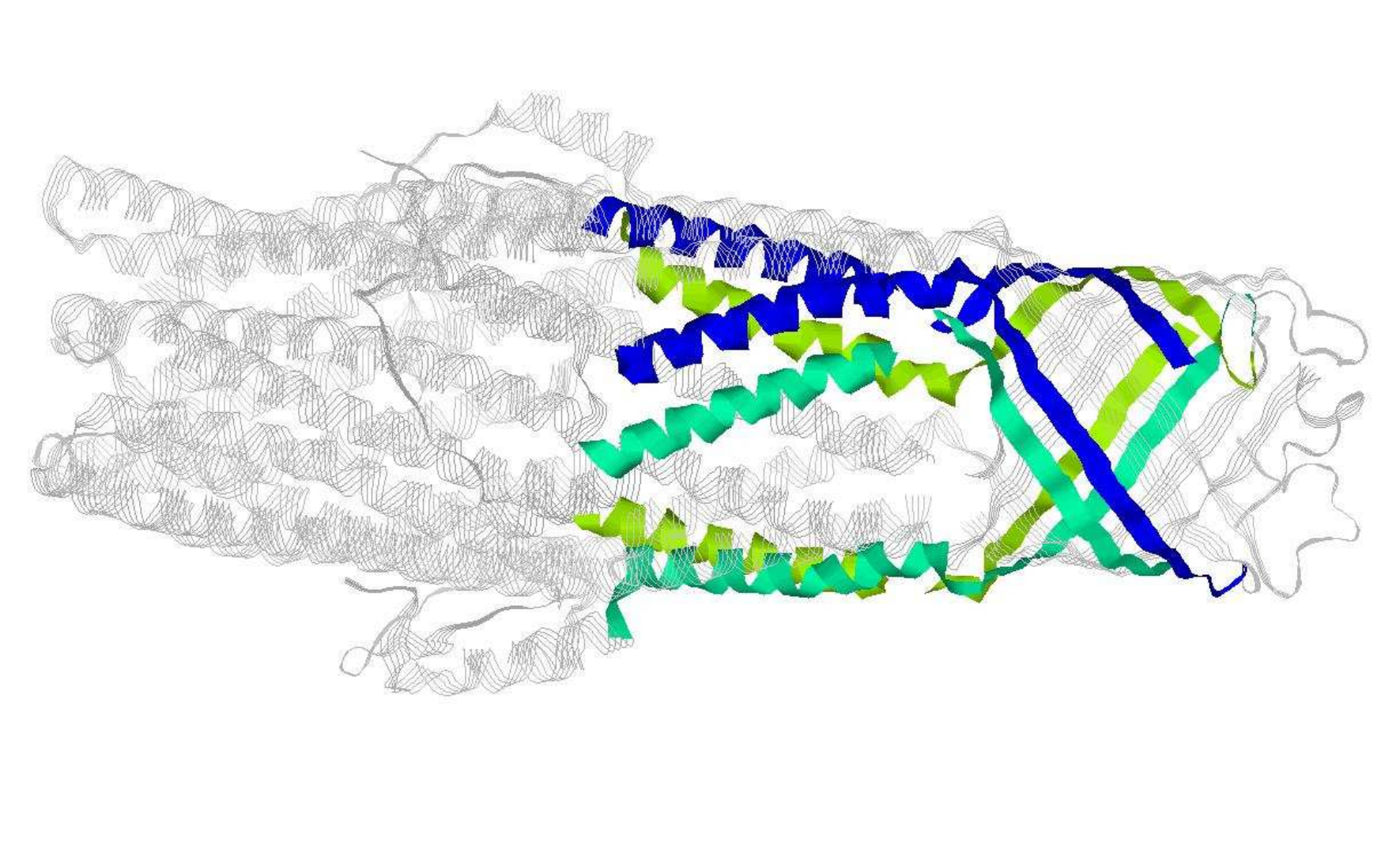}\\[-7mm]
\includegraphics[height=5cm]{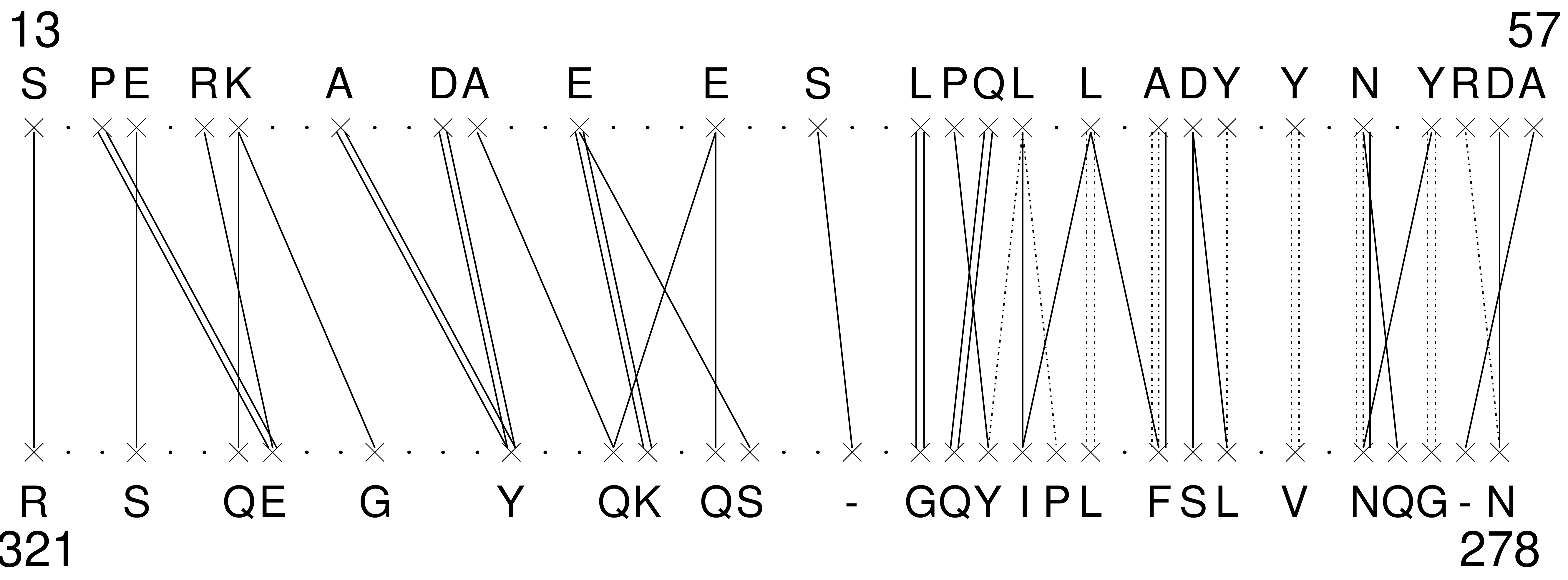}
\caption{\label{1ek9-1-48}The top image enlights the $\alpha$ and $\beta$ structures of the interface
of the trimeric membrane protein TolC (PDB: 1EK9) of Escherichia coli.
The bottom image contains the Gemini graph of the interface. 
The right part shows the same ladder structure of Figure~\ref{1efi-1-10}; the left part instead is characterized by more separated amino acids, with a distance along the sequence that oscillates 
between 3 and 4 residues, 
and the presence of structures like the letter ``V'', with angular separation of 3-4 residues. Indeed, 
this left part is an interface formed by two $\alpha$-helices. Actually, it is a big $\alpha$-coiled 
coil interface of 12 $\alpha$-helices winded up in a big helix.}
\end{center}
\end{figure}

By construction, a region, or an interaction network, contains the interactions expressed by the 
pairs of the amino acids that form the interface; this corresponds to the notion of graph. In 
mathematics, a graph is a set of vertices (here the amino acids) connected by a set of links 
(here the weak chemical bonds). 
Therefore, it is natural to introduce the following graphical representation, done 
with the program GeminiGraph.
Vertices are the amino acids; for reader's ease, those involved in a weak chemical bond are 
symbolised by a cross ``$\times$'' whereas those not involved in weak chemical bonds are 
symbolized by a dot ``$\cdot$''. Dashed-dotted lines indicate backbone-backbone interactions, 
solid lines indicate side chain-backbone or side chain-side chain. 
Amino acid type and number are indicated. See Figures~\ref{1efi-1-10},~\ref{1ek9-1-48}.

The interaction networks have been extensively compared with known cases in literature, observing a
good assessment of the amino acids involved in protein interfaces. The comparison shows that
Gemini detects rather accurately the amino acids geometrically and chemically involved in the 
interfaces. The chemical accuracy is particularly remarkable since the GeminiDistances selection is 
based on the geometry of the interface and no chemical selection is done.
This recalls Crick's concept that he formulated observing $\alpha$-coiled coil interfaces: the analysis 
of the geometry of a protein interface leads to its chemical specificity \cite{crick}.

\section{Developments}
There are basically three lines of development that emerge from the Gemini interaction networks. I 
will briefly present them.

Firstly, the interaction networks must be analysed and systematically compared, looking for 
patterns. Many parameters can matter. The polarity of the residues has been preliminarily studied in 
\cite{poster}. Some statistics on the length of the side chain and the differential use of the amino 
acids has been presented in \cite{risson} and, previously, in \cite{ivan}.
This research continues with the main focus on the interfaces formed by the 
alignment of two parallel or antiparallel $\beta$ strands. Nearly 60 representatives have been 
collected for this geometry that I will call $2\beta$.
Indeed, the ladder structure of Figure~\ref{1efi-1-10} is extremely frequently observed in this type 
of interface but is not observed in other interface geometries, thus it is a candidate to be a 
distinctive feature. 
This is a good example of the patterns that I would like to trace in the Gemini graphs: features that 
allow to distinguish geometries and to characterize their chemical properties. 
The preliminary analyses cited indicate that the amino acids are ``flexible'', they adapt to play 
different roles; this suggests that specific features will not be at the amino acid level but
possibly to the lower level of atomic groups in the side chain.
Moreover, the investigations point toward the joined structure, in which both the sides
of an interface are important. Patterns or elementary blocks must appear in an interface, not 
just in a sequence.

Secondly, interaction networks can be used to propose amino acid substitution and test the
effects with in vitro experiments. Moreover, the principles itself that I have adopted in designing 
the Gemini programs can be tested against these experiments.
Given a network of interactions, it is reasonable to expect that the effect of a mutation will be 
different according to which amino acid
is modified. In particular, little effect is expected on amino acids marked by a dot, big effect is
expected for amino acids with many connections.
The experimental part is of pertinence of C. Lesieur and is performed at the facilities of the BioPark
of Archamps.
One indication that has been found is that some interfaces are ``active'' even in the absence of 
the rest of the chain. This means that the subunits can recognize each other even when the rest of 
the chain has been removed. 
An opposite result would have indicated that the whole chain is always needed for the assembly,
thus showing a marginal role of the interface. 

Thirdly, one can imagine to simulate the process of association of two subunits. For this, a software 
is available, Simulation of Diffusional Association SDA \cite{sda}. This software implements the 
Langevin equation for Brownian motion and allows to trace Brownian trajectories of two 
molecules in water. Statistics on the trajectories produces the association rates, namely the
number of encounters per time, and the residence time. Clearly, these simulations can replace in vitro
experiments of association. 
The interest, for the Gemini team researches, is to test artificially created interfaces
and study their interactions. My student J. Zrimi has invested his internship in simulating
the association of different subunits of three proteins \cite{jihad, hystid}, leading 
to the confirmation of a role of the four histidine in the association of these proteins.

I am directly involved in the first and third of these projects, the second being based on experimental 
manipulations.

\section{Discussion}

The Annecy team groups the competences and the facilities to work on both the theoretical 
and the experimental aspects of protein assembly, to understand the mechanisms of assembly, the 
sequence-structure-interface relationship, and the structural determinants of the interface geometry. 
I have created the team and I am the main responsible of the theoretical part.
The most important result has been the creation of the programs Gemini \cite{gemini}, of which I am the 
main author (80\%). For the graphical part, I asked the collaboration of my internship student
\cite{mottin}.
The creation of these program was, more than just writing lines of code, the  search of the 
correct ideas to translate a three-dimensional all atoms information into a synthetic description of 
the most relevant interactions. This process lasted for more than one year. Of course, 
the principles implemented in Gemini have been largely discussed with C.~Lesieur and also with
L. Vuillon.
I have tutored several internship students. In particular, J. Zrimi has been here for a five months 
internship. He was student of Master 2 at the Master ``Production et Valorisation des Substances 
Naturelles et Biopolym\'eres'' of the Faculty of sciences and techniques of Marrakech.
He will come again to our laboratory for his PhD studies. 

A very important event, of which I am promoter and co-organizer, is the conference 
``Theoretical approaches for the genome and the proteins'', TAGp2010, that will take place in 
Annecy-Le-Vieux in October 2010. 
This conference follows two previous meetings, 2008, 2006, that were focused on the genome.

\renewcommand\bibname{Bibliography: Biophysics}
\addcontentsline{toc}{chapter}{Bibliography: Biophysics}

\end{document}